\documentclass{aastex}
\usepackage{emulateapj5}

\newcommand{\civ}{C\,{\sc iv}}
\newcommand{\ebv}{$E(B-V)$}
\newcommand{\etal}{{\it et\,al.}}
\newcommand{\ha}{H$\alpha$}
\newcommand{\hb}{H$\beta$}
\newcommand{\hi}{H\,{\sc i}}

\newcommand{\ho}{$H_0$}

\newcommand{\kms}{km\,s$^{-1}$}

\newcommand{\mgii}{Mg\,{\sc ii}}

\newcommand{\oii}{[O\,{\sc ii}]}
\newcommand{\oiii}{[O\,{\sc iii}]}

\newcommand{\alii}{Al\,{\sc ii}}
\newcommand{\AlII}{Al\,{\sc ii}\,$\lambda$1670}
\newcommand{\aliii}{Al\,{\sc iii}}
\newcommand{\AlIII}{Al\,{\sc iii}\,$\lambda\lambda$1854,1862}

\newcommand{\cii}{C\,{\sc ii}}

\newcommand{\ciii}{C\,{\sc iii}]}
\newcommand{\CIII}{C\,{\sc iii}]\,$\lambda$1908}
\newcommand{\CIV}{C\,{\sc iv}\,$\lambda\lambda$1548,1550}

\newcommand{\caii}{Ca\,{\sc ii}}
\newcommand{\CaH}{Ca\,{\sc ii}\,$\lambda$3969}
\newcommand{\CaK}{Ca\,{\sc ii}\,$\lambda$3934}

\newcommand{\coii}{Co\,{\sc ii}}
\newcommand{\crii}{Cr\,{\sc ii}}
\newcommand{\FeI}{Fe\,{\sc i}\,$\lambda$2484}
\newcommand{\fei}{Fe\,{\sc i}}
\newcommand{\feii}{Fe\,{\sc ii}}
\newcommand{\feiii}{Fe\,{\sc iii}}
\newcommand{\feiv}{Fe\,{\sc iv}}
\newcommand{\hei}{He\,{\sc i}}
\newcommand{\HeI}{He\,{\sc i}\,$\lambda$3188}
\newcommand{\HEI}{He\,{\sc i}\,$\lambda$3889}
\newcommand{\heii}{He\,{\sc ii}}

\newcommand{\lalala}{$\lambda$$\lambda$$\lambda$}
\newcommand{\lya}{Ly$\alpha$}

\newcommand{\mgi}{Mg\,{\sc i}}
\newcommand{\mgiii}{Mg\,{\sc iii}}

\newcommand{\MgI}{Mg\,{\sc i}\,$\lambda$2852}
\newcommand{\MgII}{Mg\,{\sc ii}\,$\lambda\lambda$2796,2803}
\newcommand{\mni}{Mn\,{\sc i}}
\newcommand{\mnii}{Mn\,{\sc ii}}
\newcommand{\nai}{Na\,{\sc i}}

\newcommand{\Nv}{N\,{\sc v}}

\newcommand{\neiii}{[Ne\,{\sc iii}]}
\newcommand{\NeIIIa}{[Ne\,{\sc iii}]\,$\lambda$3869}
\newcommand{\NeIIIb}{[Ne\,{\sc iii}]\,$\lambda$3969}

\newcommand{\NIii}{Ni\,{\sc ii}}
\newcommand{\oi}{O\,{\sc i}}
\newcommand{\OI}{O\,{\sc i}\,$\lambda$1302}
\newcommand{\OII}{[O\,{\sc ii}]\,$\lambda\lambda$3727,3729}
\newcommand{\Oiii}{O\,{\sc iii}}
\newcommand{\OIII}{O\,{\sc iii}\,$\lambda$3133}

\newcommand{\ovi}{O\,{\sc vi}}

\newcommand{\Si}{S\,{\sc i}}
\newcommand{\SIi}{Si\,{\sc i}}
\newcommand{\SIii}{Si\,{\sc ii}}

\newcommand{\SIiv}{Si\,{\sc iv}}

\newcommand{\SiIII}{Si\,{\sc iii}]\,$\lambda$1892}

\newcommand{\sio}{Si\,{\sc iv}/O\,{\sc iv}]}
\newcommand{\SiO}{Si\,{\sc iv}/O\,{\sc iv}]\,$\lambda$1400}
\newcommand{\znii}{Zn\,{\sc ii}}

\slugcomment{Accepted to ApJS March 1, 2002; to appear in August 2002,
volume 141, number 2; astro-ph/0203252}
\shorttitle{SDSS UNUSUAL BAL QUASARS}
\shortauthors{Hall et al.}

\begin{document}

\title{Unusual Broad Absorption Line Quasars from the Sloan Digital Sky Survey}

\author{
Patrick B. Hall\altaffilmark{1,2},
Scott F. Anderson\altaffilmark{3},
Michael A. Strauss\altaffilmark{1},
Donald G. York\altaffilmark{4,5},
Gordon T. Richards\altaffilmark{6},
Xiaohui Fan\altaffilmark{7},
G. R. Knapp\altaffilmark{1},
Donald P. Schneider\altaffilmark{6},
Daniel E. Vanden Berk\altaffilmark{8},
T. R. Geballe\altaffilmark{9},
Amanda E. Bauer\altaffilmark{10},
Robert H. Becker\altaffilmark{11,12},
Marc Davis\altaffilmark{13},
Hans-Walter Rix\altaffilmark{14},
R. C. Nichol\altaffilmark{15},
Neta A. Bahcall\altaffilmark{1},
J. Brinkmann\altaffilmark{16},
Robert Brunner\altaffilmark{17},
A. J. Connolly\altaffilmark{18},
Istv\'an Csabai\altaffilmark{19,20},
Mamoru Doi\altaffilmark{21},
Masataka Fukugita\altaffilmark{22},
James E. Gunn\altaffilmark{1},
Zoltan Haiman\altaffilmark{1},
Michael Harvanek\altaffilmark{16},
Timothy M. Heckman\altaffilmark{20},
G. S. Hennessy\altaffilmark{23},
Naohisa Inada\altaffilmark{22},
\v{Z}eljko Ivezi\'{c}\altaffilmark{1},
David Johnston\altaffilmark{4},
S. Kleinman\altaffilmark{16},
Julian H. Krolik\altaffilmark{20},
Jurek Krzesinski\altaffilmark{16,24},
Peter Z. Kunszt\altaffilmark{20},
D.Q. Lamb\altaffilmark{4},
Daniel C. Long\altaffilmark{16},
Robert H. Lupton\altaffilmark{1},
Gajus Miknaitis\altaffilmark{3},
Jeffrey A. Munn\altaffilmark{25},
Vijay K.  Narayanan\altaffilmark{1},
Eric Neilsen\altaffilmark{8},
P. R. Newman\altaffilmark{16},
Atsuko Nitta\altaffilmark{16},
Sadanori Okamura\altaffilmark{21},
Laura Pentericci\altaffilmark{14},
Jeffrey R. Pier\altaffilmark{25},
David J. Schlegel\altaffilmark{1},
S. Snedden\altaffilmark{16},
Alexander S. Szalay\altaffilmark{20},
Anirudda R. Thakar\altaffilmark{20},
Zlatan Tsvetanov\altaffilmark{20},
Richard L. White\altaffilmark{26},
Wei Zheng\altaffilmark{20}
}
\altaffiltext{1}{Princeton University Observatory, Princeton, NJ 08544}
\altaffiltext{2}{Pontificia Universidad Cat\'{o}lica de Chile, Departamento de Astronom\'{\i}a y Astrof\'{\i}sica, Facultad de F\'{\i}sica, Casilla 306, Santiago 22, Chile}
\altaffiltext{3}{University of Washington, Department of Astronomy, Box 351580, Seattle, WA 98195}
\altaffiltext{4}{The University of Chicago, Department of Astronomy and Astrophysics, 5640 S. Ellis Ave., Chicago, IL 60637}
\altaffiltext{5}{The University of Chicago, Enrico Fermi Institute, 5640 S. Ellis Ave., Chicago, IL 60637}
\altaffiltext{6}{Department of Astronomy and Astrophysics, The Pennsylvania State University, University Park, PA 16802}
\altaffiltext{7}{Institute for Advanced Study, Olden Lane, Princeton, NJ 08540}
\altaffiltext{8}{Fermi National Accelerator Laboratory, P.O. Box 500, Batavia, IL 60510}
\altaffiltext{9}{Gemini Observatory, 670 North A`ohoku Place, Hilo, HI 96720}
\altaffiltext{10}{Department of Physics, University of Cincinnati, Cincinnati, OH 45221}
\altaffiltext{11}{Physics Department, University of California, Davis, CA 95616}
\altaffiltext{12}{IGPP-LLNL, L-413, 7000 East Ave., Livermore, CA 94550}
\altaffiltext{13}{University of California at Berkeley, Depts. of Physics and Astronomy, 601 Campbell Hall, Berkeley, CA 94720}
\altaffiltext{14}{Max-Planck-Institut f\"{u}r Astronomie, K\"{o}nigstuhl 17, D-69117 Heidelberg, Germany}
\altaffiltext{15}{Dept. of Physics, Carnegie Mellon University, 5000 Forbes Ave., Pittsburgh, PA-15232}
\altaffiltext{16}{Apache Point Observatory, P.O. Box 59, Sunspot, NM 88349-0059}
\altaffiltext{17}{Department of Astronomy, California Institute of Technology, Pasadena, CA 91125}
\altaffiltext{18}{Department of Physics and Astronomy, University of Pittsburgh, Pittsburgh, PA 15260}
\altaffiltext{19}{Department of Physics, E\"otv\"os University, Budapest, Pf.\ 32, Hungary, H-1518}
\altaffiltext{20}{Dept. of Physics and Astronomy, The Johns Hopkins University, 3701 San Martin Drive, Baltimore, MD 21218}
\altaffiltext{21}{Department of Astronomy and Research Center for the Early Universe, School of Science, University of Tokyo, Hongo, Bunkyo, Tokyo, 113-0033, Japan}
\altaffiltext{22}{University of Tokyo, Institute for Cosmic Ray Research, Kashiwa, 2778582, Japan}
\altaffiltext{23}{U.S. Naval Observatory, 3450 Massachusetts Ave., NW, Washington, DC  20392-5420}
\altaffiltext{24}{Mt. Suhora Observatory, Cracow Pedagogical University, ul. Podchorazych 2, 30-084 Cracow, Poland}
\altaffiltext{25}{U.S. Naval Observatory, Flagstaff Station, P.O. Box 1149, Flagstaff, AZ  86002-1149}
\altaffiltext{26}{Space Telescope Science Institute, Baltimore, MD 21218}

\begin{abstract}
\small 

The Sloan Digital Sky Survey has confirmed the existence of populations
of broad absorption line (BAL) quasars with various unusual properties.
We present and discuss twenty-three such objects and consider the implications
of their wide range of properties for models of BAL outflows 
and quasars in general.
We have discovered one BAL quasar with a record number of absorption lines.
Two other similarly complex objects with many narrow troughs show broad
\mgii\ absorption extending longward of their systemic host galaxy redshifts.
This 
can be explained as absorption of an extended continuum
source by the rotation-dominated base of a disk wind.
Five other objects have absorption which removes an unprecedented $\sim$90\% 
of all flux shortward of \mgii. The absorption in one of 
them 
has varied across the ultraviolet 
with an amplitude and rate of change as great as ever seen.  
This same object may also show broad H$\beta$ absorption.
Numerous reddened BAL quasars have been found, including 
at least one reddened mini-BAL quasar with very strong \feii\ emission.  
The five reddest objects have continuum reddenings of \ebv\,$\simeq0.5$,
and in two of them we find strong evidence
that the reddening curve is even steeper than that of the SMC.
We have found at least one object with absorption from \feiii\ but not \feii.
This may be due to a 
high column density of
moderately high-ionization gas, but the \feiii\ level populations 
must also be affected by some sort of resonance.
Finally, we have found two luminous, probably reddened high-redshift objects
which may be BAL quasars whose troughs partially cover different regions of the
continuum source as a function of velocity.

\end{abstract}

\small 

\section{Introduction}  \label{INTRO}

The Sloan Digital Sky Survey \markcite{yor00}(SDSS; {York} {et~al.} 2000) is using a drift-scanning
imaging camera and two multi-fiber double spectrographs on a dedicated 2.5m
telescope \markcite{gun98}({Gunn} {et~al.} 1998) to image 10,000\sq\arcdeg\ of sky on the SDSS
$ugriz$ AB magnitude system \markcite{fuk96,sdss85}({Fukugita} {et~al.} 1996; {Stoughton} {et~al.} 2002).  Spectra will be obtained for
$\sim$10$^6$ galaxies to $r=17.8$ and $\sim$10$^5$ quasars to $i=19.1$
($i=20.2$ for $z>3$ candidates). 

Quasar candidates are targeted for spectroscopy based on color criteria
\markcite{sdssqtarget}({Richards} {et~al.} 2002) or because they are unresolved objects
with radio emission detected by the FIRST survey \markcite{bwh95}({Becker}, {White}, \& {Helfand} 1995).
The essence of the color selection is simple:  target objects whose broad-band
colors are different from those of normal stars and galaxies, especially
objects with colors similar to those expected for known and simulated quasars
\markcite{sdssqtarget}({Richards} {et~al.} 2002).  Objects in the few untargeted outlying regions of color
space, plus FIRST sources with either resolved or unresolved counterparts, can
also be selected 
to $i=20.5$ 
as {\em serendipity} targets \markcite{sdss85}({Stoughton} {et~al.} 2002).
Due to these inclusive criteria, the selection of candidates using $i$ band
magnitudes rather than blue magnitudes which are more affected by
absorption and reddening, 
and its area and depth, the SDSS is proving effective at finding 
unusual quasars.  

Many of these unusual quasars are unusual broad absorption line (BAL) quasars.
BAL quasars show absorption from gas with blueshifted
outflow velocities of typically 0.1$c$ \markcite{wea91}({Weymann} {et~al.} 1991).  About 10\% of quasars
exhibit BAL troughs, but this is usually attributed to an orientation effect.
Most quasars probably have BAL outflows covering $\sim$10\% of the sky as seen
from the quasar, with mass loss rates possibly comparable to the accretion
rates required to power the quasar ($\sim1$\, $M_{\odot}$\,yr$^{-1}$).  
Therefore an understanding
of BAL outflows is required for an understanding of quasars as a whole.
Unusual BAL quasars may help in this endeavor since they delineate the full
range of 
parameter space spanned by BAL outflows.
Examples of unusual BAL quasars with extremely strong or complex absorption
whose nature is difficult to discern at first glance have been found in the
past few years through followup of FIRST radio sources \markcite{bec97,bea00}({Becker} {et~al.} 1997, 2000) and
$z>4$ quasar candidates from the Digitized Palomar Sky Survey \markcite{djo01b}({Djorgovski} {et~al.} 2001). 

In this paper we show that the SDSS has confirmed that these are not
unique objects, but members of populations of unusual BAL quasars.
These populations include extremely reddened objects 
and objects with unprecedentedly strong absorption, 
absorption from record numbers of species, 
absorption from unusual species or with unusual line ratios,
or absorption which extends longward of the systemic redshift.
We begin in \S\ref{BKGD} with an overview of our current knowledge
of BAL quasars, to put in context the implications of these unusual objects.
We discuss our sample of SDSS quasars in \S\ref{DATA}
and our selection of a sample of unusual BAL quasars from it in \S\ref{SELECT}.
In \S\ref{ANALYSIS} we present the different categories of unusual BAL quasars
we have identified.
In \S\ref{DISCUSSION} we discuss the implications of these objects
for models of BAL outflows.
We summarize our conclusions in \S\ref{CONCLUSIONS}.  In Appendix \ref{BI}
we discuss how we measure the strength of BAL troughs for SDSS quasars.

\section{Our Current Knowledge of BAL Quasars}  \label{BKGD} 

To understand why our BAL quasars are unusual and important, we briefly review
the properties of `ordinary' BAL quasars and some key current questions in BAL
quasar research.  For more extensive discussions, see \markcite{wey97}{Weymann} (1997) and other
contributions to \markcite{asw97}{Arav}, {Shlosman}, \& {Weymann} (1997), \markcite{ham99}{Hamann} (2000), 
and the many contributions to \markcite{ckg01}{Crenshaw}, {Kraemer}, \&  {George} (2001b).

\subsection{Spectra and Absorption Trough Properties}	\label{SPEC}	
Observationally, BAL quasars show troughs $\sim$2,000$-$20,000\,\kms\ wide 
arising from resonance line absorption in gas with blueshifted (outflowing)
velocities up to 66,000~\kms\ \markcite{fol83}({Foltz} {et~al.} 1983).  
The absorption troughs are {\em detached} $\sim$20\% of the 
time \markcite{tur88}({Turnshek} 1988), meaning that the onset redshift of the absorption lies 
shortward of the systemic redshift, by up to 50,000~\kms\ \markcite{jea96}({Jannuzi} {et~al.} 1996).
Note that we use positive velocities to indicate blueshifts
since BAL troughs are outflows.

The usual formal definition of a BAL quasar is a quasar with positive
{\em balnicity index}.  The BI is a measure of the equivalent width of
the \civ\ absorption defined in the seminal paper of \markcite{wea91}{Weymann} {et~al.} (1991), 
hereafter W91.  However, this criterion 
is not perfect; for example,
it ignores absorption $<$2000\,\kms\ wide
even though many such {\em mini-BALs} are now known to share most or all of
the other characteristics of BAL quasars \markcite{ham99}({Hamann} 2000).  
We discuss this issue further in Appendix \ref{BI}, where we define the
{\em intrinsic absorption index}.  The AI is a refined BI designed 
to make optimal use of SDSS spectra and to include as BAL quasars objects which
were previously excluded but are clearly related to traditional BAL quasars.

\begin{figure}
\epsscale{1.0}
\plotone{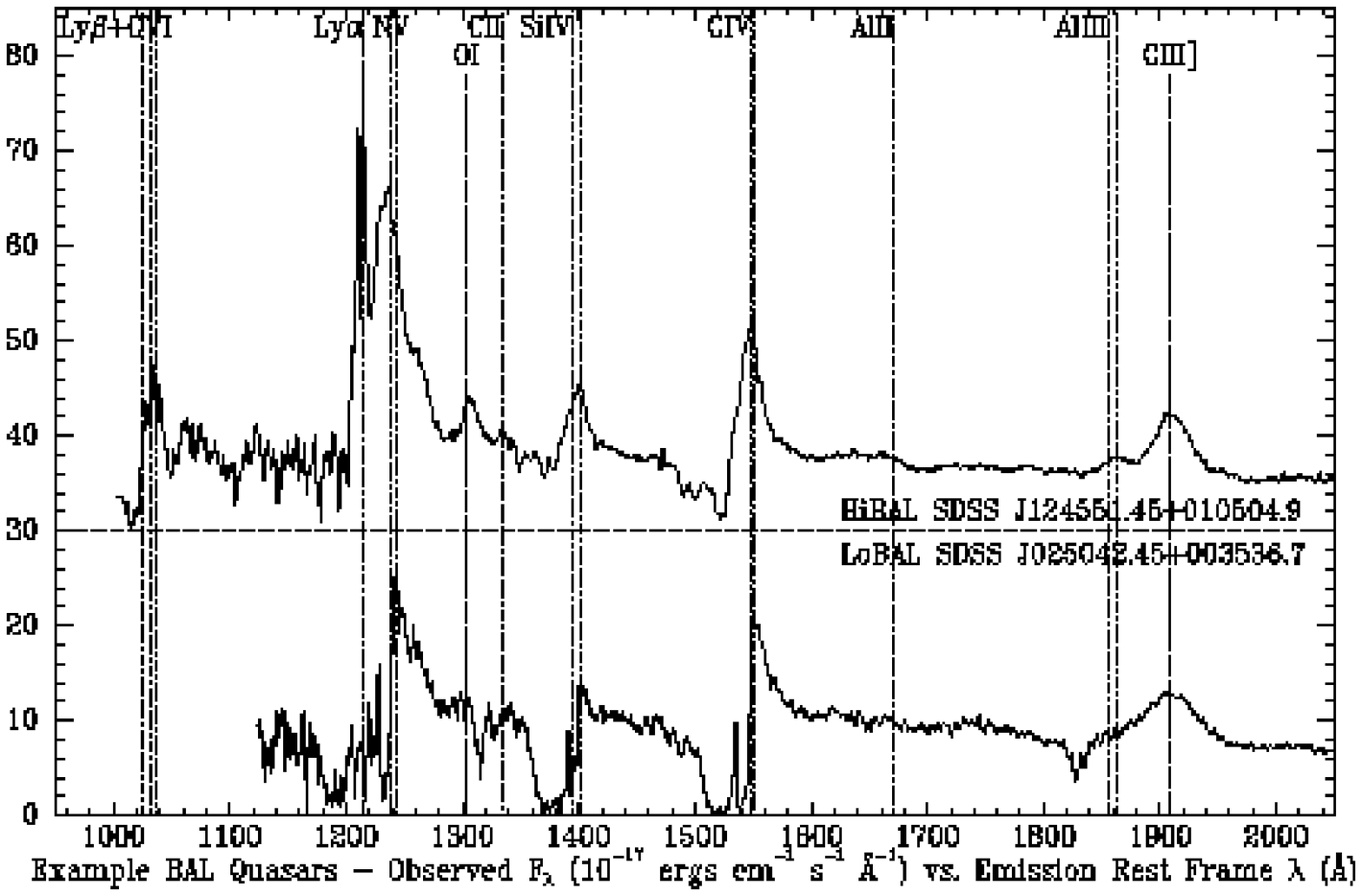}
\caption[]{ \singlespace 
Rest-frame spectra of two illustrative BAL quasars from the SDSS Early Data
Release quasar sample \markcite{sdssedrq}({Schneider} {et~al.} 2002).  The units of $F_{\lambda}$ are
$10^{-17}$\,ergs cm$^{-2}$\,s$^{-1}$\,\AA$^{-1}$ in this and all following
Figures.  The top spectrum is the HiBAL SDSS~J124551.45+010504.9
(LBQS~1243+0121), with absorption in \civ, \ovi, \Nv, \SIiv, and \lya.
The bottom spectrum is the LoBAL SDSS~J025042.45+003536.7.  Its spectral
coverage does not include \ovi, but all other species seen in absorption in
the HiBAL are present, as well as \aliii, \cii, and possibly very weak \alii.
(The spectral coverage does not include \mgii.)
These spectra help illustrate the range of properties seen in BAL outflows.  
The absorption does not go to zero in either object (partial covering), 
though it does come close.
In the HiBAL, the \civ\ and \SIiv\ trough profiles are similar but not
identical:  both appear somewhat detached from the emission line, but \civ\ is
deeper at the lowest outflow velocities.
The weak \lya\ absorption shows that the BAL outflow is highly ionized.
In the LoBAL, the \aliii\ and \cii\ troughs are narrower and weaker than the
\civ\ and \SIiv\ troughs.  Usually they appear weaker due to lower partial
covering, but without high-resolution spectra the possibility of unsaturated
low-ionization troughs cannot be ruled out.
There appear to be two systems in the outflow: a narrow low-velocity system
which is highly ionized (weak \cii, strong \Nv) and a broad high-velocity
system which includes the bulk of the low-ionization absorption, accompanied
by strong \lya\ absorption.
This LoBAL is slightly unusual in that respect: low-ionization absorption is
usually strongest at the lowest outflow velocities.
}\label{f_examples}
\end{figure}

BAL quasars are divided into three observational subtypes depending on what type
of transitions are seen in absorption. 
A list of relevant transitions with rest frame
$\lambda_{vac}$$>$1215\,\AA\ is given in Table~\ref{t_lines}, and two of
these three subtypes are illustrated and discussed in Figure\,\ref{f_examples}.
High-ionization BAL quasars ({\em HiBALs}) 
show absorption from \civ, \Nv, \SIiv, and \lya\ 
(in order of decreasing typical strength, for $\lambda_{vac}$$>$1215\,\AA).
Low-ionization BAL quasars ({\em LoBALs}; \markcite{vwk93}{Voit}, {Weymann}, \& {Korista} 1993) 
exhibit the above high-ionization absorption plus 
absorption in \mgii, \aliii, \alii\ 
and sometimes \feii, 
\feiii\ 
and other low-ionization species, again roughly in order of decreasing strength.
Note that the absorbing gas has a range of ionization states even in HiBALs.
The observed range simply extends to lower ionization states in LoBALs.
The rare Iron LoBALs ({\em FeLoBALs}; \markcite{haz87,cow94,bec97,bea00,sdss76}{Hazard} {et~al.} 1987; {Cowie} {et~al.} 1994; {Becker} {et~al.} 1997, 2000; {Menou} {et~al.} 2001)
also show absorption from excited fine-structure levels or excited atomic terms
of \feii\ or \feiii.\footnote{An atomic 
{\em term} is specified by quantum numbers $n$, $L$, and $S$, and consists of 
(2$S$+1)$\times$(2$L$+1) energy {\em states} grouped into {\em levels} of
different total angular momenta $J$=$L$+$S$.  Though it is a misnomer, we 
follow convention and use 
{\em excited-state} to refer to absorption from excited fine-structure
levels, excited terms, or both.  Note that absorption arising from excited
fine-structure levels of the ground term of \feii\ is sometimes denoted
\feii*, though technically \feii* to \feii**** should be used, depending on the
excitation \markcite{mor75}({Morton} 1975), and similarly for other ions.} 

A key feature of BAL outflows which has only recently been appreciated is that
the absorption is saturated (optical depth $\tau$ of a few) in almost all cases
where optical depths can be determined using line ratios and other saturation
diagnostics \markcite{ara01}(e.g., {Arav} {et~al.} 2001b).  This is true even though the absorption
troughs rarely reach zero flux.  This `nonblack saturation' means that the BAL
outflows typically exhibit {\em partial covering} of the continuum source 
\markcite{aea99b}({Arav} {et~al.} 1999a), 
possibly combined with infill of the troughs by host galaxy light or scattered
light which bypasses the BAL outflow \markcite{ogl99}({Ogle} {et~al.} 1999).  The BAL outflow is usually
thought to lie outside the broad emission line region since both the continuum
and the broad emission lines are absorbed in most objects \markcite{tea88}({Turnshek} {et~al.} 1988).
However, in some cases, 
only the continuum emission is absorbed \markcite{aea99b}({Arav} {et~al.} 1999a).

Species of a given ionization tend to have similar but not identical trough
shapes in velocity space, due to similar but not identical partial coverings
as a function of velocity for different ions \markcite{ara01}(e.g., {Arav} {et~al.} 2001b).
When present, low-ionization troughs are always 
narrower than high-ionization troughs, are usually strongest at low
outflow velocities (\markcite{vwk93}{Voit} {et~al.} 1993; but see \markcite{aea01}{Arav} {et~al.} 2001a),
and are often seen in relatively narrow features where the
high-ionization absorption is strongest.  This latter feature indicates that the
low and high ionization regions are closely associated \markcite{dek01}(e.g., {de Kool} {et~al.} 2001).  
Large local density gradients are probably required for gas
with ionization parameter $U$ 
ranging at least from 10$^{-2}$ to 1 \markcite{ara01}({Arav} {et~al.} 2001b)
to exist in such close proximity \markcite{aea01}({Arav} {et~al.} 2001a).

BAL troughs have never been seen to vary in velocity at a significant level.
Limits on the acceleration reach $<$0.03\,cm\,s$^{-2}$ in one case 
\markcite{wcp95}({Wampler}, {Chugai}, \& {Petitjean} 1995), and the most significant variations claimed are only 
$\Delta v=55\pm25$\,\kms\ over 5 rest-frame years
($dv/dt$=0.035\,cm\,s$^{-2}$; \markcite{vi01}{Vilkoviskij} \& {Irwin} 2001)
and $\Delta v=125\pm63$\,\kms\ over 6.6 rest-frame years 
($dv/dt$=0.06\,cm\,s$^{-2}$; \markcite{rvs02}{Rupke}, {Veilleux}, \& {Sanders} 2002).
Thus the BAL gas must either be coasting or in a stable flow pattern with
a significant transverse velocity component \markcite{aea99}({Arav} {et~al.} 1999b).
BAL troughs have been seen to vary in strength (\S\ref{VAR}),
probably due mostly to variations in covering factor rather than ionization,
since BAL troughs are typically saturated with optical depths
$\tau$ of a few (e.g., \markcite{ara01}{Arav} {et~al.} 2001b).  

\subsection{BAL Quasar Fraction and Global Covering Factor} \label{FRAC}

BAL quasars form 8\% of the optically selected 
flux-limited Large Bright Quasar Survey, but 11\% after accounting for the 
average differential optical/UV $k-$correction between BAL and non-BAL quasars
caused by the absorption troughs \markcite{wey97}({Weymann} 1997).
LoBALs comprise $\sim$15\% of BAL quasars (W91), but the differential
$k$-correction between LoBALs and HiBALs could be substantial \markcite{sf92}({Sprayberry} \& {Foltz} 1992).
If all quasars can be BAL quasars when viewed along certain lines of sight, then
the $k$-corrected BAL quasar fraction should be equal to the average global
covering factor (GCF) of those lines of sight \markcite{mor88}({Morris} 1988).  

The $k$-corrections used to estimate the true BAL quasar fraction 
must also account for any differential attenuation
between BAL and non-BAL quasars.  
It has been widely held that HiBALs as a population are not heavily reddened
(e.g., W91) but that LoBALs are \markcite{sf92}({Sprayberry} \& {Foltz} 1992), and thus that the true fraction
of LoBALs is underestimated by flux-limited optical surveys.  
Other surveys can overcome this bias:  
$\sim20_{-10}^{+15}$\% of {\em all} IRAS-selected quasars are LoBALs
\markcite{lea89,bm92}(W91; {Low} {et~al.} 1989; {Boroson} \& {Meyers} 1992) and $18_{-9}^{+14}$\% of BAL quasars are LoBALs 
in the optical and radio flux-limited FIRST Bright Quasar Survey (FBQS),
which has an observed BAL quasar fraction of 14$\pm$4\%
\markcite{bea00}(18$\pm$4\% if the balnicity criterion is relaxed; {Becker} {et~al.} 2000).
The FBQS and the SDSS also find that HiBALs as well as LoBALs are reddened
\markcite{bro01b,sdss76}({Brotherton} {et~al.} 2001b; {Menou} {et~al.} 2001).
Thus all types of BAL quasars are underrepresented in
optical flux-limited samples and the true average GCF must be $>$0.11.
A GCF of $\sim$0.3 
would help explain why 
BAL quasars are typically more polarized than non-BAL quasars \markcite{sh99,hl00}({Schmidt} \& {Hines} 1999; {Hutsem{\' e}kers} \& {Lamy} 2000)
and 
%
have generally weaker narrow \oiii\ emission \markcite{bm92}({Boroson} \& {Meyers} 1992).

An upper limit to the GCF can be estimated since broad absorption line troughs
are really broad scattering troughs: absorption of these resonance line photons
is followed by emission of an identical photon.  If the GCF was unity, then the
`absorbed' flux would be redistributed to a red wing in each line, and the
emission and absorption equivalent widths (EW) would be identical \markcite{hkm93}({Hamann}, {Korista}, \& {Morris} 1993).
Since the absorption EW is typically much greater than the emission EW, 
the BAL gas must have GCF$\lesssim$0.3.
However, this inference ignores the very real possibility of preferential
destruction of line photons by dust, particularly in LoBALs \markcite{vwk93}({Voit} {et~al.} 1993).
At least some LoBALs meet every test for being recently (re)fueled quasars
``in the act of casting off their cocoons of dust and gas" \markcite{vwk93}({Voit} {et~al.} 1993).
Such dust, combined with near-unity GCFs, helps explain many of their
unusual properties
\markcite{cs01}({Canalizo} \& {Stockton} 2001).


\subsection{Theoretical Models} \label{THEORY}
The scattering of resonance line photons can provide the radiative acceleration
which at least partially drives BAL outflows \markcite{aea95}({Arav} {et~al.} 1995).  
The difficulty arises in accelerating the gas to sufficient velocities without
completely stripping the resonance-line-absorbing ions of their electrons.
The disk wind model of Murray \& Chiang (1998, and references therein)
has been very successful in explaining this and other properties of BAL quasars.
In this model, a wind from an 
accretion disk is shielded from soft X-rays by a high column density 
($N_H\gtrsim10^{23}\,\rm cm^{-2}$) of highly ionized gas ($U\sim10$).
Being stripped of many outer electrons, the ions in this `hitchhiking' gas
transmit UV photons from resonance lines such as \civ.  
Thus the wind can be accelerated by scattering such photons without becoming too
highly ionized by higher energy photons.  The `hitchhiking' gas 
successfully explains the strong soft X-ray absorption in BAL quasars 
($N_H=10^{23-24}$\,cm$^{-2}$; e.g., \markcite{gea01}{Green} {et~al.} 2001a)
and absorption from very high ionization lines (up to Ne\,{\sc viii},
Mg\,{\sc x}, and Si\,{\sc xii} in SBS 1542+541; \markcite{tea98}{Telfer} {et~al.} 1998).
The quasar structure proposed by \markcite{elv00}{Elvis} (2000) also posits a disk wind,
but from a narrow range of radii, such that BAL quasars are only observed
when the line of sight is directly aligned with the wind. 
A recent summary of theoretical and computational modeling of disk winds
can be found in \markcite{psk00}{Proga}, {Stone}, \& {Kallman} (2000).
Some BAL quasars, particularly LoBALs, may be quasars cocooned by dust and gas
rather than quasars with disk winds \markcite{bea00}({Becker} {et~al.} 2000), but the only serious 
modeling relevant to this alternative has been the work of \markcite{wbp99}{Williams}, {Baker}, \& {Perry} (1999).


It is clear that 
broad absorption line quasars remain an active area of research.
Disk wind models explain many properties of
BAL quasars, but 
it is unclear if they can explain the full range of BAL trough profiles and
column densities.
The average global covering factor of BAL gas
and the range of global covering factors remain uncertain.  
Thus it is an open question how many BAL quasars are normal quasars seen along 
special lines of sight, and how many are young quasars emerging from dusty 
`cocoons'.  BAL quasars with unusual and extreme
properties may be useful in answering some of these questions.


\section{SDSS Observations and Data Processing}  \label{DATA} 

Whenever possible, the spectral and photometric data presented here are taken
from the SDSS Early Data Release (EDR), since it was produced with essentially
a single uniform version of the SDSS data processing pipeline.  A detailed
description of the EDR observations and reductions is given by \markcite{sdss85}{Stoughton} {et~al.} (2002).  
Here we review only a few points relevant to our primarily spectroscopic
analysis.  SDSS spectra are obtained using plates holding 640 fibers, each of
which subtends 3\arcsec\ on the sky, and cover 3800--9200\,\AA\ with resolution
of $1800<R<2100$ and sampling of $\simeq$2.4 pixels per resolution element.  
The relative spectrophometric calibration is good to $\sim$10\%,
but the absolute spectrophometric calibration to only $\sim$30\%.
Many objects have spectra from multiple plates for quality assurance purposes.
All spectra presented here are coadditions of all available spectra using 
inverse variance weighting at each pixel.  
Unless otherwise indicated, all spectra are from the SDSS; the
exceptions are from the Keck, CFHT, and ARC 3.5m telescopes. 

All magnitudes in this paper are {\em PSF magnitudes} calculated by fitting a
model PSF to the image of the 
object and correcting the resulting magnitude to a 7\farcs4
aperture (\markcite{lup01}{Lupton} {et~al.} 2001 and \S4.4.5 of \markcite{sdss85}{Stoughton} {et~al.} 2002).
The SDSS uses asinh magnitudes \markcite{sdss26}({Lupton}, {Gunn}, \& {Szalay} 1999) which differ from
traditional magnitudes by $<$1\% for signal-to-noise ratios SNR$\gtrsim$10.
Because the photometric calibration is still uncertain at the $\sim$5\% level,
all magnitudes are provisional and are denoted using asterisks.
The astrometric calibration is good to 0\farcs1 RMS per coordinate.  IAU
designations for each object presented here (e.g., SDSS~J172341.09+555340.6)
are given in the Tables, but are shortened in the text for brevity
(e.g., SDSS~1723+5553).  Full IAU designations given in the text
refer to quasars in the EDR quasar catalog \markcite{sdssedrq}({Schneider} {et~al.} 2002).

\section{BAL Quasar and Unusual BAL Quasar Selection}  \label{SELECT} 

Visual inspections were carried out to flag unusual spectra which were not 
immediately identifiable as normal quasars, HiBALs, LoBALs, or any other
known type of object or spectral reduction problem.  SFA and/or PBH inspected
all 95 spectroscopic plates in the EDR, all 43 other plates with numbers less
than 416 --- the highest number in the EDR --- 
completed on or before modified julian date (MJD) 52056, 
and a selection of 69 `post-EDR' plates numbered 417 and up.
Also, all EDR spectra not targeted and spectroscopically confirmed as
galaxies were inspected by DEV and/or AEB, and all EDR objects 
spectroscopically classified as quasars were re-inspected by PBH.  
Many other workers also inspected varying quantities of plates 
so that each spectrum has likely been inspected three times.

Most of the unusual objects found were unusual BAL quasars.  From $\sim$120,000
spectra on 207 spectroscopic plates, including $\sim$8,000 quasar spectra, we
have identified eighteen unusual BAL quasars and two mysterious objects which
might be unusual BAL quasars (\S\ref{GRADUAL}).  Three additional
unusual BAL quasars were selected from SDSS images but were confirmed with
spectra obtained with other telescopes. 
All these unusual BAL quasars can be divided into a handful of categories, 
as discussed in the next section.  We are confident that there are no further
examples of these categories of unusual BAL quasars on the inspected plates
among spectra which have SNR$\geq$6 per pixel in at least one of the $g$, 
$r$ or $i$ bands.  This is the only sense in which 
our unusual BAL quasar sample could be considered a complete sample.
A quasar sample from all the plates we inspected has not yet been defined,
and the EDR quasar sample \markcite{sdssedrq}({Schneider} {et~al.} 2002) is incomplete because not
all quasar candidates in the EDR area have been observed spectroscopically and
because the quasar candidate selection criteria were not the same for all
spectroscopic observations in the EDR.  
Nonetheless, since it is of interest to know what fraction of the quasar
population unusual BAL quasars comprise, in \S\ref{PERCENT} 
we estimate this fraction for the EDR.

Table~\ref{t_info} gives general information on the unusual BAL quasars
presented in this paper.  
%
In particular, the Target Code shows whether or not
the object was targeted by the Quasar, FIRST, ROSAT, Serendipity and star
selection algorithms (see the Note to the Table, and \markcite{sdss85}{Stoughton} {et~al.} 2002).
None of our objects were selected as galaxy targets, or as ROSAT targets.
Only three objects were not selected as Quasar targets, and one of them
(SDSS~1730+5850) was not selected only because of its faintness.
Only one target (SDSS~0127+0114) was completely overlooked by color
selection and was selected only as a FIRST target.

\begin{figure}
\epsscale{1.00}
\plotone{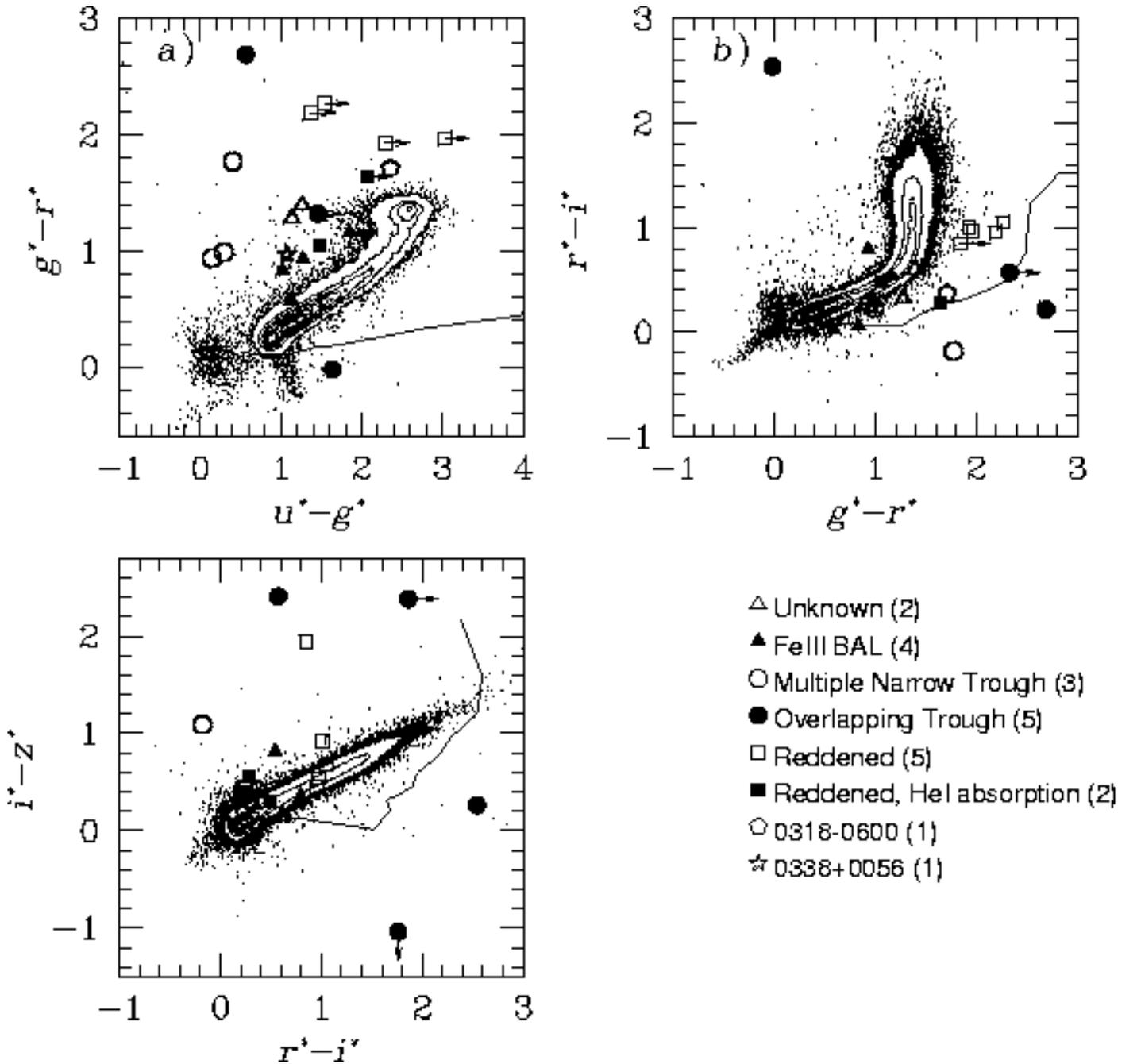}
\caption[]{ \singlespace 
The distribution of unusual BAL quasars in SDSS color space.  Each panel
is a projection of SDSS color-color space with dots and contours showing the
locus of unresolved objects (mostly stars).  
The line in each panel shows the mean color of simulated quasars as a function
of redshift \markcite{fan99}({Fan} 1999), with lower redshifts to the left.  
Most of the change in color with redshift is due 
to the influence of the Ly$\alpha$ forest; for ordinary quasars with $z<2.5$,
the colors are almost unchanging \markcite{sdss55}(see {Richards} {et~al.} 2001).  Thus most of the
objects in the cloud with $-0.2 < u^* - g^* < 0.6$ and $-0.2 < g^* - r^* < 0.5$ 
are low redshift quasars.  
The unusual BAL quasars presented in this paper are shown with the large
symbols, coded according to category as shown at the lower right,
with the number of objects of each category in parentheses.  Arrows denote
detections at lower than 5$\sigma$ significance.  Most of the quasars are at
$z<2.5$, and thus most of the scatter in colors is due to the BAL troughs.
}\label{f_ccdiag}
\end{figure}

The efficiency of SDSS color selection in finding unusual BAL quasars is shown
in Figure\,\ref{f_ccdiag}.  
The projections of the stellar locus in the three SDSS color-color
diagrams are shown, along with the unusual BAL quasars presented in this paper,
coded according to category.  Most of the scatter in colors is due to the BAL
troughs.  Most of these objects have colors quite distinct from ordinary quasars
as well as ordinary stars.  The SDSS is sensitive to these unusual objects
because the SDSS quasar target selection algorithm selects objects with unusual
colors even if they lie far from the locus of ordinary quasars in color space.

All SDSS and followup spectra presented herein are available for download from
the contributed data 
section of the SDSS Archive at {\tt http://archive.stsci.edu/sdss/}.
Flux densities $F_{\lambda}$ in these spectra, and in all our figures,
are given in units of $10^{-17}$\,ergs cm$^{-2}$\,s$^{-1}$\,\AA$^{-1}$.

\section{Categories of Unusual BAL Quasars} \label{ANALYSIS} 

All of the unusual BAL quasars found in SDSS to date are LoBALs, and many are
FeLoBALs.  Most can be placed into one of five categories, which we now present
and analyze separately.
First we present three FeLoBALs with {\em many narrow troughs} 
(\S\ref{SCALLOPED}). 
Objects with similar numbers of troughs but higher terminal velocities could
have their continua almost totally absorbed; we present five of these
{\em overlapping-trough} FeLoBALs in \S\ref{ABRUPT}. 
After that we present nine heavily reddened LoBALs (\S\ref{REDDENED}) and
one slightly less reddened LoBAL with stronger \feiii\ than \feii\ absorption
(\S\ref{FE3}), along with several possibly similar objects with lower SNR.
Lastly, we present two mysterious objects which may be very unusual BAL quasars 
(\S\ref{GRADUAL}).
The implications of all these objects are discussed in \S\ref{DISCUSSION}.

Absorption from many different ions is present in these objects.  To identify
most lines, we used the wavelengths tabulated in \markcite{cem62}{Moore} (1962), 
\markcite{myj88}{Morton}, {York}, \& {Jenkins} (1988), \markcite{mor91}{Morton} (1991) and \markcite{sdss73}{Vanden Berk} {et~al.} (2001).  
Absorption and occasionally emission from many
multiplets of \feii, \feiii, and possibly \fei\ is present in many of these
objects as well.  (Recall that a {\em multiplet} is the set of all transitions
between two atomic terms.)  To identify these lines we also used wavelengths
tabulated in \markcite{cem50}{Moore} (1950), \markcite{gra81}{Grandi} (1981), \markcite{haz87}{Hazard} {et~al.} (1987) and \markcite{gcc96}{Graham}, {Clowes}, \& {Campusano} (1996).  
These wavelengths are generally accurate enough for use with the low-resolution
spectra presented here, but more accurate wavelengths are available in the 
\feii\ and \feiii\ literature referenced by \markcite{vw01}{Vestergaard} \& {Wilkes} (2001).

A short word about multiplet notation is in order.  The standard notation of
\markcite{cem50}{Moore} (1950) designates \feii\ multiplets either as ultraviolet 
(UV; $\lambda\lesssim3000\,$\AA) or optical (Opt; $\lambda\gtrsim3000\,$\AA).
Both lists are ordered separately, first by increasing energy of the lower
term and then by increasing energy of the upper term.  
Thus UV2 multiplet transitions are to a more energetic upper term than UV1, 
though they arise from the same lower term.  The higher the multiplet number, 
the higher the excitation potential (EP) of its lower term (this applies to all
ions, not just \feii).  For example, \feii\ multiplets UV1 to UV9 arise from
the ground term.  When identifying which \feii\ multiplets are responsible
for any observed absorption, we always begin with the lowest-numbered
multiplet present in that wavelength range.  Absorption from higher-numbered
multiplets (i.e., more highly excited terms) may be present as well
(or instead, if selective pumping is at work),
but we prefer to be conservative in estimating the \feii\ excitation.

Where useful, BAL troughs are plotted in velocity space using
\begin{equation}                 \label{e_beta}
\beta\equiv v/c=(R^2-1)/(R^2+1) {\rm ~~where~~} R\equiv (1+z_{em})/(1+z_{abs})
\end{equation}
\markcite{fol86}({Foltz} {et~al.} 1986).  The uncertainty on $\beta$ is given by
\begin{equation}                 \label{e_sigbeta}
\sigma_{\beta}=\frac{4R^2}{(R^2+1)^2}
\sqrt{\left(\frac{\sigma_{z_{em}}}{1+z_{em}}\right)^2+\left(\frac{\sigma_{z_{abs}}}{1+z_{abs}}\right)^2}
\end{equation}
We define the zero velocity for each line using its 
laboratory vacuum rest wavelength. 

\subsection{BAL Quasars with Many Narrow Troughs}  \label{SCALLOPED} 

Our inspection revealed three FeLoBALs with many narrow absorption troughs,
similar to FIRST~0840+3633 \markcite{bec97}({Becker} {et~al.} 1997).  We discuss the spectra of the 
high-redshift and low-redshift objects separately, and then in comparison, but
defer discussion of the implications of all these objects to \S\ref{IMPSCAL}.

\subsubsection{High Redshift Example: SDSS~1723+5553} \label{HISCAL}

\begin{figure}
\plotone{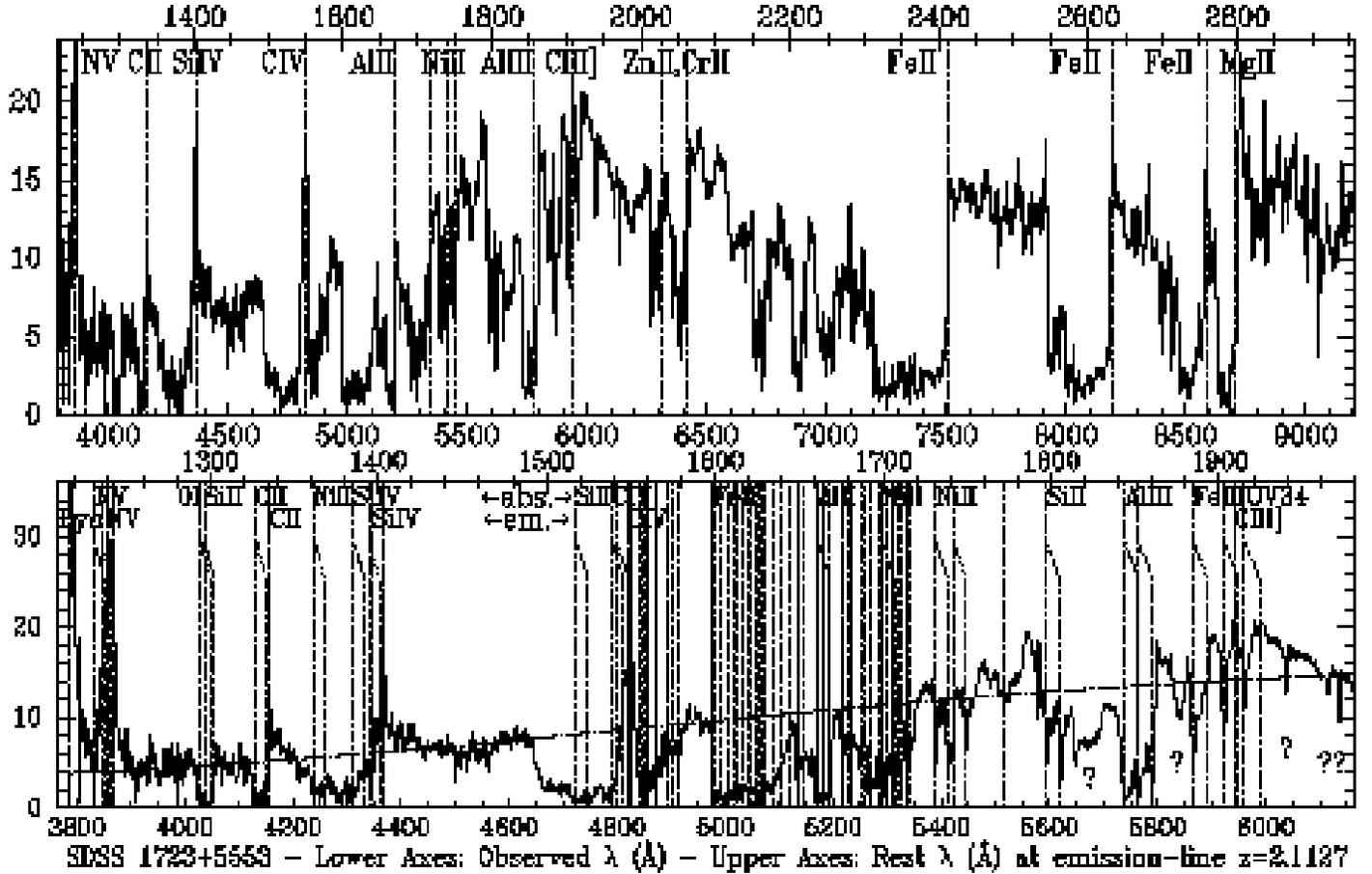}
\caption[]{ \singlespace 
SDSS~1723+5553 
at our adopted $z=2.1127\pm0.0006$.
The top panel shows the full spectrum
and the bottom panel the spectrum shortward of 1980\,\AA\ in the rest frame.
In both panels the observed wavelength is given along the bottom axis
and the rest wavelength along the top axis.  The units of $F_{\lambda}$ are
$10^{-17}$\,ergs cm$^{-2}$\,s$^{-1}$\,\AA$^{-1}$ in this and all following
Figures.  In the top panel, wavelengths of notable emission and absorption 
features are marked with dotted lines in the adopted rest frame.  
In the bottom panel, dashed lines show the wavelengths of emission lines
and dotted lines the wavelengths of absorption lines in the BAL outflow.
Bifurcated dotted lines show absorption from the same transition in two
different redshift systems.
Transitions are labelled at the top of the panel; most of the transitions from
1550\,\AA\ to 1720\,\AA\ are due to \feii\ multiplets.  Question marks denote
absorption features of unknown origin.
The smooth dot-dashed line shows the continuum fit used to calculate the 
BI and AI from the \civ\ trough.
See \S\ref{HISCAL} for further discussion.
}\label{f_hiscal}
\end{figure}

This object (Figure \ref{f_hiscal}) may have the most
absorption troughs ever observed at this resolution in a single quasar.
We identify absorption from 15 ions of 10 elements: \alii, \aliii, \cii, \civ,
\crii, \fei, \feii, \feiii, \mgii, \NIii, \Nv, \oi, \SIii, \SIiv, and \znii.
\feii\ absorption is seen from multiplets up to at least UV79, whose lower and
upper terms are 1.66\,eV and 7.37\,eV above ground, respectively.  
Only a small amount of neutral gas is probably present, since the only neutral
absorption we detect is \OI\ (intrinsically one of the strongest neutral-atom
transitions in the ultraviolet longward of \lya) and weak \FeI.  
%

We adopt an emission-line redshift of
$z=2.1127\pm0.0006$ by fitting Gaussians to the
apparent narrow emission lines of \SIiv\,$\lambda$1402.77 and 
\mgii\,$\lambda$2803.53.  Both lines used to derive the redshift are the 
longer-wavelength lines of doublets, but the apparent emission is narrower than
the doublet separations, allowing a single line to be fitted. The systematic 
uncertainty on our adopted redshift dwarfs the statistical uncertainty.

The bottom panel of Figure \ref{f_hiscal} shows the object's spectrum at
$\lambda<1980$\,\AA\ in the rest frame at the emission-line redshift;
the spectrum at $\lambda>1980$\,\AA\ is discussed in \S\ref{COMPSCAL}.
There are at least three different redshift systems in the absorbing region.
The lowest redshift system has $z=2.0942\pm0.0003$ as measured by \feiii\,UV34,
which is much stronger in this system than in either of the other two.
The intermediate redshift system, the weakest, has $z=2.100\pm0.001$ from
\NIii\ and \SIii.
The highest redshift system has $z=2.1082\pm0.0002$ from \NIii, \SIii, \znii,
and \crii; it also has weak \FeI\ and probably \OI. 
Absorption from transitions seen in both the lowest and highest redshift 
systems is indicated by the bifurcated dotted lines in Figure \ref{f_hiscal}b.
The high-ionization \SIiv\ and \civ\ troughs show some features corresponding
to the low-ionization redshift systems, but overall extend smoothly to outflow
velocities of 11300\,\kms.  

The narrow troughs in this object enabled the identification of the absorption
just longward of \civ\ as \feii\,UV44,45,46 absorption from 
terms with EP=0.15$-$0.25\,eV.
Such absorption has been seen before in 
the original FeLoBAL Q~0059$-$2735 \markcite{wcp95}({Wampler} {et~al.} 1995), but is much stronger here.  

\subsubsection{Low Redshift Examples: SDSS~1125+0029 and SDSS~1128+0113} \label{LOSCAL}

\begin{figure}
\epsscale{1.85}
\plottwo{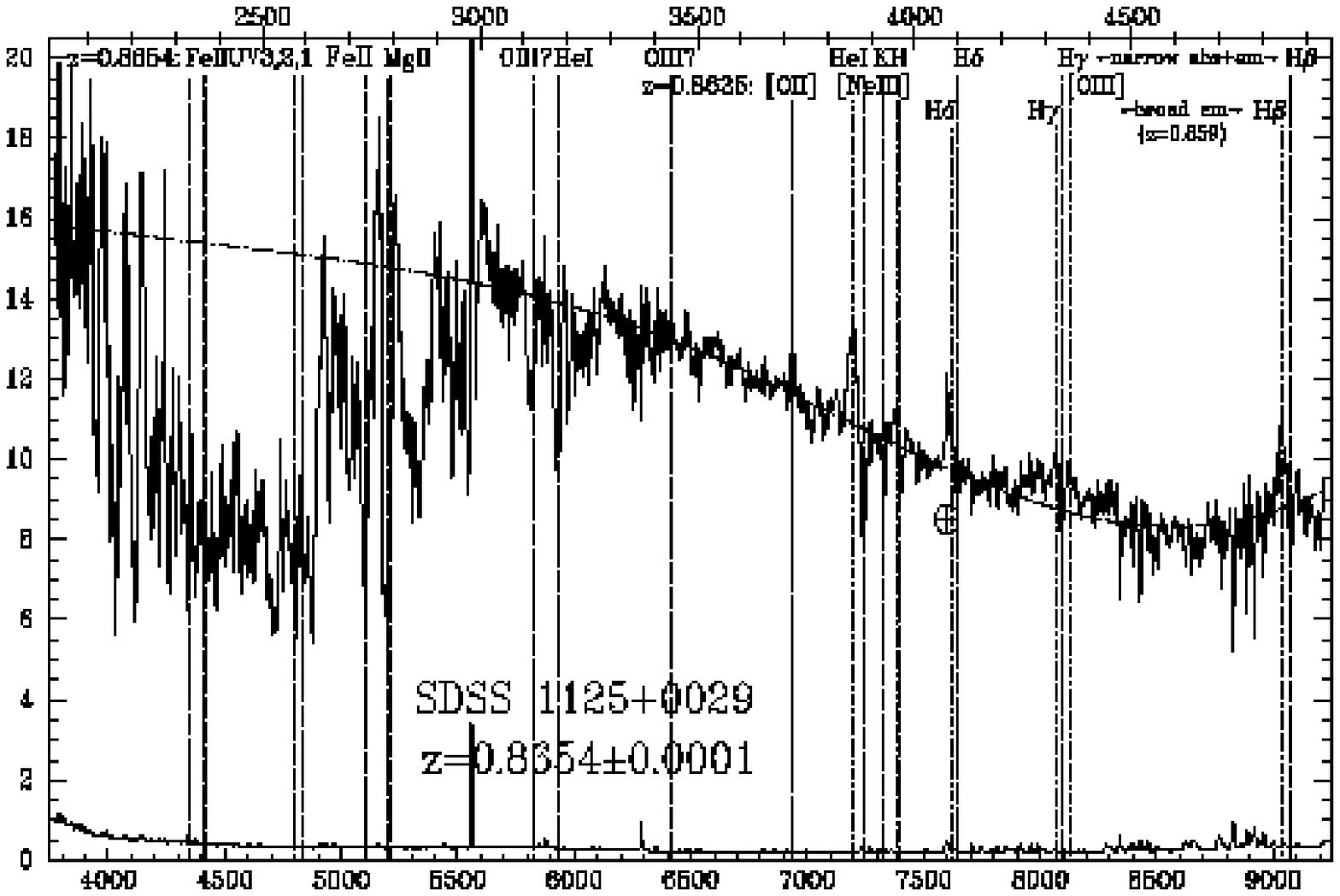}{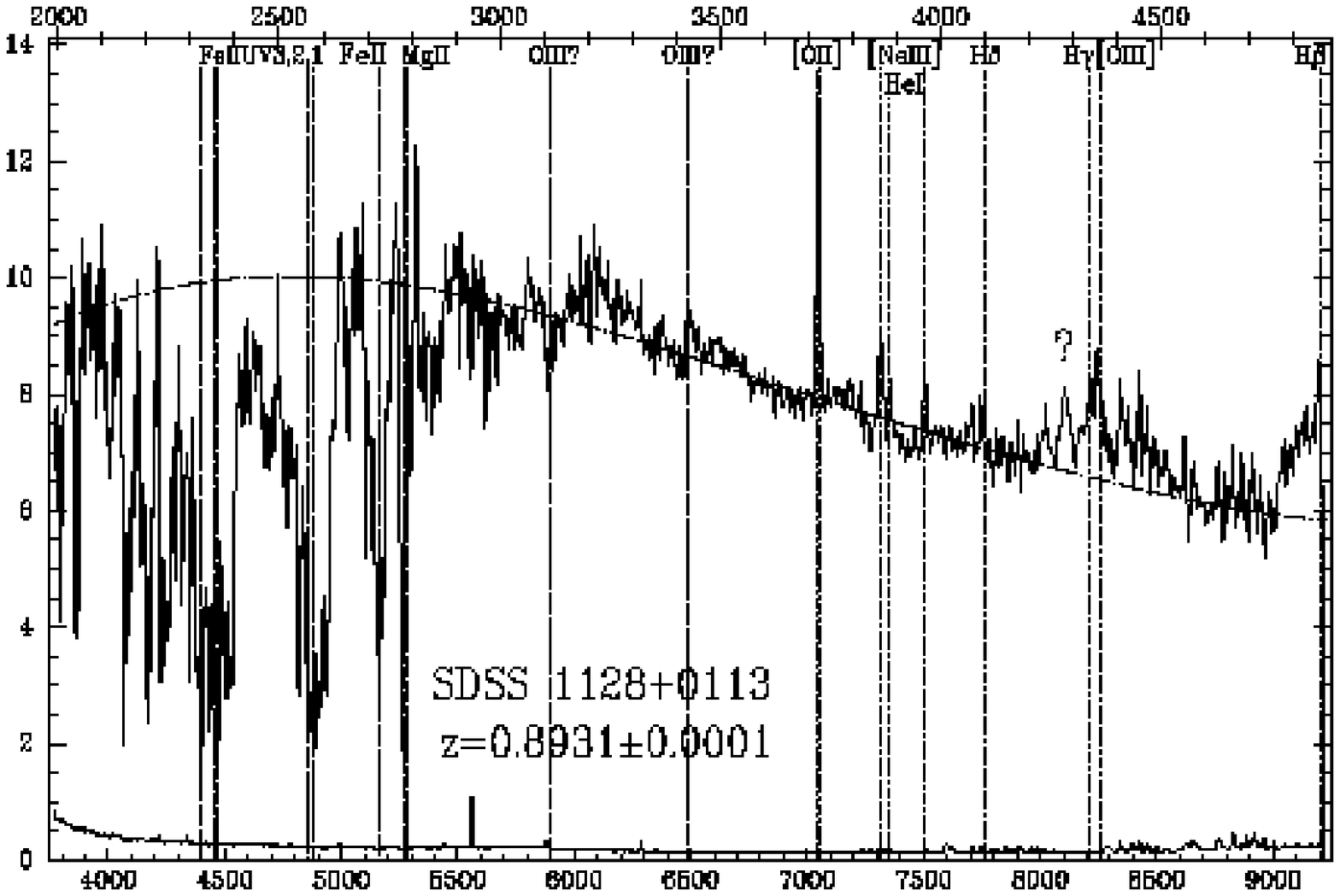}
\caption[]{ \singlespace 
SDSS spectra and error arrays of a) SDSS~1125+0029 and b) SDSS~1128+0113
(Observed $F_{\lambda}$ vs. rest frame $\lambda$ in \AA\ at the top
and observed frame $\lambda$ at the bottom).
All spectra and error arrays (lower curves) have been smoothed by a 5-pixel
boxcar filter.  The dot-dashed lines show the continuum fits used to calculate
the BI and AI (Appendix \ref{BI}).
Dotted vertical lines show the wavelengths of emission lines 
and dashed vertical lines show the wavelengths of absorption lines.  Between
2300$-$2600\,\AA\ rest frame, the strongest lines of \feii\ multiplets UV1, UV2,
and UV3 have been marked, but each is surrounded by many unmarked \feii\ lines.
Most of the absorption at $\lambda_{rest}<3500$\,\AA\ can be identified with
\feii, but the only other \feii\ absorption specifically marked is from UV62,63
at 2750\,\AA.  In both objects \mgii\ absorption is plotted at the
systemic redshift, but the \mgii\ trough extends longward of this redshift.
See \S\ref{LOSCAL} for further discussion.
}\label{f_scalloped2}
\end{figure}

Figure~\ref{f_scalloped2} shows the full SDSS spectra of both low redshift 
BAL quasars with many narrow troughs.  

\paragraph{SDSS~1125+0029} 
This object has narrow \OII\ emission at $z=0.8635\pm0.0008$,
in excellent agreement with the peak of the broader \NeIIIa\ emission.  
Broad emission is visible in H$\gamma$ and H$\beta$,
at $z=0.859\pm0.002$ (possibly affected by superimposed absorption);
the apparent broad emission at 7600$-$7640\,\AA\ (marked $\earth$ in
Figure~\ref{f_scalloped2}a) is just poorly removed telluric absorption.
Visible in the red half of the spectrum are relatively narrow absorption lines
from 
\hei\,$\lambda$3889+H8, \CaK\,(K), H$\epsilon$+\CaH\,(H)
blended with \NeIIIb\ emission, H$\delta$, H$\gamma$, and H$\beta$.
All of these lines except \caii\,K show 
reversed profiles: narrow emission in the center of a broader absorption trough.
These narrow Balmer emission lines give $z=0.8654\pm0.0001$, which we adopt as
the systemic redshift.
This is in excellent agreement with the \caii\,K redshift and agrees
with the \caii\,H redshift within the errors due to blending with \hei.
Thus the \oii\ and \neiii\ emission lines are blueshifted
by 310$\pm$130\,\kms\ and the broad Balmer lines by 1030$\pm$320\,\kms.
The BAL outflow might contribute to the \caii\ absorption 
(cf. \S\ref{ABRUPT}), but if the \caii\ is due to starlight it implies
a host galaxy with $M_{i^*}\sim-24.5$, which is extremely luminous.

This object has broad emission in H$\beta$ but not in \mgii.
This could be due to preferential destruction of \mgii\ line photons
by dust, since resonant scattering can greatly increase the path
length for line photons \markcite{vwk93}({Voit} {et~al.} 1993).
BAL troughs present in this object include \mgii\ and possibly \OIII\ 
and \hei\,$\lambda$3188.  The vast majority of the remaining absorption at
$2050<\lambda_{rest}<3500$\,\AA\ is from \feii.  
%

\paragraph{SDSS~1128+0113} 
This object has $z=0.8931\pm0.0001$ as measured by narrow \OII\ emission
and somewhat broader \NeIIIa, \hei\,$\lambda$3889.74 and \NeIIIb.
H$\gamma$ is present and may be slightly redshifted, but the SNR is low and it
may be blended with \oiii\,$\lambda$4363.  H$\beta$ is also present, but we
cannot determine its peak observed wavelength with any certainty since it
is at the extreme red end of our spectrum. 
We cannot identify the apparent moderately broad emission line just shortward of
H$\gamma$, at 4295\,\AA\ (marked ? in Figure~\ref{f_scalloped2}b); there should
be other lines present if it is \feii\ emission.
The BAL troughs in this object are not quite as omnipresent as those in
SDSS~1125+0029.

\subsubsection{Comparison of Many-Narrow-Trough BAL Quasars} \label{COMPSCAL}

\begin{figure}
\epsscale{1}
\plotone{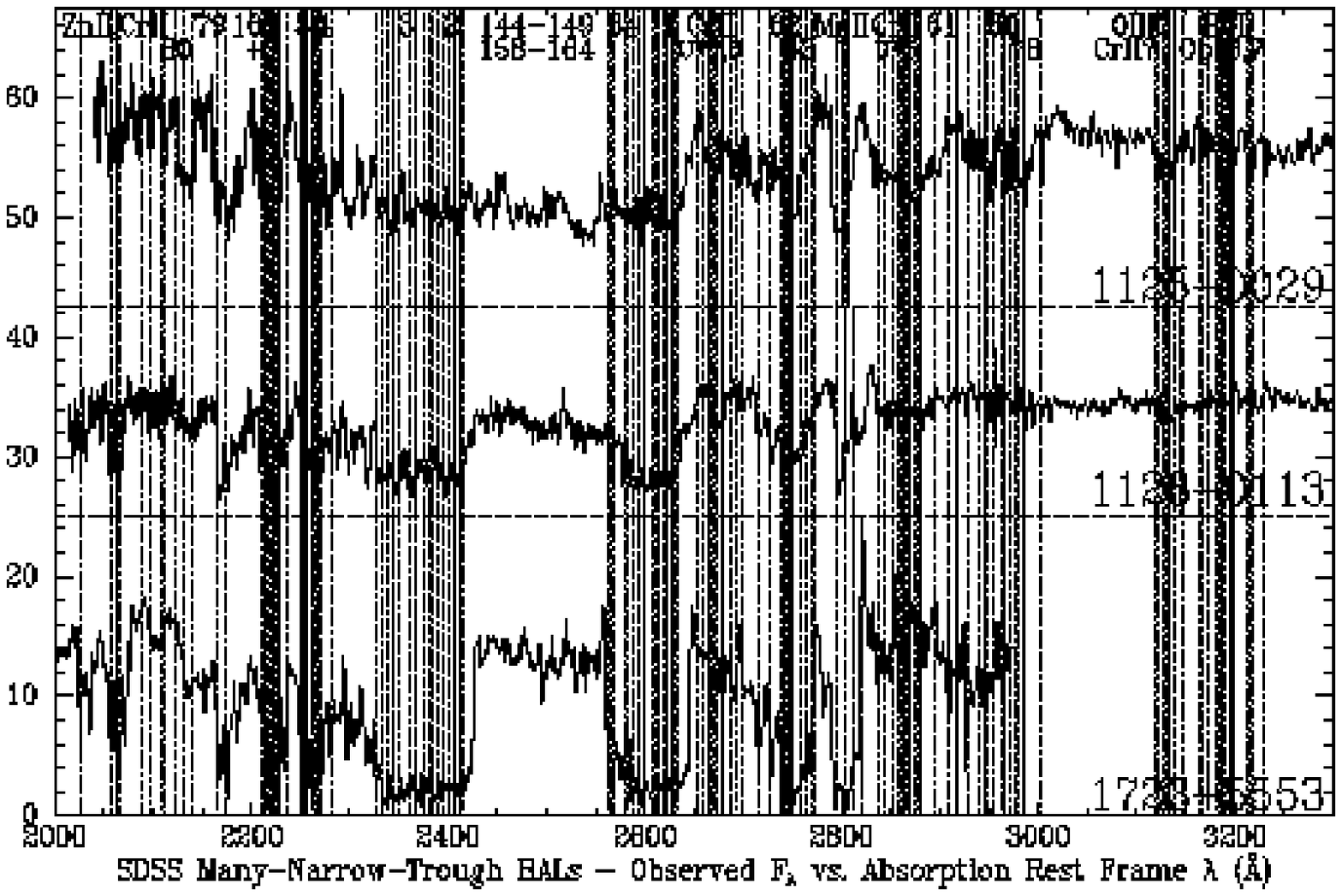}
\caption[]{ \singlespace 
The spectral region near \mgii\ in the three BAL quasars with many narrow
troughs
discussed in \S\ref{COMPSCAL}.  SDSS~1125+0029 has been smoothed by a 3-pixel 
boxcar filter, while unsmoothed spectra are displayed for the other two objects.
Horizontal dashed lines show the zero flux levels for the top two spectra.
Each spectrum is plotted
in the rest frame of its deepest absorption trough, and dotted vertical lines
show the wavelengths of selected absorption lines in this frame.
Dashed vertical lines show the wavelengths of \MgII, \OIII\ 
and \HeI\ in the adopted {\em systemic} rest frame of each object (see text).
This shows that in the top two (lower-$z$) objects,
the \mgii\ trough extends longward of the systemic redshift.
Absorption complexes are labelled just to the left of the longest wavelength
transition in
the complex whenever possible.  Optical \feii\ multiplets are labelled Opt\#,
while UV \feii\ multiplets are labelled just by number; for clarity, lines for
multiplets 144$-$149 and 158$-$164 near 2500\,\AA\ are not plotted.
Not every absorption line matches an obvious feature because the lines are
not all equally strong.  Finally, in SDSS~1723+5553 (bottom),
there is a second absorption system at a redshift between that used for the plot
and the slightly higher emission redshift
(best seen in \znii\,2026\,\AA, at the far left).  \feii\ UV2 and UV1 multiplets
from this system explain the long-wavelength ends 
of the absorption troughs near 2400\,\AA\ and 2600\,\AA.
}\label{f_scalloped3}
\end{figure}

The similarity of our many-narrow-trough BAL quasars to each other is shown in
Figure~\ref{f_scalloped3}, where we plot the wavelength range 2000---3300\,\AA\ 
in the rest frame of the deepest absorption trough for each object.  Dotted
vertical lines show the wavelengths of the labelled absorption lines or
numbered 
\feii\ multiplets in this frame.
There is no sign of \MgI\ absorption in any of the objects.
Dashed vertical lines show the wavelengths of \MgII, \HeI, and two \Oiii\ 
transitions (see next paragraph) in the adopted {\em systemic} rest frame of
each object.  
Note that if they are present, \HeI\ is blended with \feii\,Opt6 and Opt7
and the weaker \hei\,$\lambda$2945 line, is blended with \feii\,UV60 and UV78.
No absorption trough reaches zero flux in any object, indicating partial 
coverage of the continuum source by the absorbing region, and possibly a
contribution from scattered light which bypasses the absorbing region.
SDSS~1125+0029 (top) has the lowest partial covering but
the highest \feii\ excitation,
since it shows absorption which arises only from highly excited
\feii, seen longward of \mgii\ and between the more common lower excitation
\feii\ absorption troughs at 2400\,\AA\ and 2600\,\AA\ \markcite{cem50}({Moore} 1950).

Most troughs can be identified with \feii\ multiplets, but some
have uncertain identifications, notably:\\
$\bullet$ 
We initially identified the $\sim$3100-3135\,\AA\ trough with \OIII, which was
first identified as a BAL in CNOC2~J022509.6+001904 \markcite{hal99cnoc2agn}({Hall} {et~al.} 2000).  
This transition has a lower state 
36.9~eV above ground, but can be indirectly populated via Bowen
resonance-fluorescence with \heii\ \lya\ (section 4.7 of \markcite{ost89}{Osterbrock} 1989).
However, in this case we should see \Oiii\,$\lambda$3445 
with $\sim$30\% the strength of \OIII.
Such absorption is not present in either object in Figure~\ref{f_scalloped2},
so \OIII\ absorption can at best explain only part of their
$\sim$3100-3135\,\AA\ troughs.  
The lowest-numbered \feii\ multiplet which matches this trough is Opt82
(EP 3.87\,eV), but if that identification were correct
we should also see absorption near 2150\,\AA\ 
from multiplet UV213 (EP 3.14-3.22\,eV).  
Another possibility is \crii\ absorption from terms $\sim$2.45\,eV above ground
\markcite{dek02}({de Kool} {et~al.} 2002), but in SDSS~1125+0029 such absorption would be improbably strong
relative to the ground-term \crii\ trough at $\sim$2060\,\AA.
Thus the origin of the $\sim$3100-3135\,\AA\ trough remains unclear.\\
%
%
$\bullet$
Explaining the $\sim$2220\,\AA\ trough as \feii\ requires absorption from 
at least UV168 (lower term EP 2.62$-$2.68\,eV).  The 
2400$-$2550\,\AA\ absorption can then be attributed to UV144$-$149,158$-$164.
Such absorption is weak in SDSS~1128+0113 and absent in SDSS~1723+5553, however,
so their 2220\,\AA\ troughs are unlikely to be due solely to \feii\,UV168.
A contribution from \NIii\ (Table~\ref{t_lines}) is a poor fit, 
and \SIi\ and \mni\ can be ruled out because many other lines from neutral atoms
would also be seen and are not.\\ 
%
$\bullet$ 
The strong absorption near $\sim$2665\,\AA\ 
may require more than just \crii\,UV7,8 to explain it.\\
The first two cases above might be explained by selective pumping of the lower
terms of multiplets Opt82 and UV168, respectively, a possibility which
can only be tested via detailed modeling of \feii.

\subsubsection{Redshifted Absorption Troughs} \label{ZSYS}

The most interesting feature of both low-redshift objects presented in
\S\ref{LOSCAL} is that the broad \mgii\ doublet absorption trough extends 
longward of the systemic redshift
(dashed vertical lines in Figure\,\ref{f_scalloped3}).  
In SDSS~1128+0113, the \mgii\ absorption appears split into
two components, but this structure may be some combination of narrow \mgii\ 
emission and partial covering which varies with velocity.

The presence of longward-of-systemic broad absorption is quite surprising, since
the absorption in BAL quasars has always previously been seen in outflow.
A search for BAL troughs longward of the systemic redshift in all
BAL quasars in the EDR (\S\ref{PERCENT}) turned up several additional
candidates, but none stood up to scrutiny.
SDSS~J115852.87$-$004302.0 has $z=0.9833\pm0.0010$ from narrow \OII\ emission.
\mgii\ absorption extending slightly longward of this $z$ cannot be ruled out
due to noise from the 5577\,\AA\ night sky line, but there is no evidence of
longward-of-systemic absorption from other lines. 
SDSS~J131637.27$-$003636.0 has $z=0.930\pm0.001$ from narrow \oii\ emission.
There is weak \mgii\ emission in the middle of what appears to be an absorption
trough but is actually just a gap between emission from \feii\ multiplets.
SDSS~J235238.09+010552.4 has \civ\ absorption longward of the peak of \civ\ 
emission, but only because the \civ\ peak is blueshifted $\sim$4000\,\kms\ from
the \ciii\ and \mgii\ redshift of $z=2.156$.

We discuss the implications of longward-of-systemic absorption in
\S\ref{IMPSCAL}.

\subsection{BAL Quasars with Overlapping Troughs}  \label{ABRUPT} 

\begin{figure}
\epsscale{1.0}
\plotone{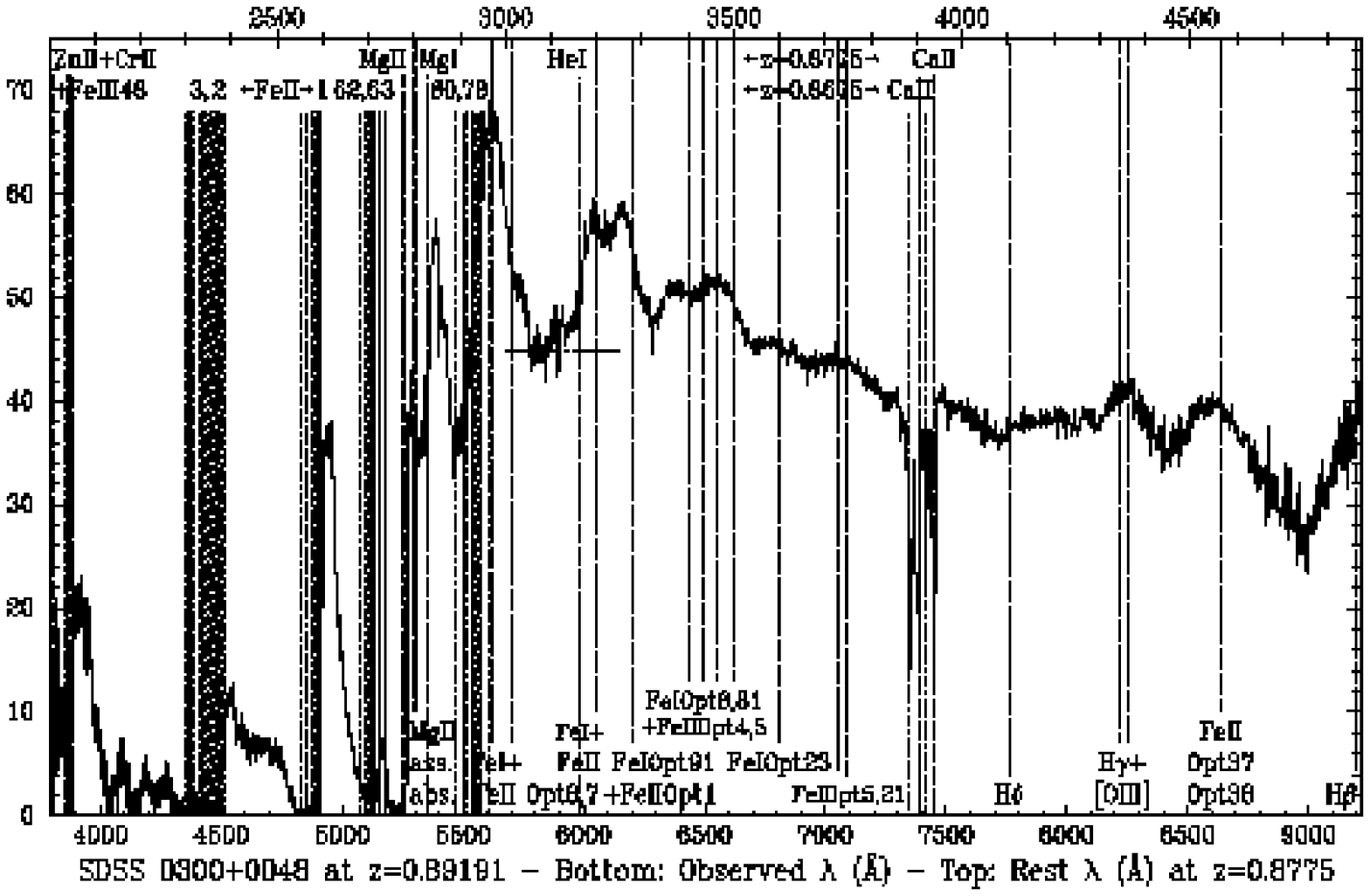}
\caption[]{ \singlespace 
SDSS~0300+0048, an overlapping-trough FeLoBAL at $z=0.89191\pm0.00005$.
This assumed systemic redshift is taken from the associated \mgii\ system,
but accurately predicts the wavelengths of broad emission features longward of
\mgii\ (dashed lines are features at this redshift, labelled along the bottom
of the plot).
The onset of broad \mgii\ absorption is at $z=0.8775$, coincident with the
highest redshift \caii\ system (dot-dashed lines show absorption at this
redshift, labelled along the top of the plot).
However, the onset of \feii\ absorption from various UV multiplets is at
$z=0.8675$, coincident with the lowest redshift \caii\ system
(features at this redshift are denoted by dotted lines
and the second row of labels across the top).
The horizontal dot-dashed line segment shows the continuum level adopted for
calculation of the AI and BI.
}\label{f_0300}
\end{figure}

\begin{figure}
\epsscale{1.0}
\plotone{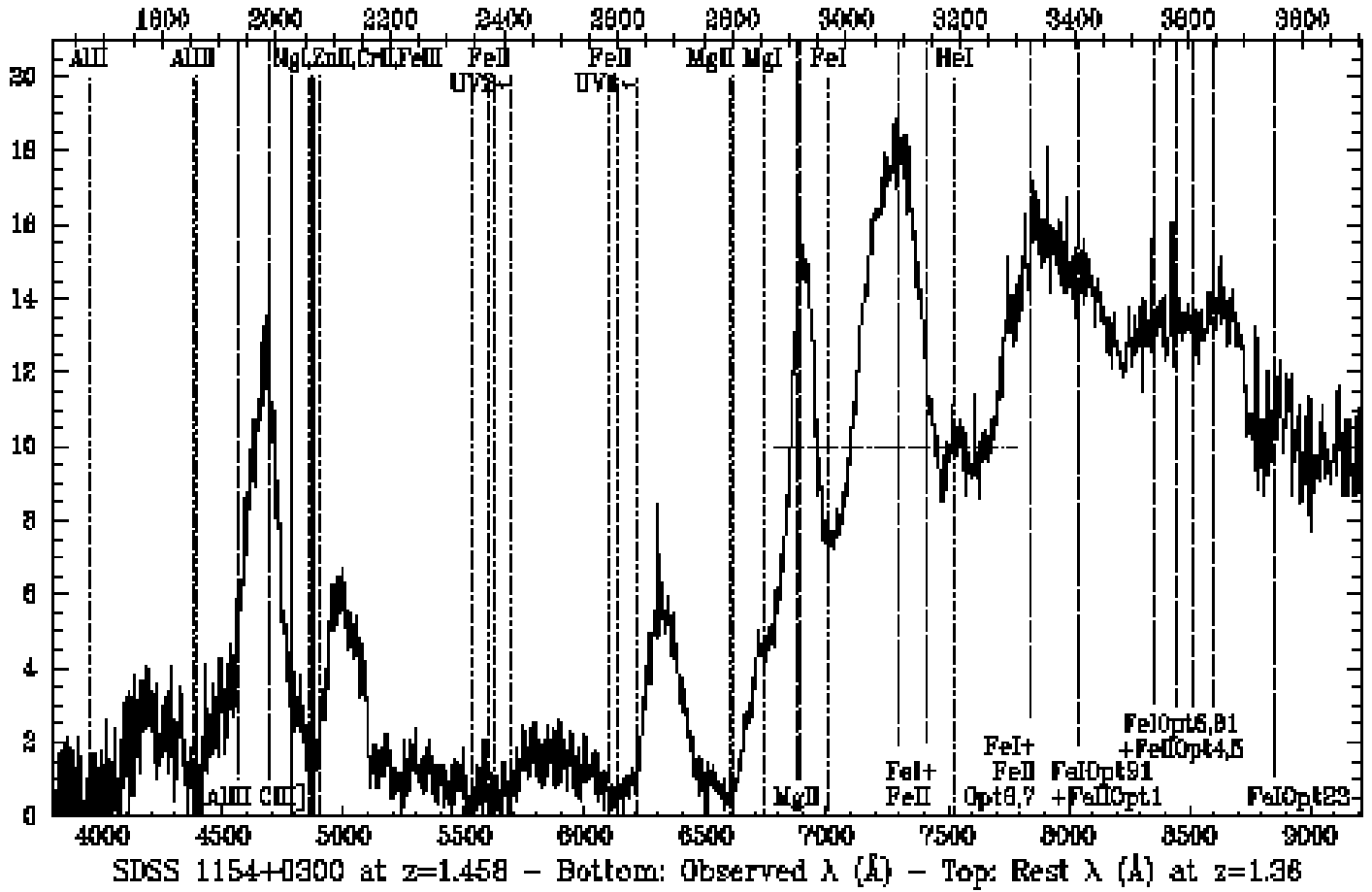}
\caption[]{ \singlespace 
SDSS~1154+0300, an overlapping-trough FeLoBAL at $z=1.458\pm0.008$.  
Dashed lines labelled along the bottom of
the plot show emission features at this redshift, while absorption features at
the redshift of peak absorption ($z=1.36$) are labelled along the top of the
plot.  Only the ground level lines and longest-wavelength
excited level lines are plotted for \feii\ multiplets UV1 and UV2.
The horizontal dot-dashed line segment shows the continuum level
adopted for calculation of the AI and BI.
}\label{f_1154}
\end{figure}

\begin{figure}
\epsscale{1.75}
\plottwo{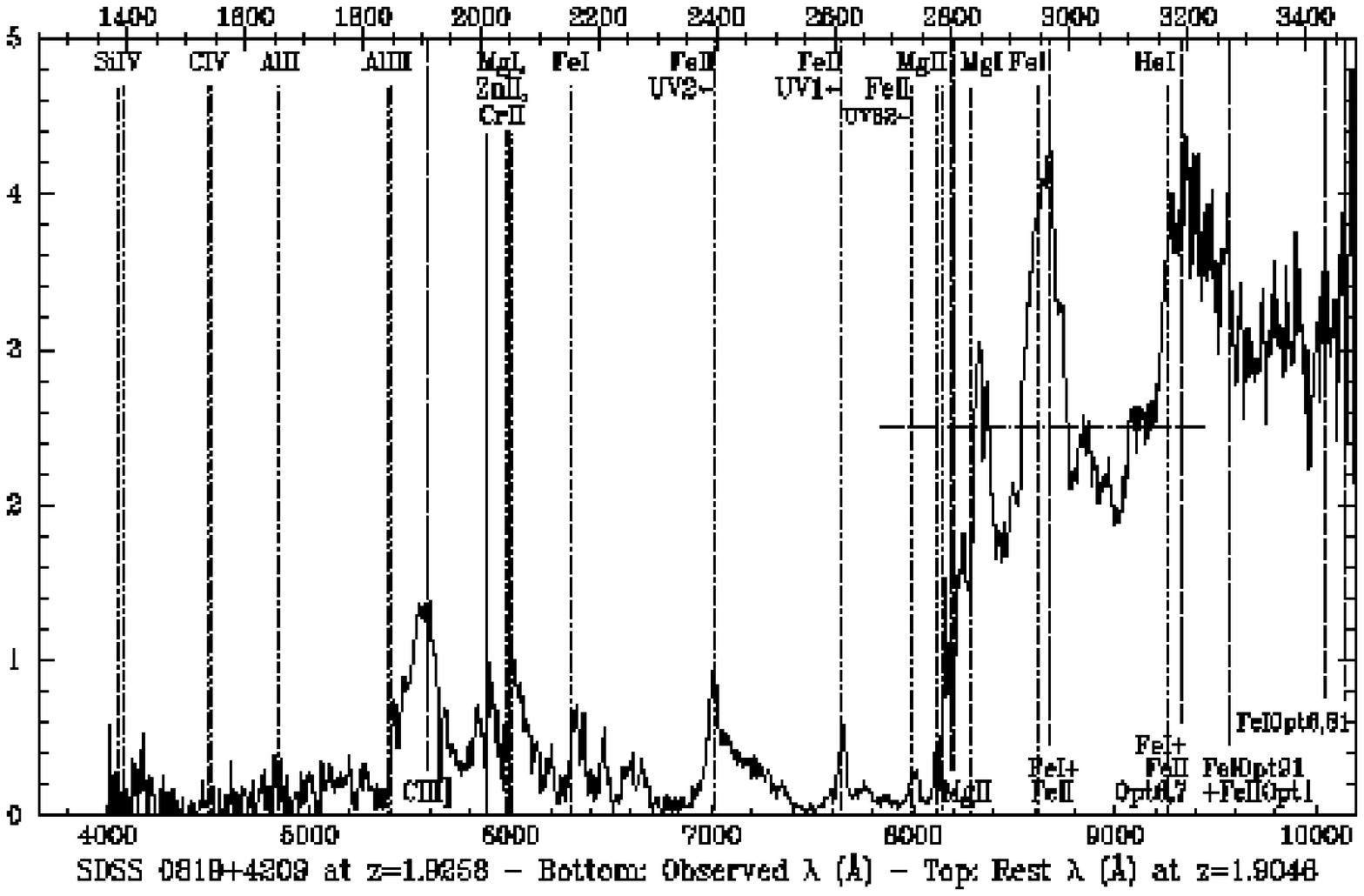}{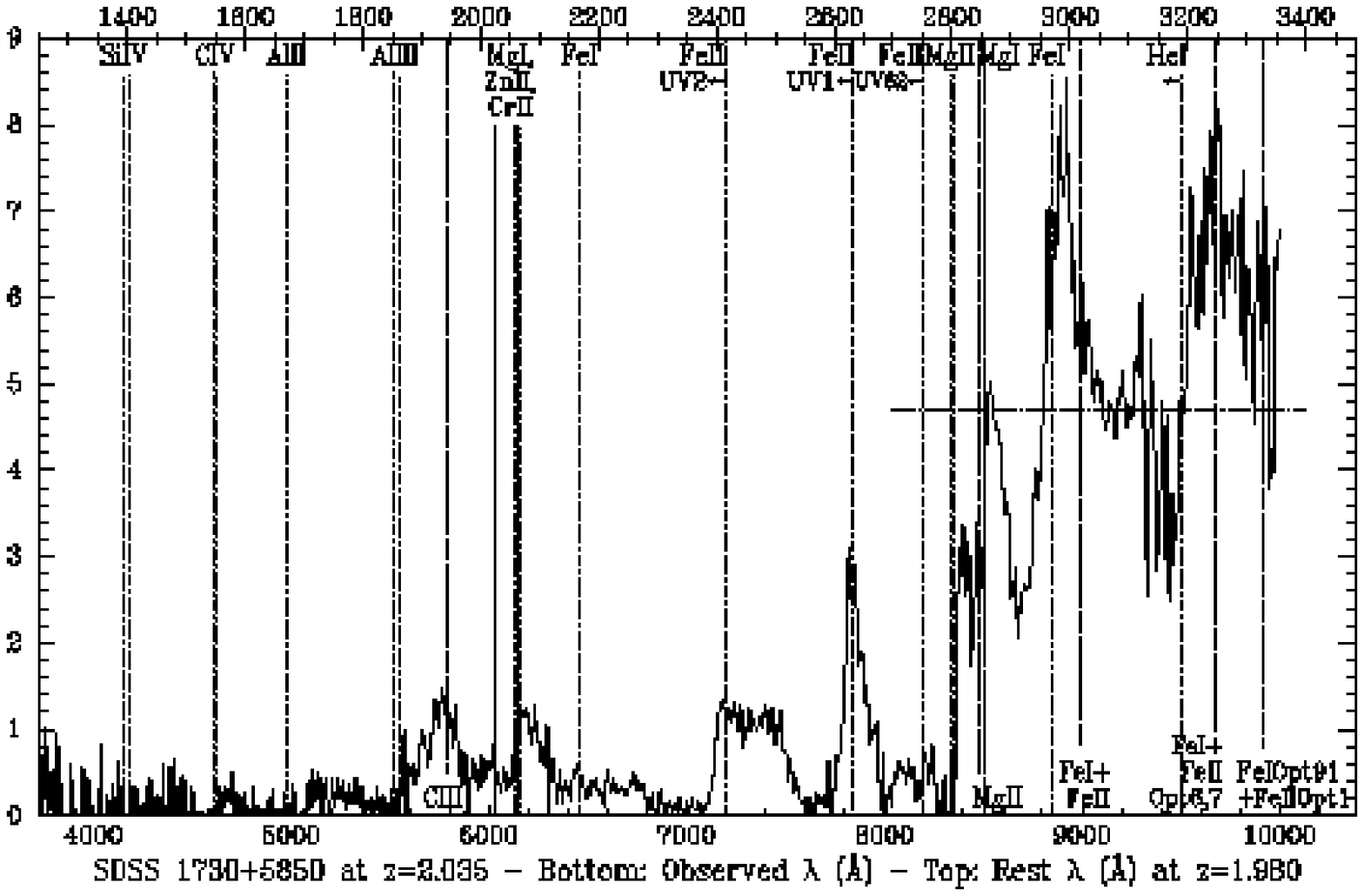}
\caption[]{ \singlespace 
Two high redshift overlapping-trough FeLoBAL quasars. In each plot, dashed lines
show emission features at the systemic redshift, labelled along the bottom of
the plot.  Dotted lines show absorption features at the redshift of the red
edge of the \mgii\ trough, labelled along the top of the plot.  
Only the longest-wavelength lines are plotted for \feii\
multiplets UV1, UV2, and UV62.  In each plot, the horizontal dot-dashed line
segment shows the continuum level adopted for calculation of the AI and BI.
a) SDSS~0819+4209 at $z=1.9258\pm0.0006$.  This assumed systemic redshift is
taken from associated \mgi\ and \mgii, but accurately predicts the wavelengths
of various broad emission features.  The redshift of the long-wavelength edge
of the \mgii\ trough is $z=1.9046\pm0.0005$.  Only the longest-wavelength
lines are plotted for \feii\ multiplets UV1, UV2, and UV62.
b) SDSS~1730+5850 at $z=2.035\pm0.005$.  This assumed systemic redshift is taken
from \CIII\ and \HeI\ emission. There may be extensive \HeI\ absorption starting
at a redshift just below $z=1.980\pm0.005$ (the latter is the onset redshift of
the \mgii\ trough).
}\label{f_abruptbals}
\end{figure}

\begin{figure}
\epsscale{1.0}
\plotone{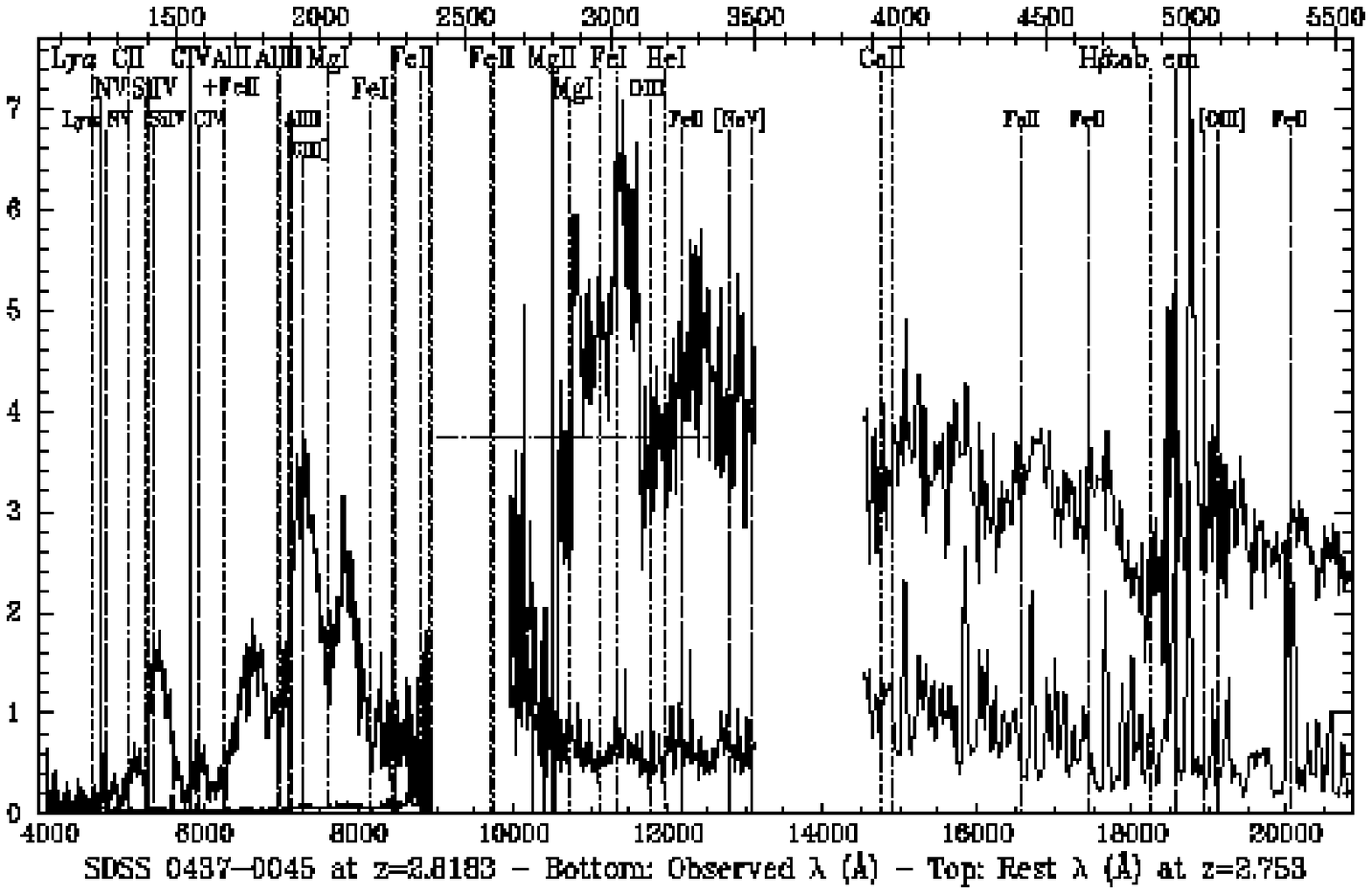}
\caption[]{ \singlespace 
Optical (Keck) plus NIR (UKIRT) spectra and error arrays of SDSS~0437$-$0045 at
$z=2.8183\pm0.0009$.  
The top two rows of labels show the wavelengths of absorption lines
at the redshift of peak absorption, $z=2.753$.  
Below that, emission lines are shown at the adopted systemic redshift.  
There is no evidence for \oiii\ emission at $z=2.8183$, but we plot its
wavelengths for reference.
Note the nearly complete absorption from \feii\ near the
expected wavelength of C\,{\sc iv}, and the possible broad H$\beta$ absorption.
(The apparent emission feature at 5000\,\AA\ in the $z=2.753$ frame is a
sky line residual; note the peak in the error spectrum just below it.)
The horizontal dot-dashed line shows the 
continuum level used in calculating the AI and BI.
}\label{f_weird2}
\end{figure}

Several FeLoBALs have been discovered from the SDSS with abrupt drops in 
flux near \MgII\ caused by overlapping absorption troughs.  That is, the troughs
remain deep at velocities comparable to the spacing of absorption troughs from
different transitions, so that there are no continuum windows between the
troughs (Figures \ref{f_0300}$-$\ref{f_weird2}).  Troughs 19000\,\kms\ wide may
be required, so that \mgii\ overlaps \feii\,UV1 at $\lambda$$\leq$2632\,\AA;
however, a width of only 13000\,\kms\ is needed if \feii\,UV62,63 absorption is
present.  These `overlapping-trough' BAL quasars resemble Mrk~231 \markcite{smi95}({Smith} {et~al.} 1995),
FIRST~1556+3517 \markcite{bec97}({Becker} {et~al.} 1997) and especially FBQS~1408+3054 \markcite{wea00,bea00}({White} {et~al.} 2000; {Becker} {et~al.} 2000),
including the complex emission and absorption near \mgii\ and the strong \MgI\ 
absorption.  However, the BAL region in our overlapping-trough objects covers
the emission region almost totally, instead of only partially as in 
FBQS~1408+3054 or Mrk~231.  (The spectrum of FBQS~1408+3054 can be seen
in \S\ref{GRADPC}). 


We first discuss the rest frame features common to all these objects
(Figures \ref{f_0300}$-$\ref{f_weird2}), and then present them individually,
in order of increasing redshift.

The apparent emission lines shortward of \mgii\ are identified as regions near
the high-velocity (blue) ends of various absorption troughs.  At high outflow
velocities, the partial covering of the absorption, or possibly its optical
depth, typically decreases so that the observed flux begins to recover to
the continuum level.  The troughs are broad enough, however, that before this
recovery is complete, another absorption trough is encountered and the 
observed flux drops abruptly.  These abortive recoveries toward the continuum
level can mimic the appearance of broad emission lines.

Longward of \mgii, the spectra are mostly free from absorption and we can
identify emission features by comparison to the SDSS composite quasar spectrum
\markcite{sdss73}({Vanden Berk} {et~al.} 2001), allowing for emission from strong \fei\ and \feii\ multiplets
whose lower terms are $\lesssim$3\,eV above ground.  In our 
object with the best coverage
longward of \mgii\ (Figure \ref{f_0300}), we see H$\beta$ in emission,
\feii\,Opt37,38 at 4570\,\AA, 
H$\gamma$+\oiii\ at 4350\,\AA,
\fei\,Opt5,21 at 3735\,\AA,
\fei\,Opt23 emission at 3600\,\AA,
a feature at 3350$-$3550\,\AA\ including
\fei\,Opt6,81 at 3465,3435\,\AA\ 
and possibly \feii\,Opt4,5 at 3500,3400\,\AA,
a feature at 3150$-$3300\,\AA\ including
\fei\,Opt91+\feii\,Opt1 at 3280\,\AA\ 
and \hei+\feii\,Opt6,7+\fei\,Opt155-158 at 3200\,\AA, 
and a blend of \fei\,Opt9,30 at 3015\,\AA\ 
with \feii\,UV60,78+\fei\,UV1 at 2970\,\AA.
Many of these features are also visible in Figures \ref{f_1154}-\ref{f_weird2}.

Near \mgii\ the spectra are blends of emission and absorption.  
There may be \hei, \fei\ and/or \feii\ 
absorption shortward of the 3200\,\AA\ emission
feature in some or all of the objects, most notably in SDSS~1730+5850
(Figure \ref{f_abruptbals}b).  There is almost certainly \fei\ and/or \feii\ 
absorption shortward of the 2970\,\AA\ emission feature, 
because the observed flux dips lower than essentially any
point in the continuum longward of \mgii.  
Dust may contribute to this drop in flux, but the effect is too sudden
to be entirely due to dust.  Also,
in all objects except SDSS~0819+4209 (Figure \ref{f_abruptbals}a), the flux
recovers to this level 
at least once shortward of \mgii, which would not be the case if dust reddening
were already affecting the spectrum strongly at 3000\,\AA.
\MgII\ 
emission must also contribute to the spectrum in this region, 
but \MgI\ and \mgii\ absorption are so strong and 
abrupt that only a narrow sliver of probable \mgii\ emission remains,
just longward of the onset of those troughs.

Shortward of \mgii, the overlapping troughs make detailed line identifications
quite difficult.  However, we know that these objects are FeLoBALs because they
show \feii\,UV62,63 absorption at 2750\,\AA\ (e.g., Figure \ref{f_0300}) and
because they show \feii*\,UV1 absorption.  The latter can be 
identified because absorption from excited levels in the UV1 multiplet extends
to 2632\,\AA, while ground level UV1 absorption extends only to 2600\,\AA.
Given low-ionization troughs 13,000$-$19,000\,\kms\ wide or wider, \feii\ 
absorption can blanket the spectrum down to at least 2100\,\AA.  A recovery of
flux is detectable in all these objects around the \CIII\ line, from 2100\,\AA\ 
down to the onset of \aliii\ absorption at 1860\,\AA.  \aliii\ blended with
\feii\ and other low-ionization lines (cf. Figure \ref{f_hiscal}) can overlap
with \alii\ and still more \feii\ all the way down to \civ.  Then, given that
high ionization troughs are typically broader than low ionization troughs, it is
not surprising that the \civ\ trough overlaps with \SIiv\ and \SIiv\ with \Nv,
so that the entire spectrum down to \lya\ is essentially extinguished.

We assume the continuum in these objects shortward of \mgii\ 
is flat in $F_{\lambda}$ at the level of the 3100\,\AA\ window (indicated
by a short dot-dashed line in each Figure).  This window may be affected by
absorption from \HeI, \feii\ or even \OIII, as well as by reddening, but in
the three objects with significant coverage longward of it, this window is
not a bad match to the continuum.
%
We use 19000\,\kms\ as the upper velocity limit when calculating the AI and BI
from \mgii\ in most of these objects (Appendix~\ref{BI}).  The maximum AI in 
that case is 19000\,\kms, while the maximum BI depends 
on the detachment velocity of the BAL trough.  For SDSS~0437$-$0045,
we use \civ\ to measure the AI and BI (maximum values 25000\,\kms\ and 
20000\,\kms\, respectively) 
since our spectra fully cover the \civ\ region 
in that object.

\subsubsection{SDSS~0300+0048} \label{sdss0300} 

In addition to an extremely broad \mgii\ BAL trough, SDSS~0300+0048
(Figure~\ref{f_0300}) shows a strong \caii\ H\&K BAL trough.  This \caii\ 
absorption appears split into two relatively narrow systems, at $z=0.8675$ and
$z=0.8775$ ($\Delta v$=1550\,\kms), although there is also broad \caii\ 
absorption extending a further 2000\,\kms\ shortward.  There is associated
\mgii\ and \MgI\ absorption 2300\,\kms\ longward of the highest redshift
\caii\ system, at $z=0.89191 \pm 0.00005$.  
We have adopted this latter value as the systemic redshift.  Associated narrow
\mgii\ systems with $z_{abs} > z_{em}$ do exist, but 2300\,\kms\ 
would be an extreme velocity for such a system \markcite{fol86}({Foltz} {et~al.} 1986), whereas
BAL troughs detached by 2300\,\kms\ shortward of $z_{em}$ are unremarkable.
\caii\,H\&K absorption in BAL outflows has been seen before only in the
Seyfert\,1/LoBAL Mrk~231 \markcite{bok77}({Boksenberg} {et~al.} 1977) and the FeLoBALs FBQS~1044+3656
\markcite{wea00}({White} {et~al.} 2000) and Q~2359$-$1241 \markcite{aea01}({Arav} {et~al.} 2001a).
SDSS~0300+0048 is also a FeLoBAL, with \feii\ absorption 
at 2750\,\AA\ and near 2400 and 2600\,\AA.  However, the \feii\ BAL trough 
is associated only with the $z=0.8675$ \caii\ system 
($v=3900$\,\kms), while the \mgii\ BAL trough 
begins at $z=0.8775$, the redshift of the other \caii\ system ($v=2300$\,\kms).

Note that SDSS~0300+0048 is a binary quasar with SDSS~J025959.69+004813.5,
a non-BAL quasar 
located 19\farcs5 away at $z=0.894\pm0.001$ ($\Delta v = 330 \pm 160$\,\kms).

\subsubsection{SDSS~1154+0300} \label{sdss1154} 

The smooth troughs in SDSS~1154+0300 (Figure~\ref{f_1154}) make its redshift
difficult to pin down.  We adopt $z=1.458\pm0.008$ from various emission
features.  The absorption probably begins at this redshift (e.g., in \AlIII),
but does not reach its full depth 
until $z=1.36$.  Rest wavelengths at the latter redshift are plotted along the
top axis of Figure~\ref{f_1154}.  The more rapid onset of absorption in 
\aliii\ compared to \mgii\ suggests that the BAL region may cover the 
\aliii\ broad line region but not the \mgii\ broad line region.

\subsubsection{SDSS~0819+4209} \label{sdss0819} 

SDSS~0819+4209 (Figure~\ref{f_abruptbals}a) was discovered in a search for 
$z\gtrsim 5.8$ quasars among $i$-dropouts in SDSS images \markcite{sdss103}({Fan} {et~al.} 2001).
The sharp drop due to \mgii\ absorption mimicks the onset of the
Ly$\alpha$ forest at high redshift. 
The spectrum was obtained with ESI \markcite{esi98}({Epps} \& {Miller} 1998) at Keck II on UT 19 Mar 2001.
We adopt $z=1.9258\pm0.0006$ from unresolved \mgi\ absorption, accompanied by
broader \mgii\ absorption, located 2180$\pm$80\,\kms\ longward of the onset of
the absorption troughs at $z=1.9046\pm0.0005$.

\subsubsection{SDSS~1730+5850} \label{sdss1730} 

SDSS~1730+5850 (Figure~\ref{f_abruptbals}b) has $z=2.035\pm0.005$ 
and shows no flux, within the errors, below \alii\,$\lambda$1670.
It may have very extensive \HeI\ absorption starting just shortward of
$z=1.980\pm0.005$, the onset redshift of the \mgii\ and \mgi\ BAL troughs,
though this needs confirmation given the strong telluric absorption 
at those observed wavelengths.  
This object was discovered using the Double Imaging Spectrograph (DIS) at the
APO 3.5m on UT 27 May 2000 during exploratory spectroscopy of SDSS objects with
odd colors.  Spectra were later obtained
twice by the SDSS, on UT 23 Aug 2000 and 18 Apr 2001. 
No significant variability was detected between those observations 
(not surprising given the low SNR of the spectra).  The coadded SDSS spectrum
is a factor of $\sim$2.25 fainter than the discovery spectrum. This discrepancy,
while large, could be due to the uncertainties in the SDSS fluxing and in
comparing slit and fiber spectra.  The scaled
SDSS spectrum shows good agreement with the discovery spectrum at
$\lambda_{rest}>2600$\,\AA, but the `peaks' at 1900\,\AA, 2100\,\AA\ and
2500\,\AA\ are stronger by a factor of 1.4--2.  That is, the absorption
at those wavelengths appears weaker than in the discovery spectrum.  
Nonetheless, the uncertainties are so large that we do not feel this is a firm
detection of variability, though
it does suggest that careful monitoring of this object might be worthwhile.
In Figure~\ref{f_abruptbals} we have therefore summed the scaled discovery
spectrum and the coadded, smoothed SDSS spectrum 
to achieve the best SNR and wavelength coverage.

\subsubsection{SDSS~0437$-$0045}	\label{WEIRD2}

SDSS~0437$-$0045 was selected from early SDSS images by XF and MAS as having
unusual colors \markcite{fan99c}({Fan} {et~al.} 1999).  Optical spectra were obtained at the APO 3.5m
using DIS on UT 22 Mar 1999, 29 Dec 1999, and 04 Jan 2000, at Keck II using the
Low Resolution Imaging Spectrograph \markcite{lris}(LRIS; {Oke} {et~al.} 1995) on UT 15 Oct 1999,
and at the Canada-France-Hawaii Telescope (CFHT) using the 
Subarcsecond Imaging Spectrograph \markcite{cra92}(SIS; {Crampton} {et~al.} 1992) on UT 27 Jan 2001.  
Standard IRAF\footnote{The Image Reduction and Analysis Facility 
is distributed by the National Optical Astronomy Observatories.} 
reduction procedures were used to obtain flux-calibrated spectra.
Near infrared spectra covering 1.45-2.09~$\mu$m at resolution 0.0025~$\mu$m and
1.0-1.32~$\mu$m at resolution 0.0050~$\mu$m were obtained using the Cooled
Grating Spectrometer \markcite{mou90}(CGS4; {Mountain} {et~al.} 1990) at the United Kingdom Infrared 
Telescope (UKIRT) on UT 27 and 30 Sep 1999 respectively.  
The longer wavelength spectrum was obtained under non-photometric conditions.  
The data were reduced using the Starlink Figaro package.  Near-IR (NIR)
photometry obtained by HWR at Calar Alto in Nov. 1999 yields 
$J=17.2$, $H=16.7$ and $K=16.2$, 
making $J-K$ somewhat bluer than average for a quasar.
The spectra at 1.1-1.312\,\micron\ and 1.5-1.75\,\micron\ were
normalized to the fluxes expected from the $J$ and $H$ photometry, respectively.
A further correction of 1.33 was applied to the longer-wavelength spectrum to
make the spectral slope continuous between $J$ and $H$.  The true slope is
probably bluer than shown, since the more trustworthy photometry indicates the
object is bluer than the the long-wavelength spectrum is.

Figure \ref{f_weird2} shows the combined optical (Keck) and NIR (UKIRT) spectrum
of SDSS~0437$-$0045, which finally enabled us to identify the object after being
stumped by it for some time.  It is a strongly absorbed quasar with troughs that
reach peak depth at $z=2.753\pm0.002$.  There is evidence for \aliii\ absorption
at $z=2.8183\pm0.0009$.  Since the lack of abrupt long-wavelength edges to the
absorption troughs in the optical suggests that the systemic $z$ is higher than
the peak absorption $z$, we adopt $z=2.8183\pm0.0009$ as our systemic redshift.
In Figure\,\ref{f_weird2} we plot emission lines at this redshift and absorption
lines at the peak absorption redshift. There seems to be considerable absorption
from neutral gas, namely several \fei\ lines, \MgI, and \mgi\,$\lambda$2026, the
latter possibly blended with \znii.  There is no evidence for \caii\ absorption,
but we plot its wavelengths for reference. As for emission, \civ\ appears absent
because it is completely eaten away by \feii\ multiplets UV44,45,46.  
Note that the lines we identified as \oiii\ at $z=2.74389$ in \markcite{sdss90}{Hall} {et~al.} (2001)
are at the wavelengths of night sky lines, and are almost certainly not real.

This object may have H$\beta$ absorption nearly 10$^4$\,\kms\ wide,
with rest-frame EW$\sim$100~\AA. H$\beta$ absorption in AGN has previously been
seen only in NGC~4151 \markcite{ak69,ser99}({Anderson} \& {Kraft} 1969; {Sergeev} {et~al.} 1999), and there with $\leq$1000\,\kms\ width
and $\leq$3\,\AA\ rest-frame EW.  
However, the SNR is not high enough to be sure this dip
in the continuum is real. Even if so, it may just be a gap between \hb\ and 
\feii\,Opt37,38 emission (e.g., Figure 4 of \markcite{lip94}{Lipari} 1994; see also the
spectrum of FBQS~1408+3054 in \S\ref{GRADPC}).  The expected wavelengths of
emission from this and three other strong \feii\ emission multiplets are plotted
in Figure\,\ref{f_weird2}.  The features $\sim$75\,\AA\ to the red of each
of them argue that SDSS~0437$-$0045 has strong \feii\ emission at a slightly
higher systemic redshift than we have assumed, and not an \hb\ BAL trough.
A better spectrum is needed. 

\begin{figure}
\epsscale{1.75}
\plottwo{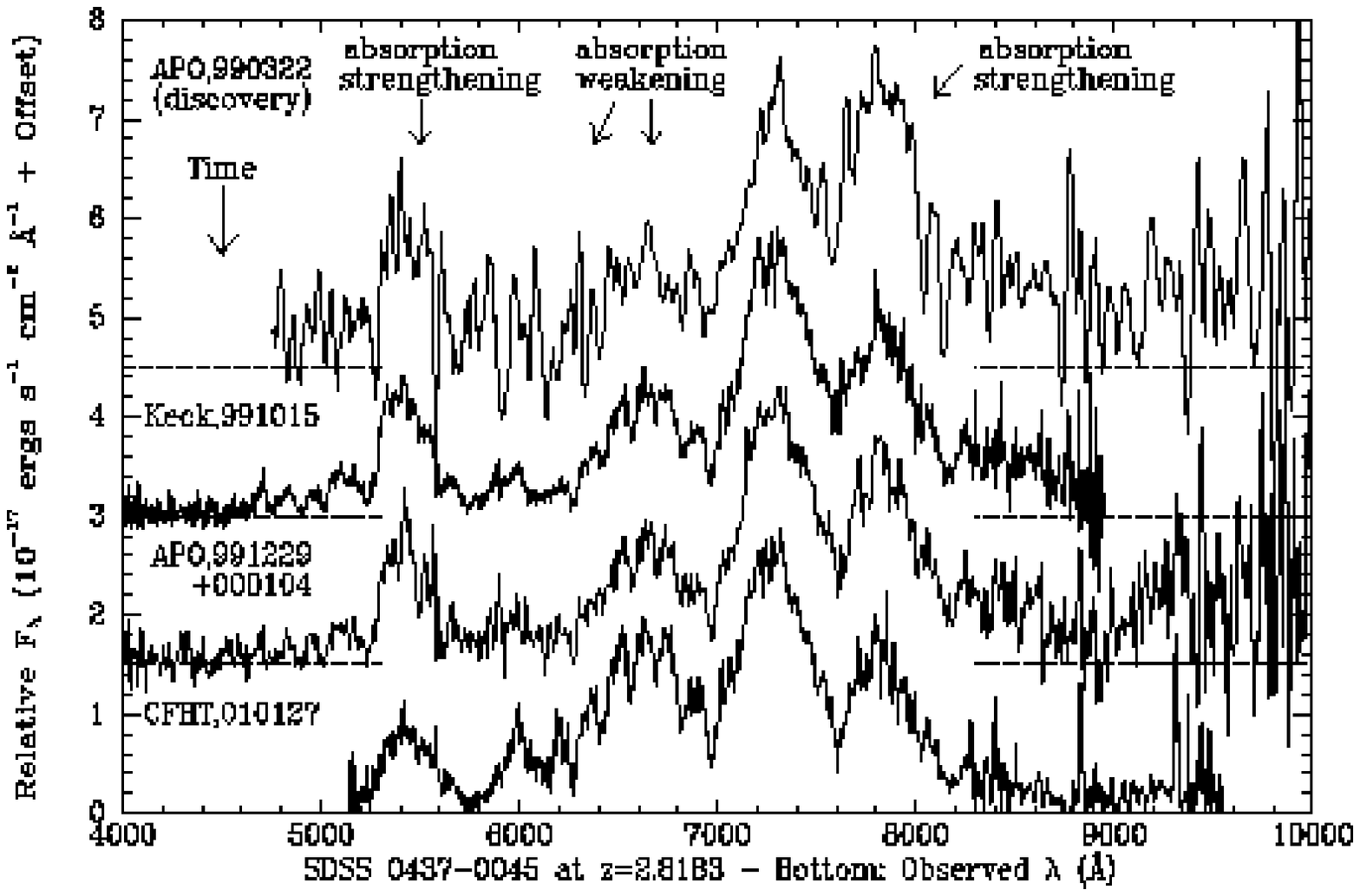}{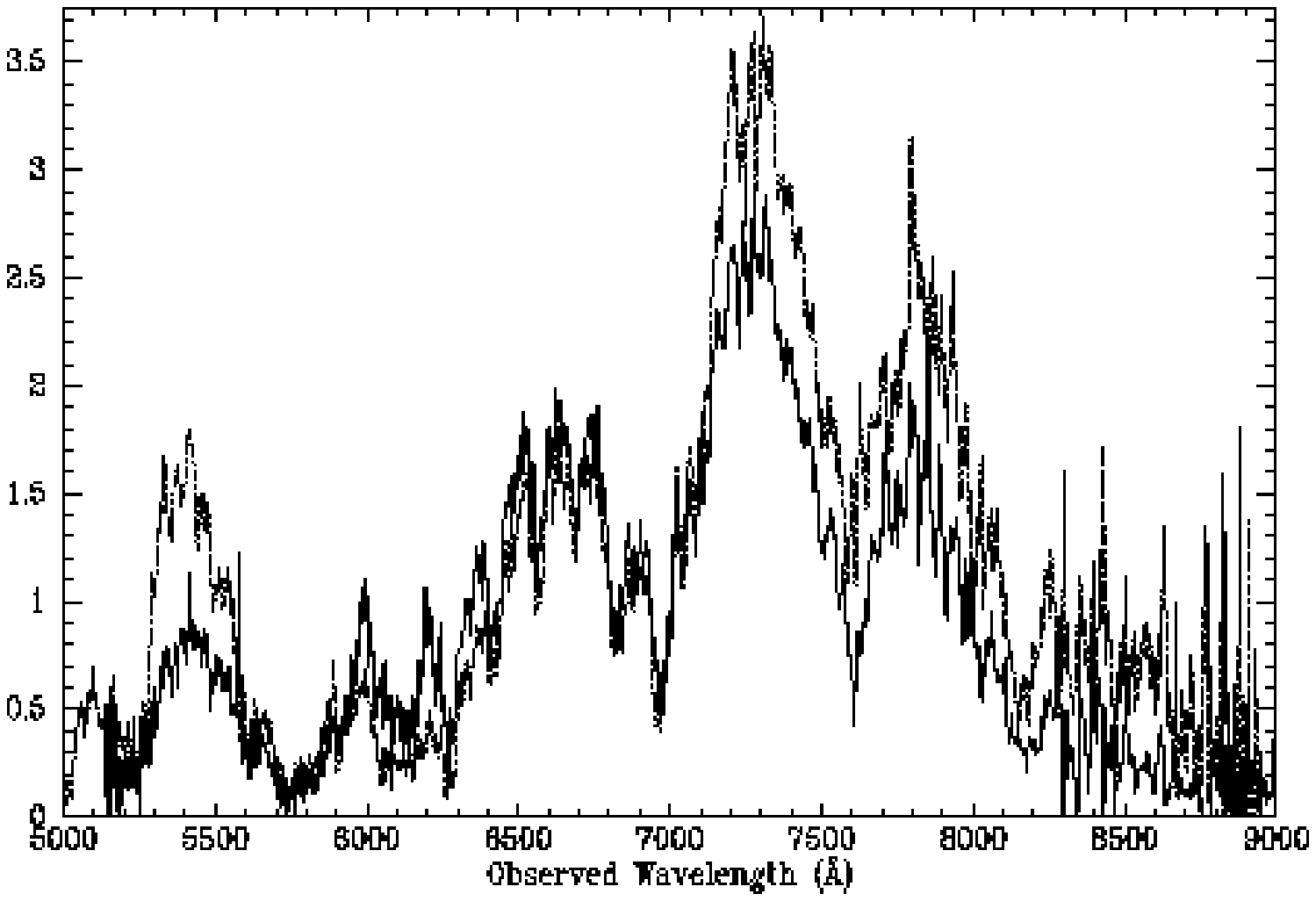}
\caption[]{ \singlespace 
a) Optical spectra of SDSS~0437$-$0045 at four epochs, with time increasing
downwards.  All spectra have been normalized relative to the flux in the CFHT
spectrum at 7000$-$7600\,\AA. The upper 3 spectra have been offset by +1.5 units
in $F_{\lambda}$ from each other; the dashed lines show the zero flux levels.
The APO discovery spectrum
has been smoothed by a 5-pixel boxcar and the CFHT spectrum by a 3-pixel boxcar.
The labels note the behavior of the absorption with time in the regions of the
spectrum indicated, relative to the absorption in the 7000$-$7600\,\AA\ region.
b) The Keck spectrum (dotted 
line) compared to the 3-pixel-smoothed CFHT spectrum (solid line), with no
mutual normalization applied.  This demonstrates the typical absolute flux
calibration uncertainties among our spectra, which prevent us from determining 
at exactly which wavelengths the absorption has varied.  
However, it is still clear that relative to the absorption at 7000$-$7600\,\AA,
the absorption at 5400\,\AA\ strengthened and that at 5900$-$7000\,\AA\ 
weakened between the epochs of the Keck and CFHT spectra.
Only $\sim$90 rest-frame days separate the two epochs.
}\label{f_w2time}
\end{figure}

The absorption shortward of 2500\,\AA\ in this object has varied with an 
unusually high
amplitude and rate of change for BAL quasars.  Figure\,\ref{f_w2time}a shows the
optical spectrum of SDSS~0437$-$0045 at four different epochs.  The spectra
have been normalized at 7000$-$7590\,\AA, where the SNR is highest, to match
the CFHT spectrum. 
For reference,
Figure \ref{f_w2time}b compares the CFHT and Keck spectra without normalization,
illustrating the typical variation in absolute flux among our spectra.  
Our normalization accounts for the uncertainty in the absolute flux calibration
of these narrow slit observations.  The relative wavelength-dependent flux
calibrations are trustworthy; for example, the normalized Keck and second-epoch
APO spectra, taken within 20 rest-frame days of each other,
agree completely within the errors across the wavelength range 4500$-$9000\,\AA.
This also reassures us that the variability is not due to fluctuations in
system sensitivities or flatfields.  

To calculate the variation of the absorption strength in SDSS~0437$-$0045, we
assume the true continuum is flat in $F_{\lambda}$ at $5\times10^{-17}$
ergs cm$^{-2}$\,s$^{-1}$\,\AA$^{-1}$.  Then, relative to the 7000$-$7590\,\AA\ 
region, in the 90 rest-frame days between the Keck and CFHT spectra epochs the
absorption weakened by 5$\pm$1\% at 5200$-$5600\,\AA\ (high-velocity \civ) and
strengthened by 8$\pm$1\% at 5900$-$7000\,\AA\ (where narrow troughs of \feii,
\alii\ and \aliii\ are visible) and 5$\pm$2.5\% at 7590$-$8150\,\AA\ 
(\mgi\ and \feii, plus \znii, \crii\ and \feiii?).
This last spectral region contains the strong telluric O$_{\rm 2}$ absorption
band at 7590$-$7700\,\AA, but the increase in absorption is larger than the
telluric correction applied and the region of apparently increased absorption is
wider than the absorption band.  
This increase in the relative absorption at 7590$-$8150\,\AA\ is somewhat less
than the 15$\pm$3\% increase in this region in the 54 rest-frame days between
the discovery and Keck spectra (Figure\,\ref{f_w2time}a). 

The variable absorption in SDSS~0437$-$0045 is discussed further in \S\ref{VAR}.


\subsection{Heavily Reddened BAL Quasars} \label{REDDENED}

SDSS has discovered a number of heavily reddened BAL quasars.
After a short discussion of how we determine the reddening in these objects,
we present two reddened mini-BALs with strong \feii\ emission,
then several extremely reddened objects with no strong emission,
and finally a bright reddened FeLoBAL.
The implications of all these objects are discussed in \S\ref{IMPRED}.

\subsubsection{Estimating the Reddening} \label{GETRED}

Since the `typical' quasar has a very blue spectrum, reddened quasars are easy
to identify.  Determining the amount of reddening is more difficult.  We assume
all reddening occurs at the quasar redshift with a Small Magellanic Cloud (SMC)
extinction curve \markcite{smc84}({Prevot} {et~al.} 1984) and $R_V=3.1$ (see below).  We deredden the
spectrum until the continuum slope matches that of the composite SDSS quasar
of \markcite{sdss73}{Vanden Berk} {et~al.} (2001).  The value of the color excess \ebv\ needed to achieve this
match is our estimated reddening.  The uncertainties on \ebv\ for each
object denote the range for which an acceptable match can be found.


We use the SMC extinction curve because the 2200\,\AA\ bump present in the LMC
and Milky Way extinction curves has never been detected from dust around quasars
\markcite{pcg00}(e.g., {Pitman}, {Clayton}, \& {Gordon} 2000), although it has been detected from dust in intervening
\mgii\ systems \markcite{mal97,coh99}({Malhotra} 1997; {Cohen} {et~al.} 1999).  The SMC curve was also used by \markcite{sf92}{Sprayberry} \& {Foltz} (1992)
and \markcite{bro01b}{Brotherton} {et~al.} (2001b), both of whom found \ebv=0.1 for a `typical' LoBAL spectrum.
The other commonly used extinction curve is the Calzetti formula \markcite{cks94}({Calzetti}, {Kinney}, \&  {Storchi-Bergmann} 1994).
This formula was derived for active star formation regions and 
empirically incorporates the `selective attenuation' effects of dust, 
including extinction, scattering, and geometrical dust distribution effects.
The Calzetti extinction curve does not have a 2200\,\AA\ bump, but like the LMC
and MW curves it is much greyer (less steep) than the SMC extinction curve.
Our use of the SMC curve instead of the Calzetti curve is conservative in the
sense that it requires a lower \ebv\ (and thus lower extinction) for a given
observed ultraviolet slope.


Two more caveats to our dereddening procedure are needed.  First, we assume a
single value of \ebv\ over all sightlines to the emission regions.
A range of \ebv\ \markcite{hw95}(e.g., {Hines} \& {Wills} 1995) or a wavelength-dependent contribution
from scattered light \markcite{bro01}({Brotherton} {et~al.} 2001a) 
will complicate the interpretation of our derived single \ebv.  However,
sometimes we can tell when these effects are important (e.g.,
SDSS~0342+0045; see below).
Second, if an accretion disk is producing the observed emission and the amount
of extinction is correlated with the inclination of the disk, then dereddening
to match the SDSS composite will introduce a systematic error in the derived
\ebv\ since the intrinsic continuum of the disk is likely to be a function of
viewing angle \markcite{hub00}(e.g., {Hubeny} {et~al.} 2000).


\subsubsection{Two Reddened Strong \feii-emitting Mini-BAL Quasars}  \label{WEIRD1}

\begin{figure}
\epsscale{1.75}
\plottwo{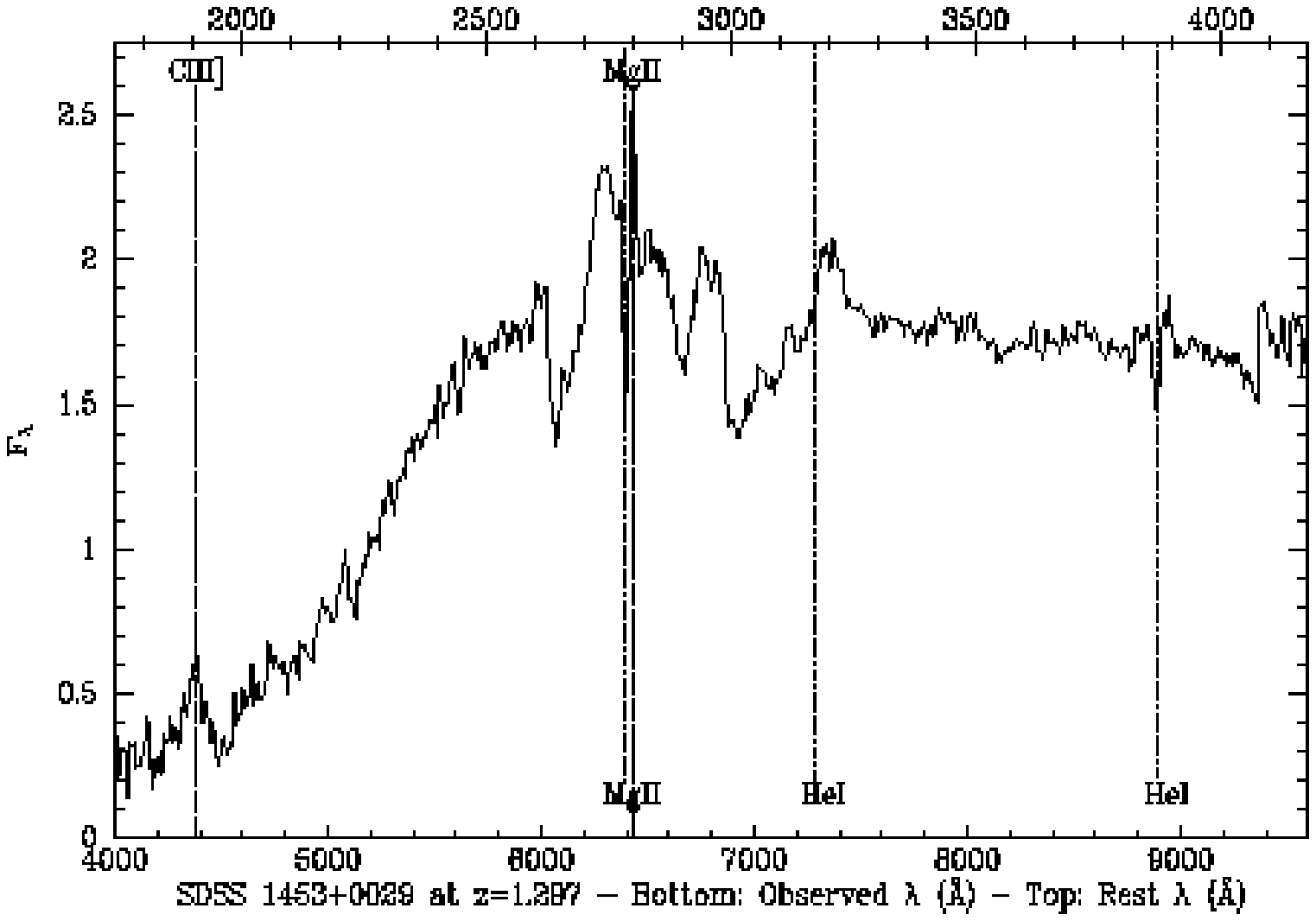}{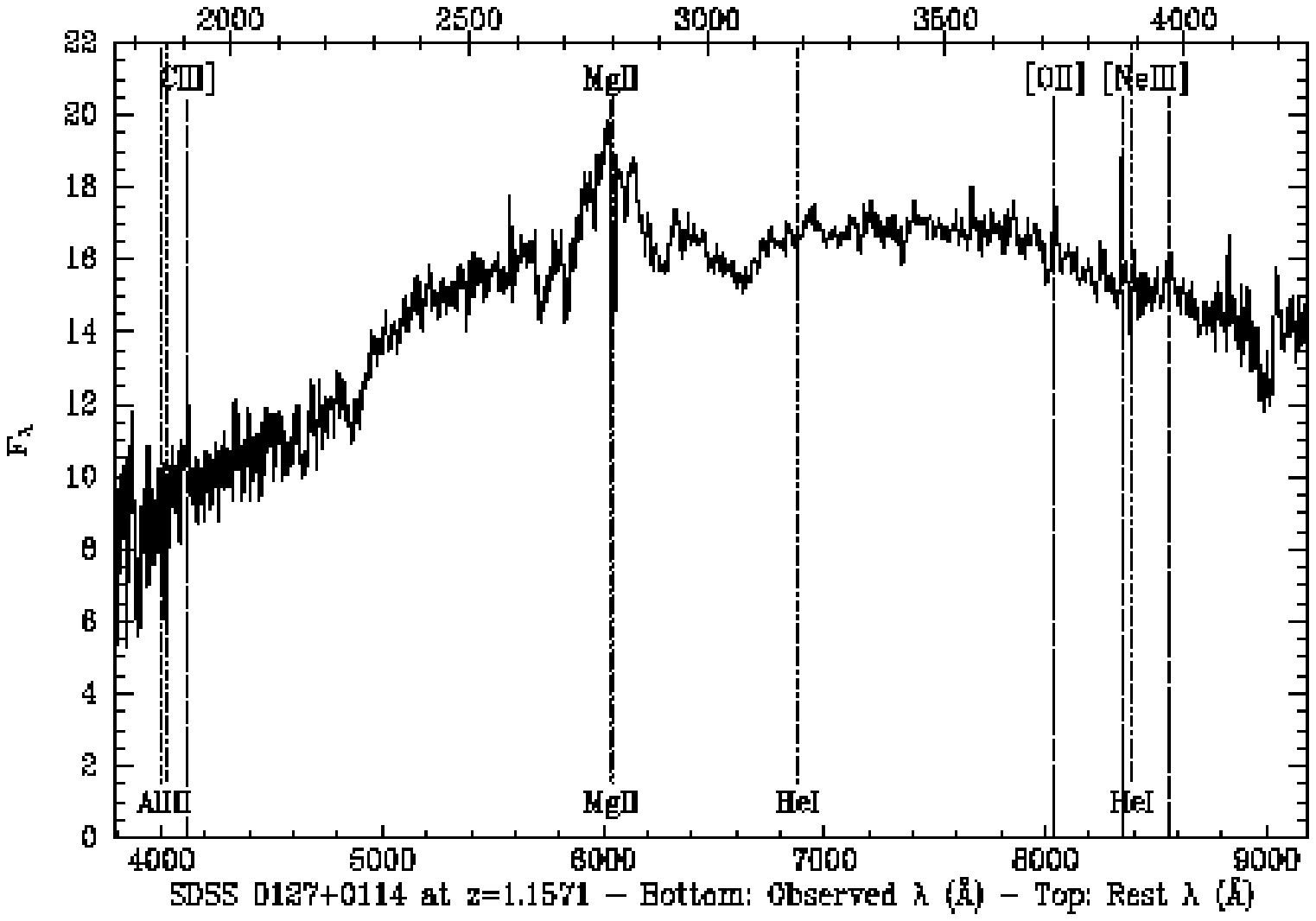}
\caption[]{ \singlespace 
a) SDSS~1453+0029,
a reddened extreme \feii-emitting mini-LoBAL quasar at $z=1.297$.
b) SDSS~0127+0114,
a reddened strong \feii-emitting mini-LoBAL quasar at $z=1.1571$.
In both plots, dashed lines denote emission, and dotted lines absorption.
The absorption lines are identified along the bottom of each plot
and the emission lines along the top.
}\label{f_weird1}
\end{figure}

SDSS~1453$+$0029 was identified by XF and MAS 
as having unusual colors in early SDSS imaging \markcite{fan99c}({Fan} {et~al.} 1999).
A discovery spectrum was obtained at the APO 3.5m, 
and a 1200s followup spectrum 
at Keck II on UT 05 Apr 2000 (Figure~\ref{f_weird1}a).  The object was observed
numerous times as a quasar candidate in the SDSS spectroscopic survey.
We adopt the redshift $z=1.297\pm0.001$ determined by the pipeline from one of
these observations (the other yielded no redshift), since it 
agrees well with the Keck spectrum.

The redshift comes from narrow \mgii\ emission and weak \ciii\ emission.
The object is similar to Q~2359$-$1241 \markcite{bro01}({Brotherton} {et~al.} 2001a); both are reddened quasars
with narrow \mgii\ emission and absorption atop a complex broader structure.
Both objects also show \HeI\ and \HEI\ in absorption \markcite{aea01}({Arav} {et~al.} 2001a).
SDSS~1453+0029 has BI=0, but AI(\mgii)=253$-$477\,\kms\ (the lower
value is from the Keck spectrum and the higher value from the coadded SDSS 
spectrum), with $v_{max}=1940-2210$\,\kms.

A similar object is SDSS~0127+0114, 
which has \aliii, \mgii, \HeI\ and \HEI\ absorption and narrow \ciii, \oii\ and
\neiii\ emission, as well as broader \mgii\ emission, at $z=1.1571\pm0.0002$
(Figure~\ref{f_weird1}b).  SDSS~0127+0114 has zero balnicity, 
but the presence of \hei\ argues that the absorption is a mini-BAL trough and
not simply an associated narrow absorption line system.

\begin{figure}
\epsscale{1.0}
\plotone{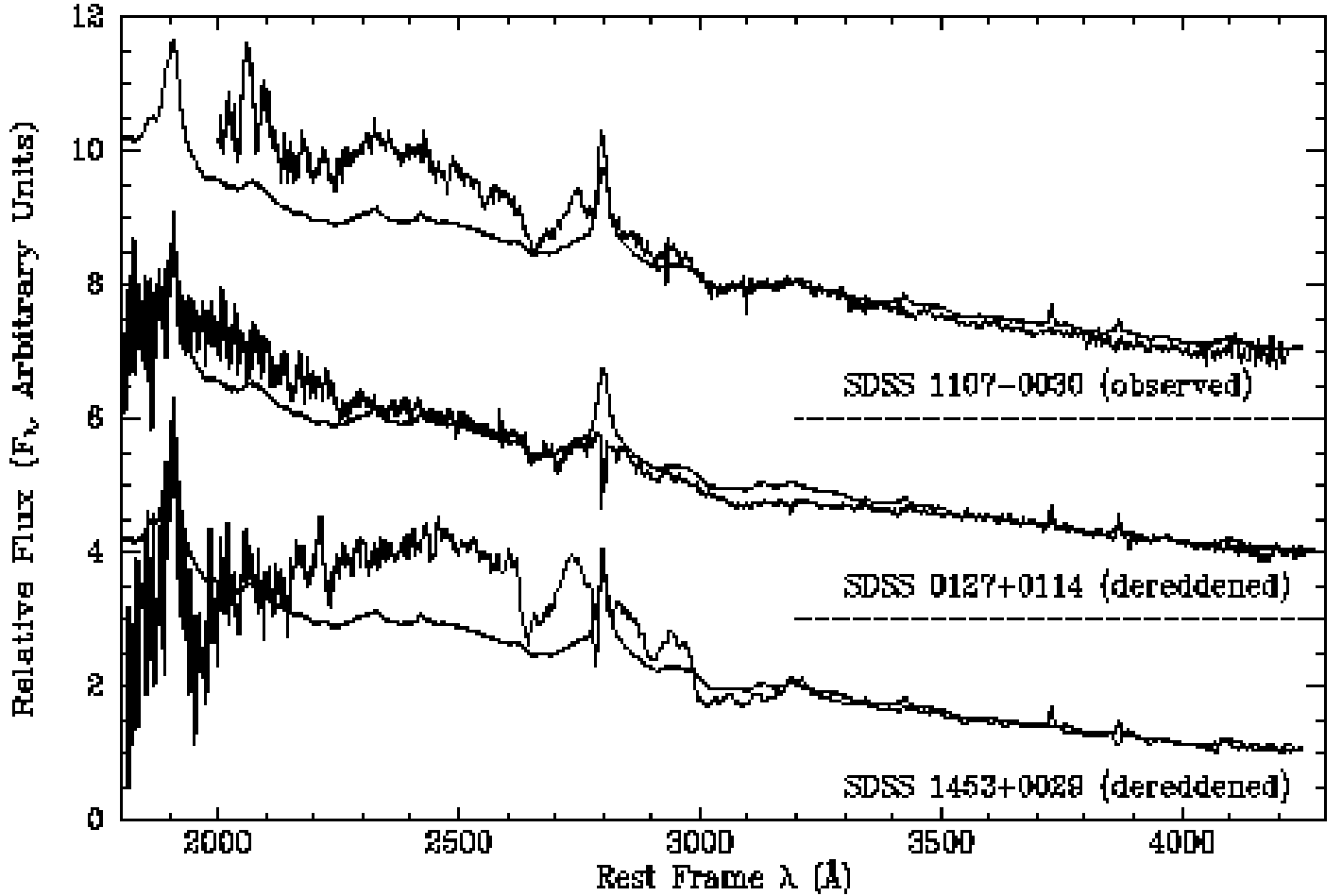}
\caption[]{ \singlespace 
Bottom: the spectrum of SDSS~1453+0029 dereddened by \ebv=0.5 and plotted atop
the composite SDSS quasar after normalizing the two spectra at $>$3350\,\AA\ in
the rest frame.
Middle: the spectrum of SDSS~0127+0114 dereddened by \ebv=0.36 and similarly
plotted atop the composite SDSS quasar.  The dotted lines indicate the zero
flux levels for this spectrum and the top one.
Top: the spectrum of the extreme \feii-emitting quasar
SDSS~J110747.45$-$003044.2 plotted atop the composite SDSS quasar to illustrate
how much stronger the \feii\ emission complexes at 2200-2630\,\AA\ and 
2700-3000\,\AA\ are in such objects.
What looks like a detached \mgii\ BAL trough at 2630-2700\,\AA\ in the observed
spectrum of SDSS~1453+0029 is actually the gap between \mgii+\feii\,UV62,63
emission and \feii\,UV1,64 emission shortward of $\sim$2630\,\AA.
Similarly, gaps between \mgii\ and \feii\,UV78 emission
and \feii\,UV78 and \feii\,Opt6,7 
produce apparent troughs at 2900\,\AA\ and 3000-3150\,\AA, respectively.
}\label{f_w1dered}
\end{figure}

We estimate SMC reddenings of \ebv=$0.50\pm0.05$ for SDSS~1453+0029
and \ebv=$0.36\pm0.03$ for SDSS~0127+0114.
These dereddened spectra are plotted atop the normalized composite SDSS spectrum
in Figure~\ref{f_w1dered}, along with the observed spectrum of the extreme
\feii-emitting quasar SDSS~J110747.45$-$003044.2 \markcite{sdssedrq}({Schneider} {et~al.} 2002). 
SDSS~0127+0114 is clearly a normal-to-strong \feii\ emitter,
while SDSS~1453+0029 is intrinsically an extreme \feii-emitting
quasar similar to SDSS~J110747.45$-$003044.2 or Q~2226$-$3905 \markcite{gcc96}({Graham} {et~al.} 1996).
What looks like a detached \mgii\ BAL trough at 2630-2700\,\AA\ in the observed
spectrum of SDSS~1453+0029 is actually the gap between \mgii+\feii\,UV62,63
emission and \feii\,UV1,64 emission shortward of $\sim$2630\,\AA.
Similarly, gaps between \mgii\ and \feii\,UV78 emission
and \feii\,UV78 and \feii\,Opt6,7 
produce apparent troughs at 2900\,\AA\ and 3000-3150\,\AA, respectively.
\feii\ emission this strong remains a challenge for quasar models 
\markcite{sp98}(e.g., {Sigut} \& {Pradhan} 1998).

\subsubsection{Extremely Reddened BAL Quasars} \label{EXTRED}

SDSS has discovered several extremely reddened BAL quasars whose emission lines
in the observed optical are very weak or absent, somewhat reminiscent of
Hawaii~167 \markcite{cow94}({Cowie} {et~al.} 1994).  
For these objects we assume that the onset redshifts
of the absorption troughs are the systemic redshifts;
where present, the weak \ciii\ lines are consistent with these redshifts.
We discuss the observed spectral slopes of these objects in terms of 
$\alpha_\nu$, where $F_\nu\propto\nu^{\alpha_\nu}$.  For reference,
the SDSS quasar composite has $\alpha_\nu=-0.44\pm0.10$, and \markcite{gre01}{Gregg} {et~al.} (2002)
adopted $\alpha_\nu\lesssim-1$ as their working definition of a red quasar.

\begin{figure}
\epsscale{1.75}
\plottwo{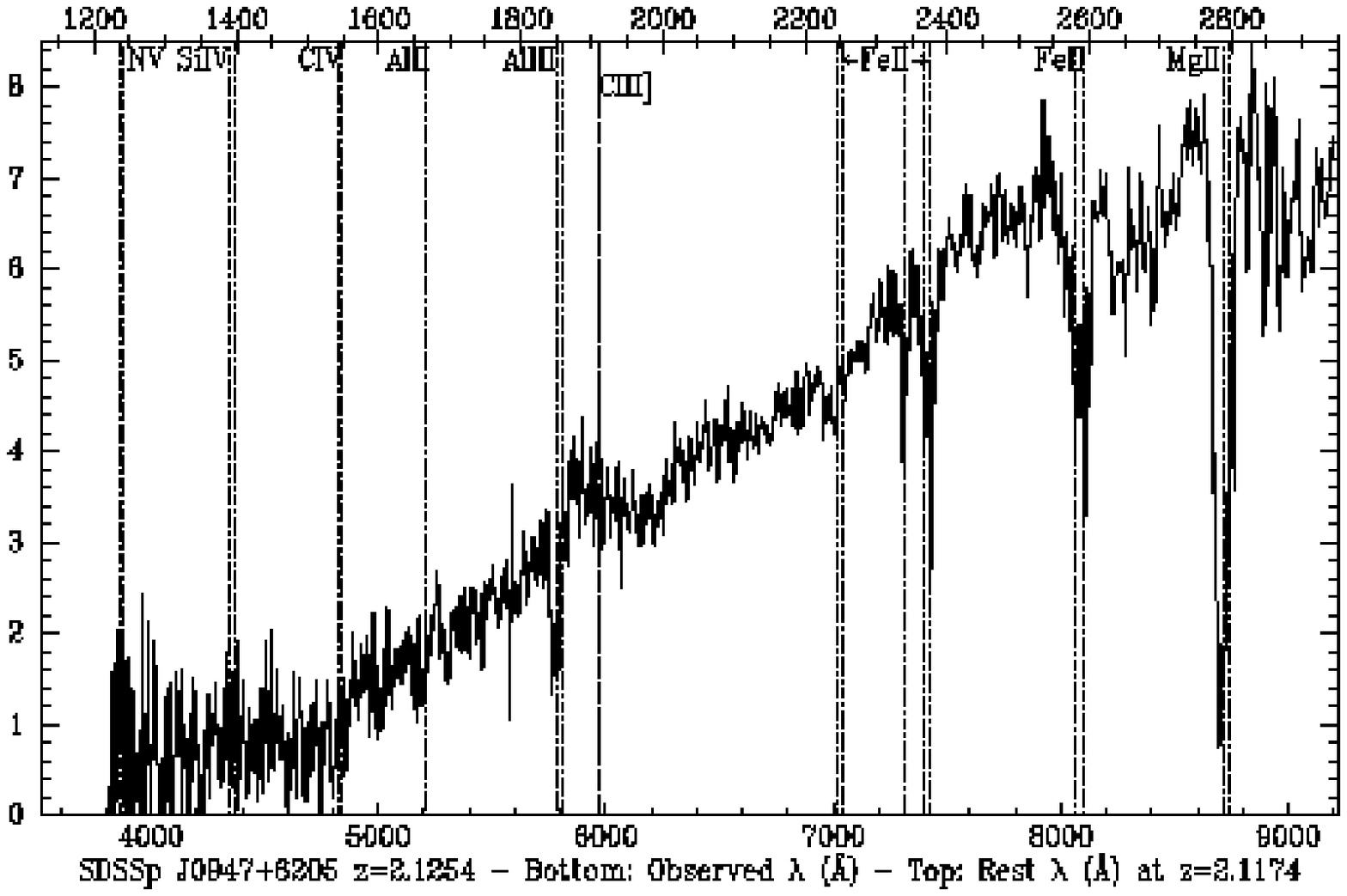}{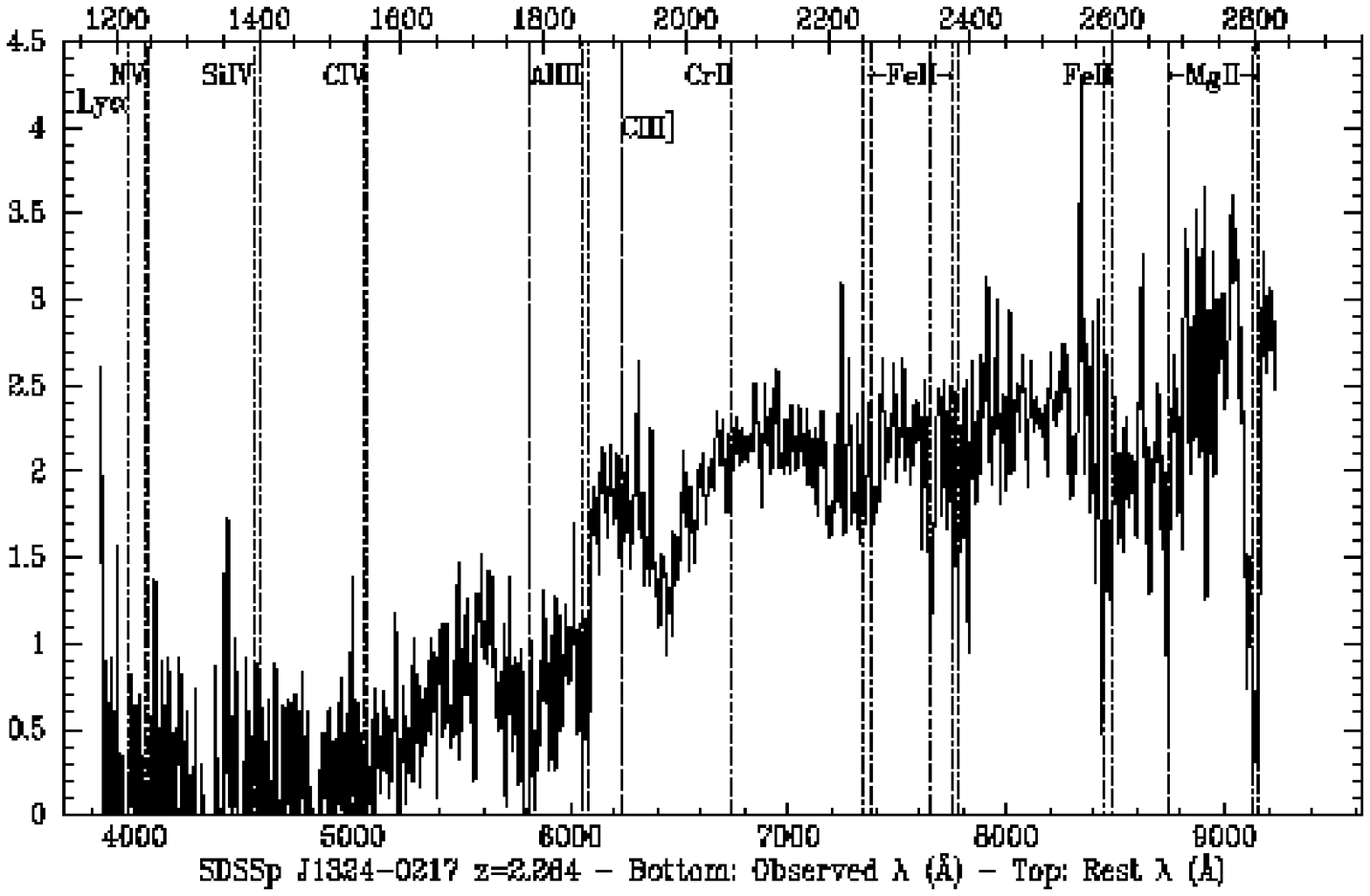}
\caption[]{ \singlespace 
Extremely reddened BAL quasar spectra, smoothed by a 7-pixel boxcar.
Observed wavelengths are shown on the bottom axes, and rest frame on the top.
Dashed lines show emission, and dotted lines absorption.
a) SDSS~0947+6205 at $z=2.1254$.  The absorption lines are identified at
$z=2.1174$, the redshift of peak absorption.
b) SDSS~1324$-$0217 at $z=2.264$. The dot-dashed lines show \aliii\ and \mgii\ 
in the broad absorption system at $z=2.123$. The absorption at $\sim$6400\,\AA\ 
observed, longward of the \ciii\ emission, may be due to \feiii\,UV48 
($\sim$2070\,\AA) absorption in this system, and \feiii\,UV34 absorption
($\sim$1910\,\AA) may be present as well, at $\sim$6000\,\AA\ observed.
}\label{f_redbals12}
\end{figure}

\paragraph{SDSS~0947+6205} \label{sdss0947} 
SDSS~0947+6205 (Figure~\ref{f_redbals12}a) is at $z=2.1254\pm0.0006$
from \feii\ and \mgii\ absorption.  The \mgii\ absorption trough spans
some 3700$\pm$200\,\kms, but the BI=0 since \civ\ region is too noisy
to measure and the \mgii\ absorption begins at the systemic redshift.
The top axis of Figure~\ref{f_redbals12}a gives the rest wavelength at the peak
absorption redshift of $z=2.1174\pm0.0007$ (detachment $v = 770 \pm 90$\,\kms).
\alii, \aliii, and \civ\ absorption are also present, detached by
1500$\pm$70\,\kms\ and with a narrower velocity width of 3700$\pm$200\,\kms.
We measure a spectral slope of $\alpha_\nu=-4.7\pm0.1$, and dereddening by
\ebv=0.43$\pm$0.03 brings the spectrum into fair agreement with the SDSS
composite.  However, there is some suggestion that shortward of \aliii\ the
object's spectral slope may steepen and the required reddening may
increase, or that the reddening curve is different from that assumed,
as we discuss below. 

\paragraph{SDSS~1324$-$0217} \label{sdss1324} 
This object 
(Figure~\ref{f_redbals12}b)
is at $z=2.264\pm0.001$ from broad \mgii, \feii, and \alii\ 
absorption; it may also have \alii\ and \civ\ absorption.
There is also broad \aliii, \mgii, and probably \feiii\,UV34,48 and weak \feii\ 
absorption at $z=2.123\pm0.001$, an outflow velocity of 13230$\pm$130\,\kms;
nonetheless, the object has BI=0. 
We measure a spectral slope of $\alpha_\nu=-2.6\pm0.1$,
but the slope appears to steepen shortward of \aliii.
Dereddening the spectrum to match the SDSS composite confirms this steepening,
which could be intrinsic or due to an extinction curve different from that of
the SMC.  We discuss this in detail in \S\ref{IMPRED}, but here we simply quote
the values of \ebv\ needed to bring the object spectrum into agreement with the
composite at different wavelengths, using the SMC extinction curve.
A reddening of \ebv=0.2$\pm$0.05 is required at 2000$-$2800\,\AA,
but \ebv=0.5$\pm$0.05 is required at 1500$-$2000\,\AA.

\begin{figure}
\epsscale{1.75}
\plottwo{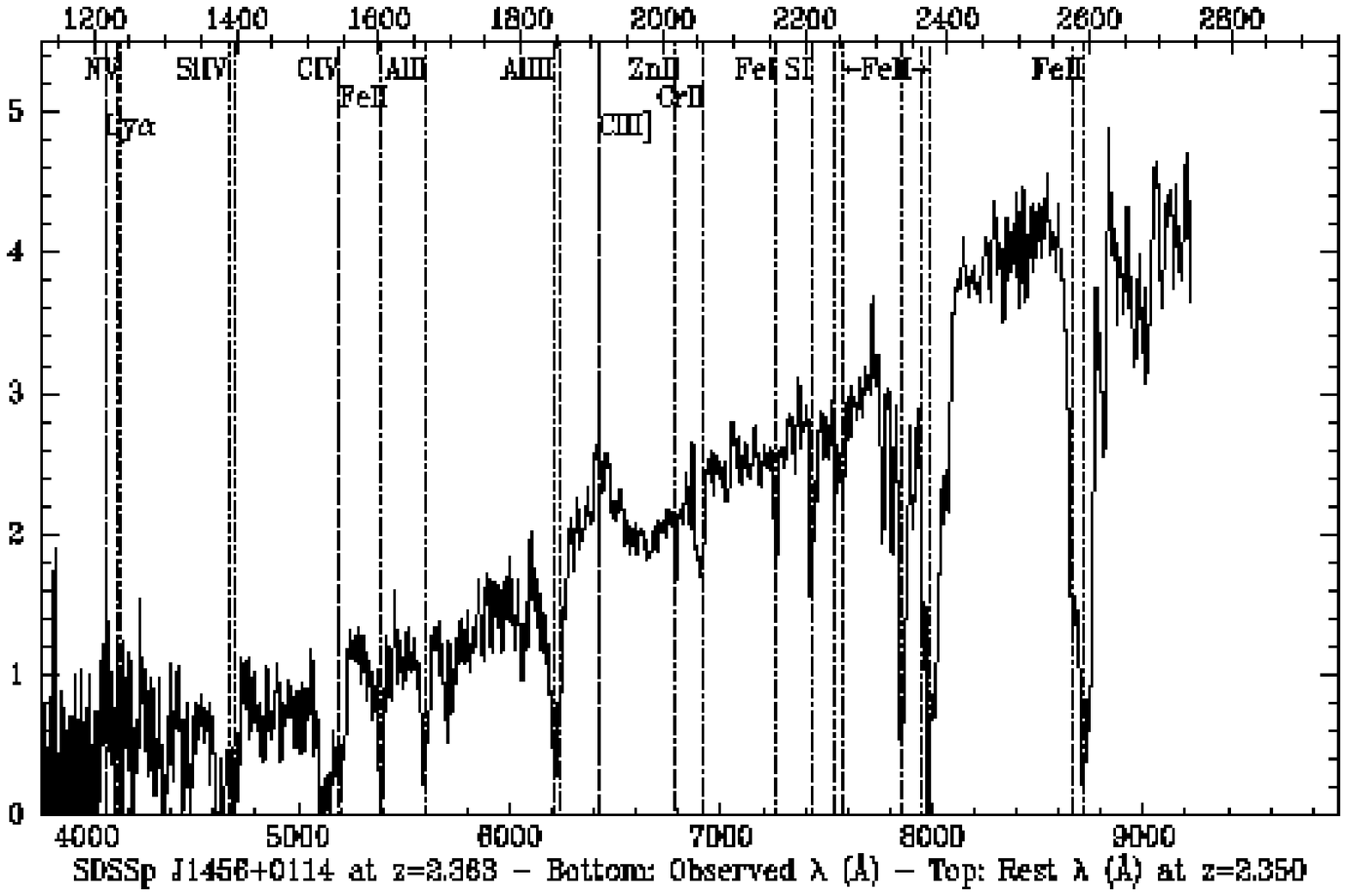}{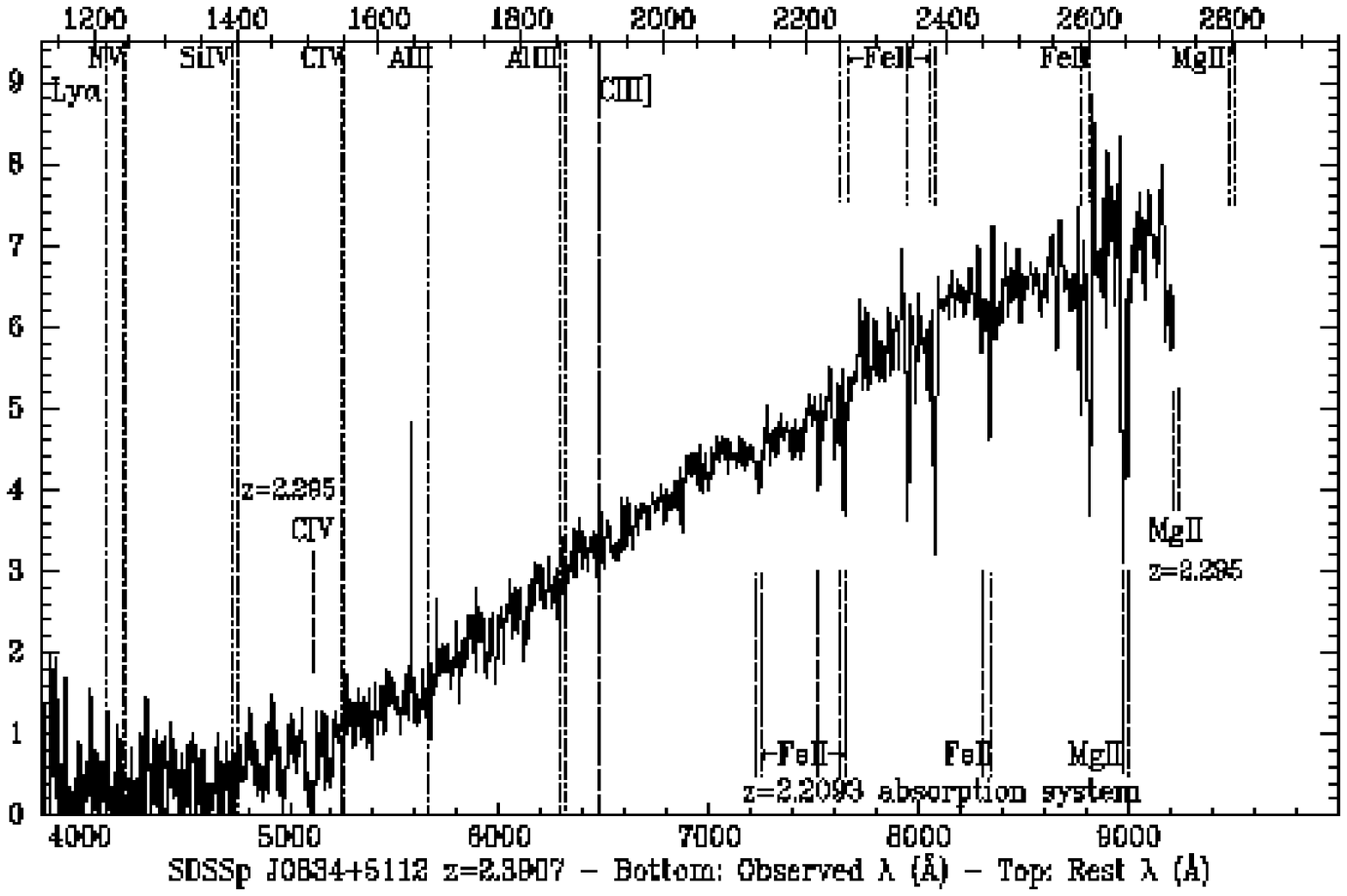}
\caption[]{ \singlespace 
Extremely reddened BAL quasar spectra, smoothed by a 7-pixel boxcar.
Observed wavelengths are shown on the bottom axes, and rest frame on the top.
Dashed lines show emission, and dotted lines absorption.
a) SDSS~1456+0114 at $z=2.363$.  The absorption lines are identified at
$z=2.350$, the redshift of peak absorption.
b) SDSS~0834+5112 at $z=2.3907$.  Absorption lines at this redshift are
identified above the spectrum, while absorption lines from the $z=2.295$ and
$z=2.2093$ systems are identified by dot-dashed lines in the bottom half of
the figure.
}\label{f_redbals34}
\end{figure}

\paragraph{SDSS~1456+0114} \label{sdss1456} 
For this extremely reddened FIRST-detected BAL quasar
(Figure~\ref{f_redbals34}a) we adopt $z=2.363\pm0.008$.
It shows broad \ciii\ emission, weak narrow \lya\ emission, and broad \civ,
\aliii\ and \feii\ absorption along with narrow \SIi, \znii, \crii, \fei\ and
\Si\ absorption at the deepest part of the trough, $z=2.350\pm0.001$ (the top
axis of Figure~\ref{f_redbals34}a gives rest wavelengths at this redshift).
We measure a spectral slope of $\alpha_\nu=-4.6\pm0.1$,   
and there is no evidence for any deviation from that slope.
Dereddening by \ebv=0.40$\pm$0.05 is required to match the SDSS composite.

\begin{figure}
\epsscale{0.65}
\plotone{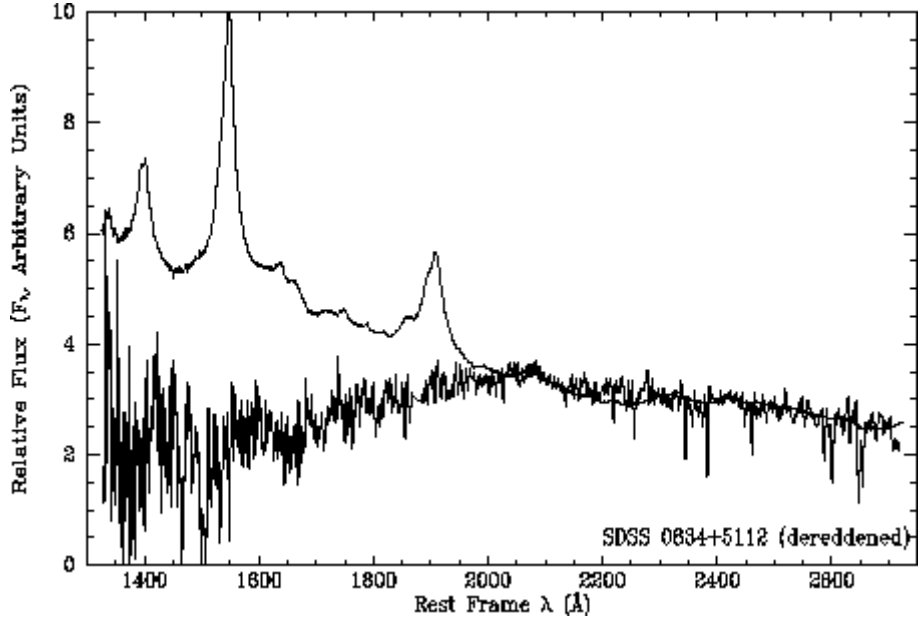}
\caption[]{ \singlespace 
Spectrum of SDSS~0834+5112 smoothed by a 7-pixel boxcar and
dereddened by \ebv=0.3 with the SMC extinction curve to match
the SDSS composite quasar from 2000-2800\,\AA.  To reproduce the composite
spectrum from 1500-2000\,\AA\ as well would require a steeper extinction curve.
SDSS~1324$-$0217 and possibly SDSS~0947+6205 also show evidence for a break in
their dereddened spectra, but SDSS~0834+5112 is the only one of our extremely
reddened objects with sufficient SNR to robustly detect it.
}\label{comp0834p5112sm7_smc0.3}
\end{figure}

\paragraph{SDSS~0834+5112} \label{sdss0834} 
This object 
(Figure~\ref{f_redbals34}b) is a FIRST source at $z=2.3907\pm0.0002$
from \mgii, \feii, \aliii, and weak \civ\ absorption.
The object has two detached absorption troughs: 
one at $z=2.295\pm0.005$ 
in broad \civ\ and \mgii,
and one at $z=2.2093\pm0.0003$ 
in narrow \mgii, \feii, \aliii, \civ\ and \SIiv.  
This latter system could be intervening; the SNR is not high enough
to tell if its \civ\ absorption is broad.  This object 
cannot be adequately fit by a single power law.
The spectrum steepens from $\alpha_\nu=-2.6\pm0.1$ at 2450$-$2725\,\AA\ to
$\alpha_\nu=-4.4\pm0.1$ at 2000$-$2450\,\AA\ and again to
$\alpha_\nu=-6.9\pm0.3$ at 1550$-$2000\,\AA.
Because of this steepening of the continuum,
dereddening by \ebv=0.3$\pm$0.05 is required to match the SDSS composite
at 2000$-$2800\,\AA, but \ebv=0.65$\pm$0.05 is required at 1500$-$2000\,\AA\ 
(Figure\,\ref{comp0834p5112sm7_smc0.3}).

\begin{figure}
\epsscale{0.65}
\plotone{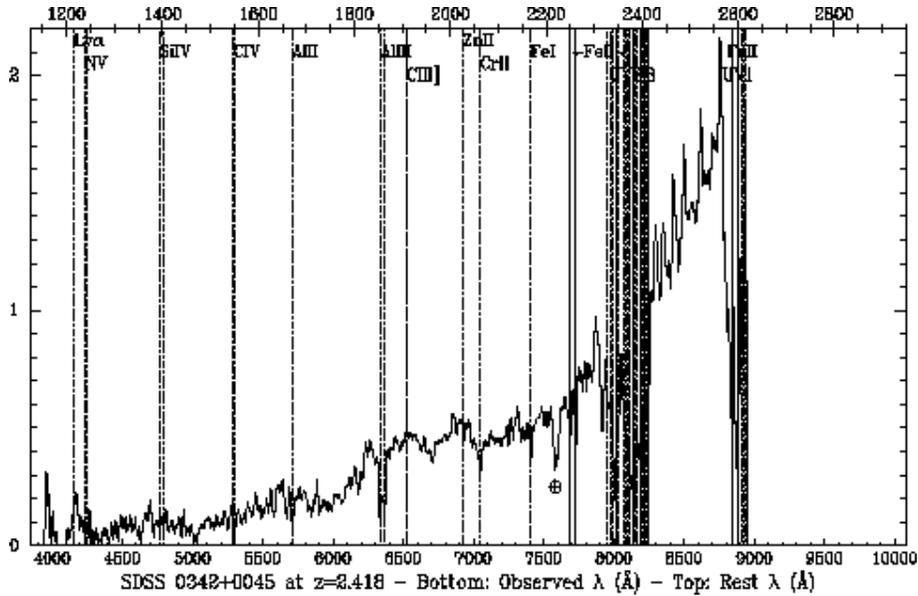}
\caption[]{ \singlespace 
Keck spectrum of SDSS~0342+0045 (smoothed by a 7-pixel boxcar),
another extremely reddened BAL quasar, at $z=2.418$.
Observed wavelengths are shown on the bottom axes, and rest frame on the top.
Dashed lines show emission, and dotted lines absorption.
Wavelengths of \feii\ absorption from multiplets UV1, UV2, and UV3
are indicated with lighter dotted lines.
}\label{f_0342p0045}
\end{figure}

\paragraph{SDSS~0342+0045} \label{sdss0342} 
We adopt $z=2.418\pm0.001$ for this FeLoBAL (Figure~\ref{f_0342p0045}) 
from numerous absorption lines.
The spectrum shown was obtained at Keck II using LRIS 
on UT 15 Oct 1999.  There is weak \civ\ absorption and maybe weak \lya+\Nv\ 
emission, but the spectrum is
dominated by strong \feii\ absorption (multiplets UV1,2,3 at least).
It is obvious from comparison to Figures~\ref{f_redbals12} and \ref{f_redbals34}
that this is a somewhat different sort of extremely reddened BAL quasar.
Instead of the spectrum steepening at shorter wavelengths,
it flattens out.  This is what is expected for a range of reddenings along the
sightlines to the emission regions:
the heavily reddened sightlines contribute only at longer
wavelengths, while less reddened sightlines dominate at shorter wavelengths.
We find that dereddening by \ebv=0.7$\pm$0.1 is required to match the SDSS
composite between 2050$-$2550\,\AA, while only \ebv=0.42$\pm$0.02 is required
between 1250$-$1850\,\AA.

\subsubsection{SDSS~0318$-$0600: A Bright, Reddened FeLoBAL} \label{sdss0318}

\begin{figure}
\epsscale{1.75}
\plottwo{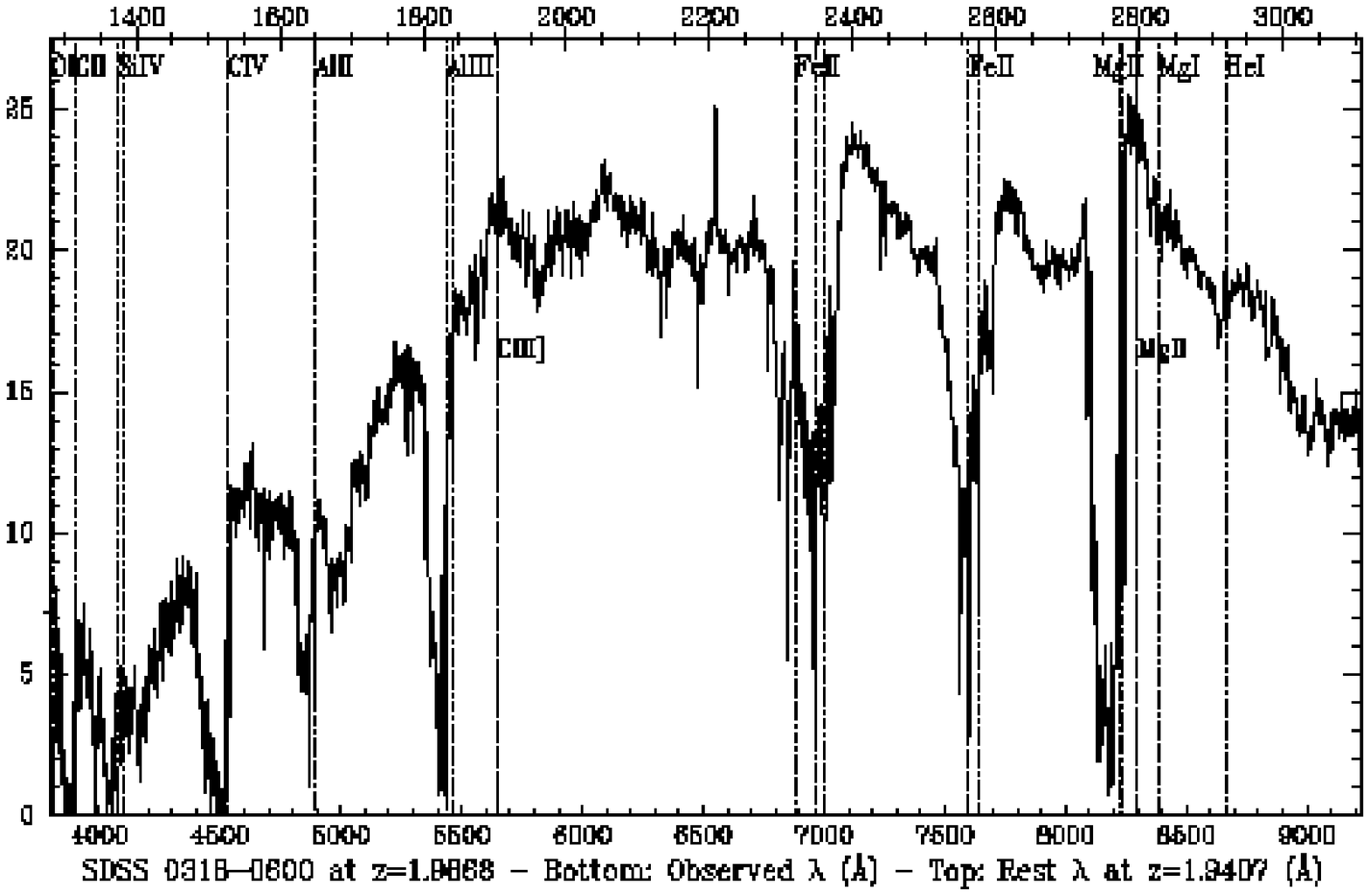}{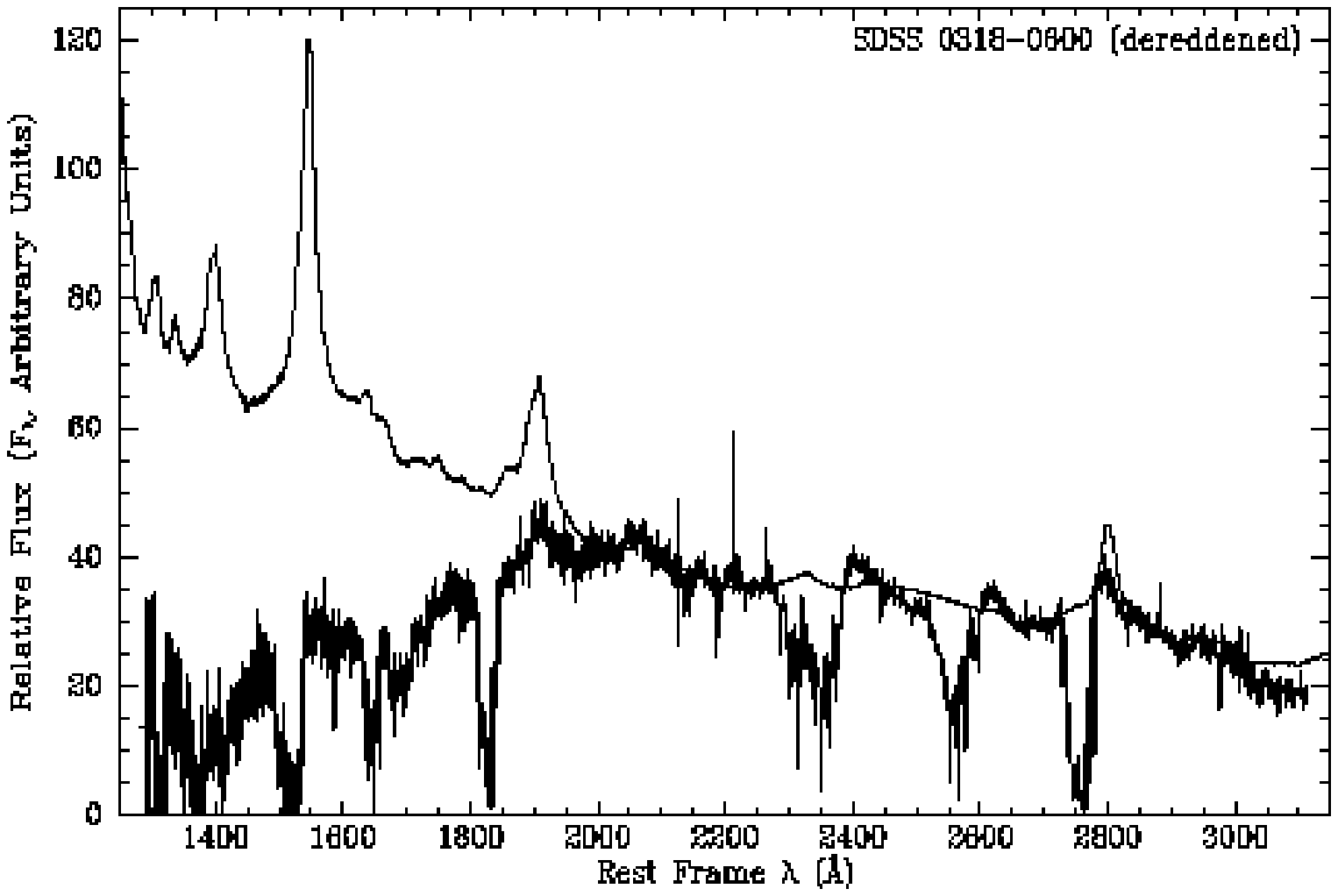}
\caption[]{ \singlespace 
a) SDSS~0318$-$0600 at $z=1.9668$, smoothed by a 3-pixel boxcar.
\ciii\ and \mgii\ emission lines at this redshift are
indicated by dashed lines and labels beneath the object's spectrum.
Dotted lines indicate absorption at $z=1.9407$ from the highest-redshift
absorption system present; the absorption just longward of \alii\ is
probably due to a combination of \NIii\ and \feii\,UV38.
Broad \feii\ emission is also visible at 2400\,\AA\ and 2600\,\AA.
Note the poorly subtracted telluric H$\alpha$ emission line from
auroral activity the night of this observation. 
b) SDSS~0318$-$0600 dereddened by \ebv$\sim$0.1 with the SMC extinction curve
to match the SDSS composite at rest wavelength $>$2000\,\AA.
To match at $<$2000\,\AA\ as well requires an extinction curve steeper 
even than that of the SMC.
}\label{f_0318}
\end{figure}

The evidence in several objects in the previous section for a reddening curve
steeper than that of the SMC is confirmed in dramatic fashion by this object.
SDSS~0318$-$0600 (Figure~\ref{f_0318}a) has $z=1.9668\pm0.0015$ from
cross-correlation with the SDSS composite quasar of \markcite{sdss73}{Vanden Berk} {et~al.} (2001).  
It has flux down to at least 1300\,\AA\ rest frame and shows absorption from a
large number of transitions, few of which appear to reach zero flux.
It is similar to FBQS~1044+3656 \markcite{dek01}({de Kool} {et~al.} 2001), except that \MgI\ 
absorption is only tentatively detected, along with \hei\,$\lambda$2945.

The long wavelength edges of the BAL troughs are at $z=1.9407\pm0.0006$ 
(a detachment velocity of 2650$\pm$160\,\kms),
with the strongest absorption at $z=1.9265\pm0.0006$.  The \civ\ trough extends
to an outflow velocity only 3000\,\kms\ higher than the \mgii\ trough.
The absorption just longward of \feii\,$\lambda$2600 from the highest-redshift
system ($z=1.9407$) shows that excited-state \feii\ absorption is present.
Multiplets UV1, UV2, and UV3 can be firmly identified, but more may be present.
The absorption just longward of \alii, at $\sim$5000$-$5100\,\AA\ observed,
is probably due to a combination of \NIii\ and \feii\,UV38.

Dereddening the spectrum by
\ebv$\sim$0.1 using the SMC extinction curve brings its slope into agreement
with that of the SDSS composite quasar at 2000-3000\,\AA\ rest frame,
but \ebv$\sim$0.4 is required to bring the slope at 1250-2000\,\AA\ into
agreement with the composite (Figure~\ref{f_0318}b).
Note that 
SDSS~0318$-$0600 is detected by 2MASS, with 
$J=16.00\pm0.08$, $H=15.58\pm0.10$ and $K_s=15.02\pm0.14$.
Its near-IR colors are {\em bluer} than those of most quasars,
indicating that the above \ebv\ may be somewhat underestimated.

The implications of our observations of SDSS~0318$-$0600 and all the other
objects presented earlier in this section are discussed in \S\ref{IMPRED}.
We end this section on heavily reddened BAL quasars 
with a discussion of 
an object possibly similar to SDSS~0318$-$0600.

\subsubsection{SDSS~0338+0056: Reddened Emission or Reddened Absorption?} \label{sdss0338} 

\begin{figure}
\epsscale{1.0}
\plotone{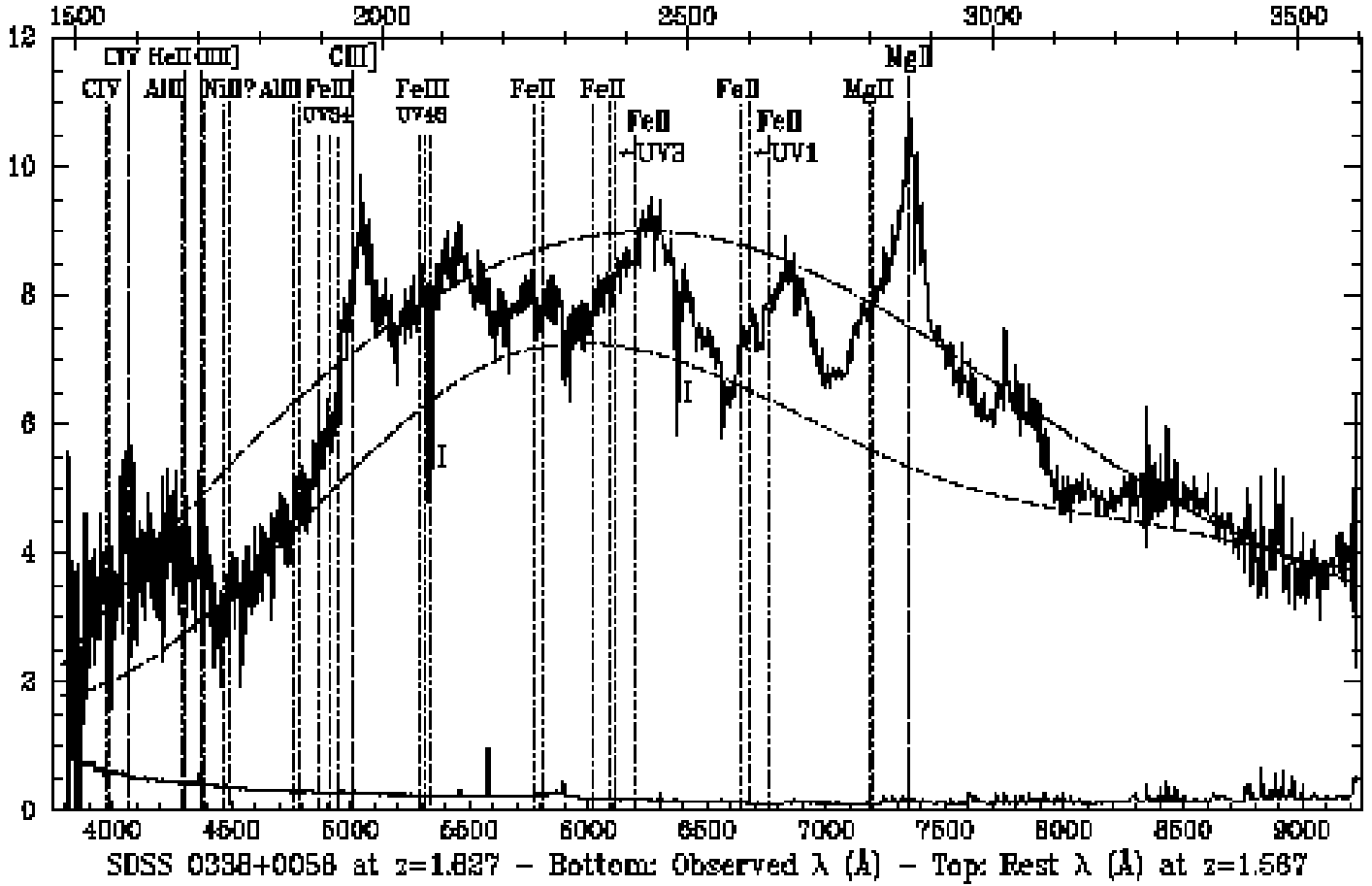}
\caption[]{ \singlespace 
SDSS~0338+0056 at $z=1.627\pm0.002$.  Emission lines at this redshift are
indicated with vertical dashed lines.  Absorption lines at $z=1.567$, the
assumed onset of the BAL troughs, are indicated with vertical dotted lines. Thin
vertical dotted lines indicate absorption lines that are not definitely present.
For \feii\ UV3 and UV1 multiplets, the wavelength of the longest-wavelength line
is plotted and arrows are drawn to indicate the presence of other lines to the
blue.  There are two intervening \mgii\ systems marked `I' beneath the spectrum.
The dot-dashed line shows the continuum fit used to calculate the BI and AI.
The dashed line shows a possible continuum fit if this object is a reddened
strong \feii-emitting quasar instead of a BAL quasar.
}\label{f_0338}
\end{figure}

In SDSS~0338+0056 at $z=1.627\pm0.002$, the spectrum declines precipitously
shortward of \ciii\ (Figure\,\ref{f_0338}).  This decline is confirmed by the
object's $u-g$ color.  The spectrum between \ciii\ and \mgii\ seems to be a 
blend of \feii\ emission and \mgii\ and \feii\ absorption, but 
the object is not yet definitively understood.
We outline two possible interpretations here.

The object may be a strong \feii-emitting quasar which is reddened with an
extinction
curve steeper than that of the SMC.  In this interpretation, the approximate
continuum is given by the dashed line in Figure\,\ref{f_0338} and the apparent
BAL troughs are in fact just gaps between strong \feii\ emission multiplets.
The smoothness of the spectrum from 1750$-$1900\,\AA\ is hard to explain in this
model, since there are \feii\ multiplets in that wavelength range.  However, the
object does resemble SDSS~0318$-$0600 (\S\ref{sdss0318}) without the BAL
troughs, so we consider this interpretation is plausible.

Alternatively, the object may be a reddened BAL quasar with strong
\aliii+\feiii\,UV34 absorption.  In this interpretation, the continuum shortward
of \mgii\ is given by the dot-dashed line in Figure\,\ref{f_0338}.  There is a
good match between the putative troughs of \feiii\,UV34, \feii\,UV1, \feii\,UV3
and \mgii.  Also, \feiii\,UV34 absorption would be stronger than \feiii\,UV48
absorption, which is normal.  
However, \alii\ absorption is weak at best and the start of a \civ\ trough
is not clearly detected.  If better SNR in the blue confirms the lack of a
\civ\ BAL, then this BAL quasar hypothesis would be ruled out.
Even if the BAL hypothesis is correct, the apparently wider trough in \aliii\ 
than in \mgii\ would require explanation.  It could be due to unusually strong
\NIii\ absorption at the high-velocity end of the apparent \aliii\ trough,
high-velocity \mgii\ absorption masked by \feii\ emission, or an overestimated
continuum level at $<$1750\,\AA\ due to strong \heii+\Oiii\ emission.  In the
latter case, the BAL quasar model continuum should match the reddened 
strong-\feii-emitter model continuum at $<$1750\,\AA.  The resulting very steep
drop in the continuum at $<$1900\,\AA\ would require either an extinction curve
steeper than that of the SMC or a red continuum with the same origin as the red
continua of the mystery objects of \S\ref{GRADUAL}, whatever that origin is.

\subsection{BAL Quasars with Strong \feiii\ Absorption} \label{FE3} 

\subsubsection{SDSS~2215$-$0045} \label{sdss2215} 

\begin{figure}
\epsscale{1.0}
\plotone{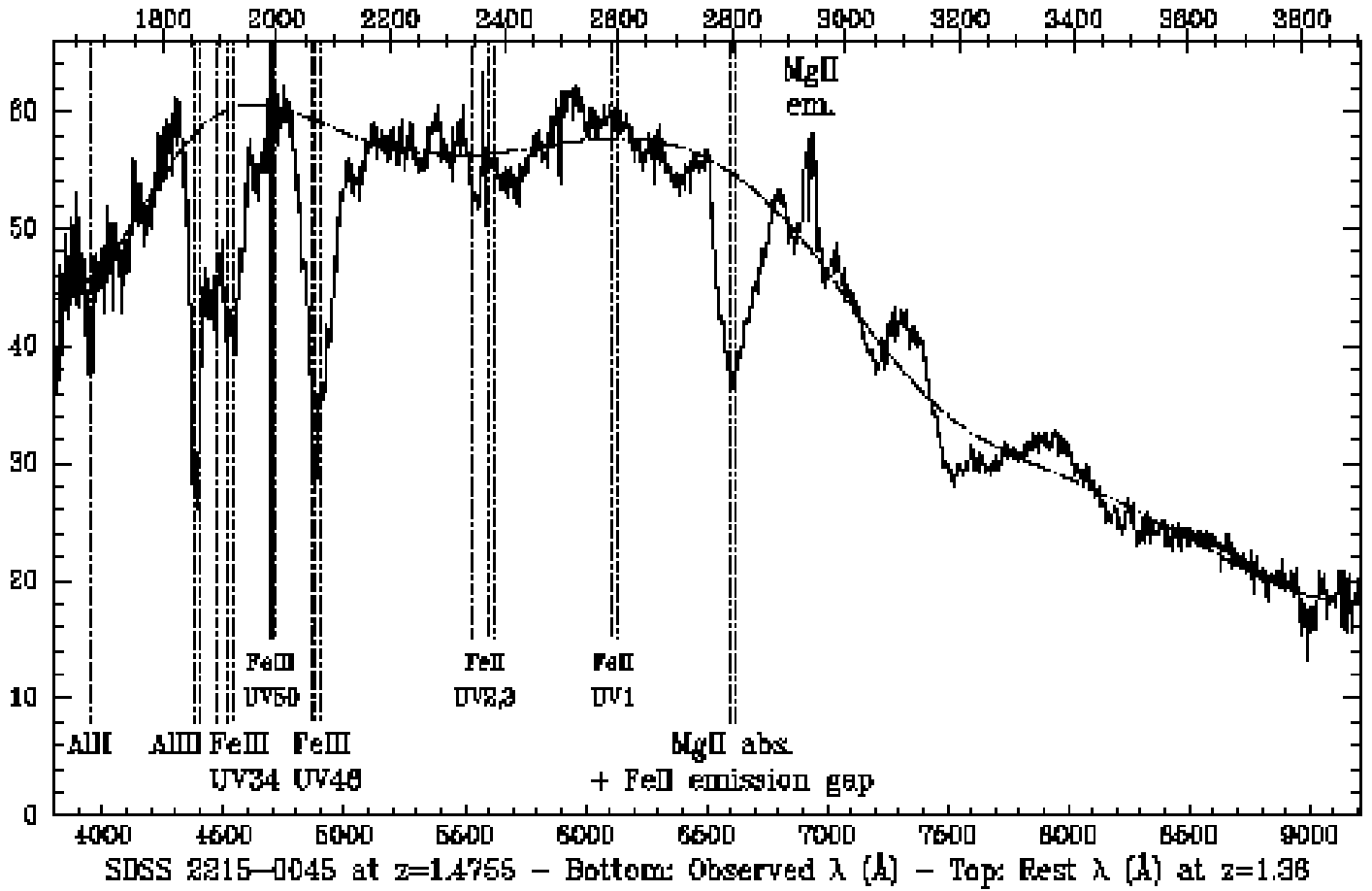}
\caption[]{ \singlespace 
SDSS~2215$-$0045 at $z=1.4755$.  The top axis plots rest wavelengths at the peak
absorption trough $z=1.36$, rather than this emission-line $z$, to 
demonstrate the absence of \feii\ absorption at 2200$-$2600\,\AA.  
The thick dotted lines show detected absorption troughs, while the
thin dotted lines show the expected positions of troughs that are not detected.
The dot-dashed line shows the continuum fit used to normalize the spectrum
for calculation of the BI and AI, and for plotting the following Figure.
}\label{f_2215}
\end{figure}

SDSS~2215$-$0045 (Figure \ref{f_2215}) 
is a reddened LoBAL with detached \MgII, \AlIII, \AlII, and \feiii\ absorption.
Our redshift $z=1.4755\pm0.0002$ is set by narrow associated \mgii\ absorption
seen atop a weak broad \mgii\ line.
SDSS~2215$-$0045 is detected by 2MASS, with $J=15.68\pm0.08$, $H=14.89\pm0.07$
and $K_s=14.71\pm0.10$.  Its near-IR colors are bluer than most quasars.
This is consistent with the spectrum longward of 3000\,\AA, which is as blue as
the SDSS composite.  The spectrum shortward of 3000\,\AA, however, appears 
to be reddened by \ebv\,$\simeq0.15\pm0.05$ compared to the SDSS composite.
This will be a lower limit to the reddening if SDSS~2215$-$0045 is indeed
bluer than the average quasar.

By comparison to SDSS~1723+5553 (Figure~\ref{f_scalloped3}, bottom left corner),
we initially identified the strong trough at $\lambda_{obs}$$\sim$4900\,\AA\ as
\crii\,\lalala2056,2062,2066.
However, the implied relative abundance of Cr is implausible,
and the expected corresponding \znii\ is missing.  This absorption is in fact
due to \feiii\,UV48 \lalala2062.21,2068.90,2079.65 (EP=5.08\,eV).
There is also \feiii\,UV34 \lalala1895.46,1914.06,1926.30 (EP=3.73\,eV)
absorption longward of Al\,{\sc iii}, at $\lambda_{obs}$$\sim$4500\,\AA.
However, \feii\ absorption is weak at best.  No \feii\ absorption troughs
are detectable to a limit of 10\% of the strength of the \feiii\ troughs.
In addition, \mgii\ absorption is probably much weaker than it appears.  There
are gaps between \feii\ emission complexes at $\sim$2670\,\AA\ (rest-frame),
the same wavelength as the detached \mgii\ absorption, and at $\sim$3100\,\AA\ 
(e.g., Figure~\ref{f_w1dered}).  Given the strength of the \feii\ emission
around the 3100\,\AA\ gap, most of what appears to be detached \mgii\ absorption
in SDSS~2215$-$0045 could be due to the 2670\,\AA\ gap.  However, the presence
of other absorption troughs at this redshift argues that some \mgii\ absorption
is present.  Both \feii\ emission modeling and spectropolarimetry would be
useful in untangling this object's absorption troughs from its \feii\ emission.

\begin{figure}
\epsscale{1.50}
\plotone{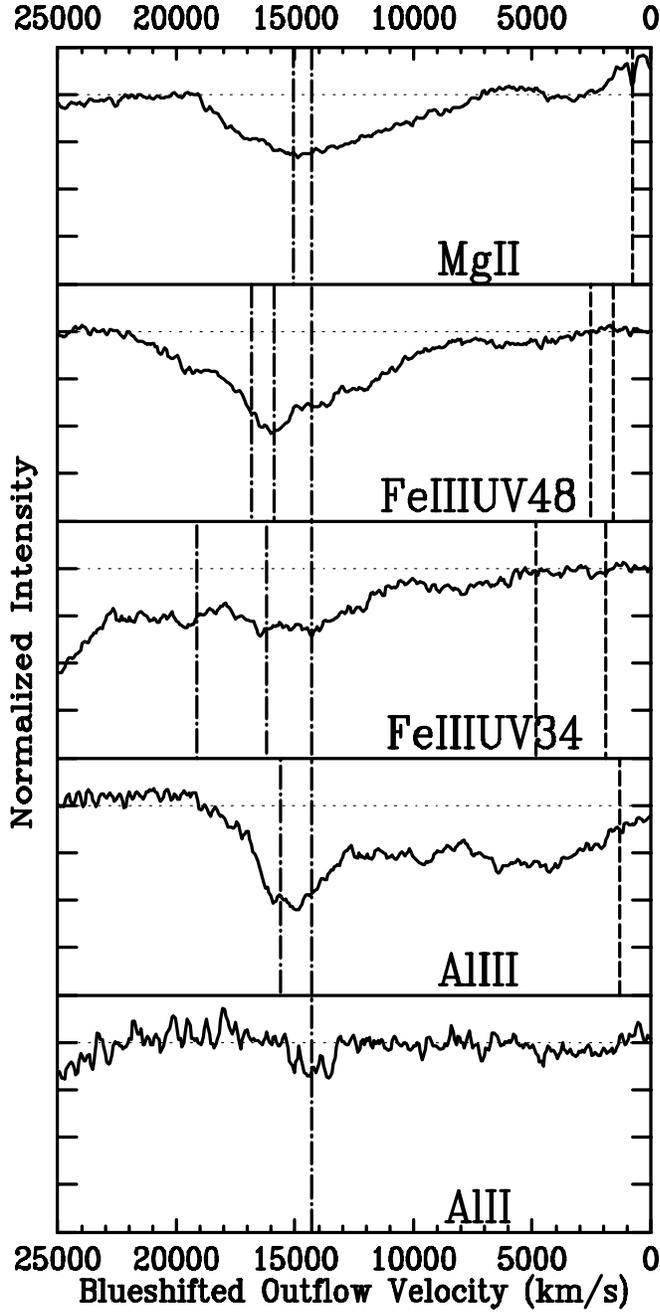}
\caption[]{ \singlespace 
Normalized spectrum of SDSS~2215$-$0045 around the troughs of \mgii,
\feiii\,UV48 \lalala2062.21,2068.90,2079.65,
\feiii\,UV34 \lalala1895.46,1914.06,1926.30, \aliii\ 
(those last two being blended together) and \alii.
Each subpanel shows the normalized intensity from 0 to 1.25,
with a dotted line drawn at a value of 1.
All troughs are plotted in terms of blueshifted outflow velocity from the 
adopted systemic $z=1.4755$ (for the longest-wavelength line of the doublets
and triplets).
The dashed vertical lines show the wavelengths of the blue members of doublets
or triplets at zero velocity.  The dot-dashed vertical lines show the 
wavelengths of each member of each singlet, doublet, or triplet 
at 14300\,\kms, the central velocity of the \alii\ absorption.
}\label{f_2215vplots5}
\end{figure}

In Figure\,\ref{f_2215vplots5} we plot the normalized spectrum of 
SDSS~2215$-$0045 around the troughs of \mgii, \feiii\,UV48, \feiii\,UV34, 
\aliii, and \alii.  All troughs were normalized using the continuum fit shown
as the dot-dashed line in Figure \ref{f_2215}, though as discussed above,
the \mgii\ trough is confused with complex \feii\ emission.
The strength of the \mgii\ trough could well be less than shown,
and its velocity structure is also untrustworthy.
The troughs are plotted in terms of blueshifted velocity from 
systemic; 
for multiplets, the velocity is for the longest wavelength line.  
The dot-dashed vertical lines show the wavelengths of every line of
each multiplet at 14300\,\kms, the central velocity of the \alii\ absorption.

Within the uncertainties, the troughs appear to have the same velocities for
their peak absorption, and all troughs except for \alii\ are consistent with
having the same starting and ending velocities 
(6000 and 18000\,\kms, respectively).  
The absorption troughs are thus unusual for a LoBAL in that they are strongest
near the high-velocity end rather than near the low-velocity end.
At the velocity of peak absorption the normalized depths of the \feiii\,UV48 and
\aliii\ lines are the same, within the uncertainties.  This suggests that they
share the same partial covering factor.  
%
The \alii\ and \feiii\,UV34 troughs are not as deep; 
this is probably because they are not saturated, rather than the depth being a
reflection of covering fraction, since \feiii\,UV34 and \feiii\,UV48 
absorption should have the same covering factor.  

Finally, 
the relative strengths of 
UV48 and 
UV34 absorption are reversed in this object compared to other BAL quasars 
with \feiii\ (e.g., SDSS~1723+5553, \S\ref{HISCAL}).
%
We discuss the implications of this 
in \S\ref{IMPFE3}.

\subsubsection{Other BAL Quasars With Possible Strong \feiii\ Absorption} \label{MOREFE3}

We have found several other candidate strong \feiii\ absorbing BAL quasars
besides SDSS~2215$-$0045, though none with \feiii\ absorption as strong
relative to \mgii\ as in that object.
The criteria we use for selecting such candidates is the presence of a
trough at $\sim$2070\,\AA\ which is stronger than \feii\,UV1,2,3 troughs near
2400\,\AA\ and 2600\,\AA.  When such \feii\ troughs are present with strength
comparable to or greater than the 2070\,\AA\ trough, \crii\ is a more likely
identification for the latter than \feiii. 

We note in passing that PC~0227+0057 \markcite{ssg99}({Schneider}, {Schmidt}, \& {Gunn} 1999) has been identified as a
similar object at $z\simeq1.52$.  It has weak \feii\ absorption, but its 
2070\,\AA\ trough probably includes a contribution from \feiii.  The trough
is too strong relative to the \feii\ troughs to be due solely to \crii.


\paragraph{SDSS~1214$-$0001}	\label{sdss1214}	

\begin{figure}
\epsscale{1.0}
\plotone{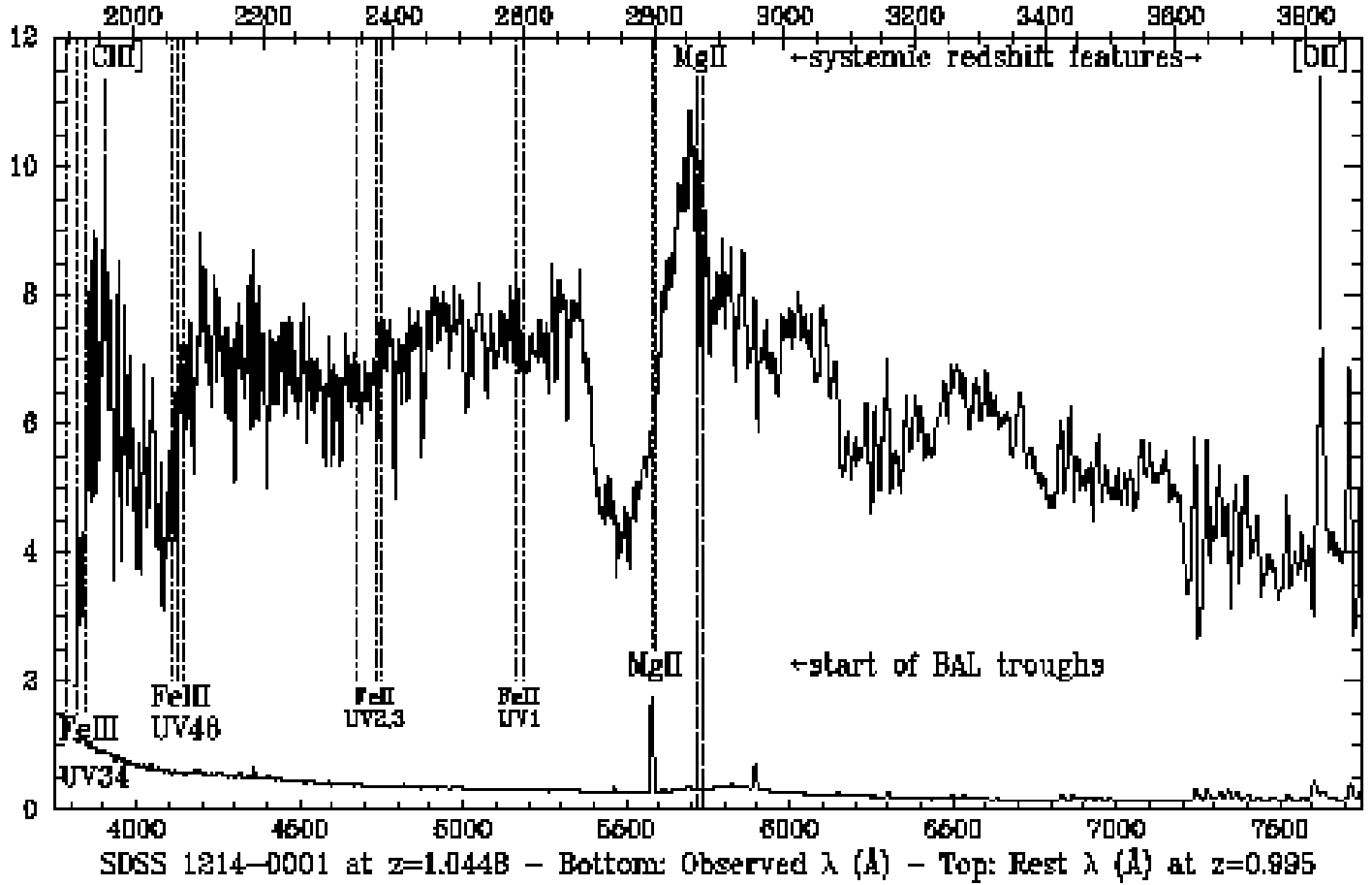}
\caption[]{ \singlespace 
Partial spectrum and error array of SDSS~1214$-$0001 (data at $>$7750\,\AA\ are
not plotted due to very poor night sky line subtraction).  
\oii\ emission and \mgii\ absorption define the systemic redshift $z=1.0448$.  
There are broad \mgii\ peaks just shortward of this $z$, 
and there is possible weak \ciii\ at this $z$ within the uncertainties
(dashed lines show features at the systemic $z$).
The top axis gives rest wavelengths at the redshift of the start of the BAL
troughs, $z=0.995$.  Given the absence of strong \feii\ absorption at this
redshift (thin dotted lines), and the good match of \feiii\,UV34 and 
\feiii\,UV48 troughs to the well-defined \mgii\ BAL trough 
(thick dotted lines), this object is likely similar to SDSS~2215$-$0045.
}\label{f_sdss1214}
\end{figure}

SDSS~1214$-$0001 (Figure\,\ref{f_sdss1214}) is at $z=1.0448\pm0.0004$
from associated \mgii\ absorption and \oii\ emission (the latter is real,
despite being near a night sky line).
The peak of the broad \mgii\ emission is shortward of this redshift.
There is a detached \mgii\ trough, but no matching \feii\ troughs.
Thus the 2070\,\AA\ trough is likely to be \feiii\,UV48.  This claim is 
bolstered by the apparent start of a \feiii\,UV34 trough at the blue edge of
the spectrum, at a redshift which matches the start of both the \feiii\,UV48
and \mgii\ troughs.  Higher SNR data extended further to the blue are needed to
study this object further.

\paragraph{SDSS~0149$-$0114}	\label{sdss0149}  

\begin{figure}
\epsscale{1.75}
\plottwo{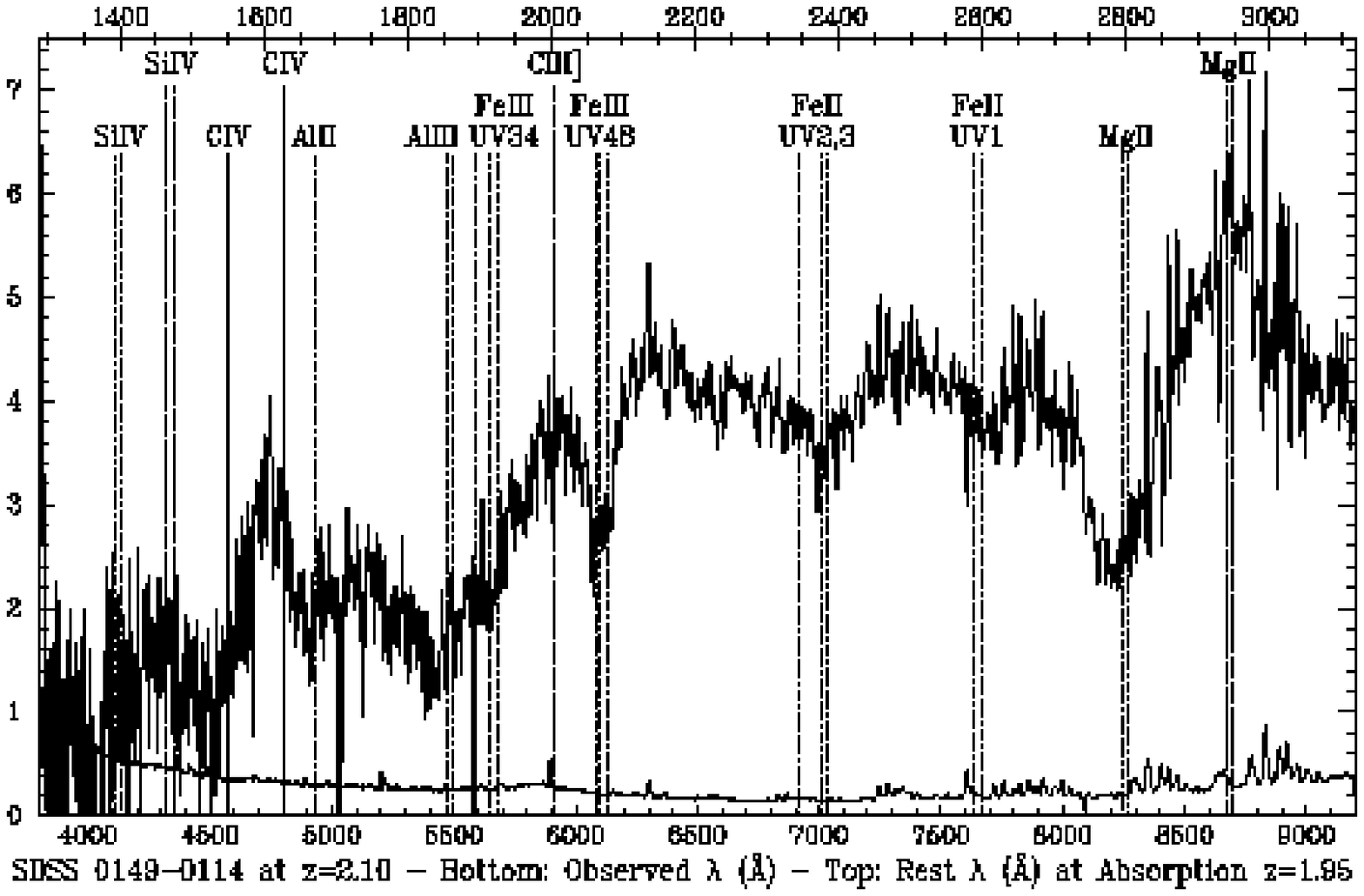}{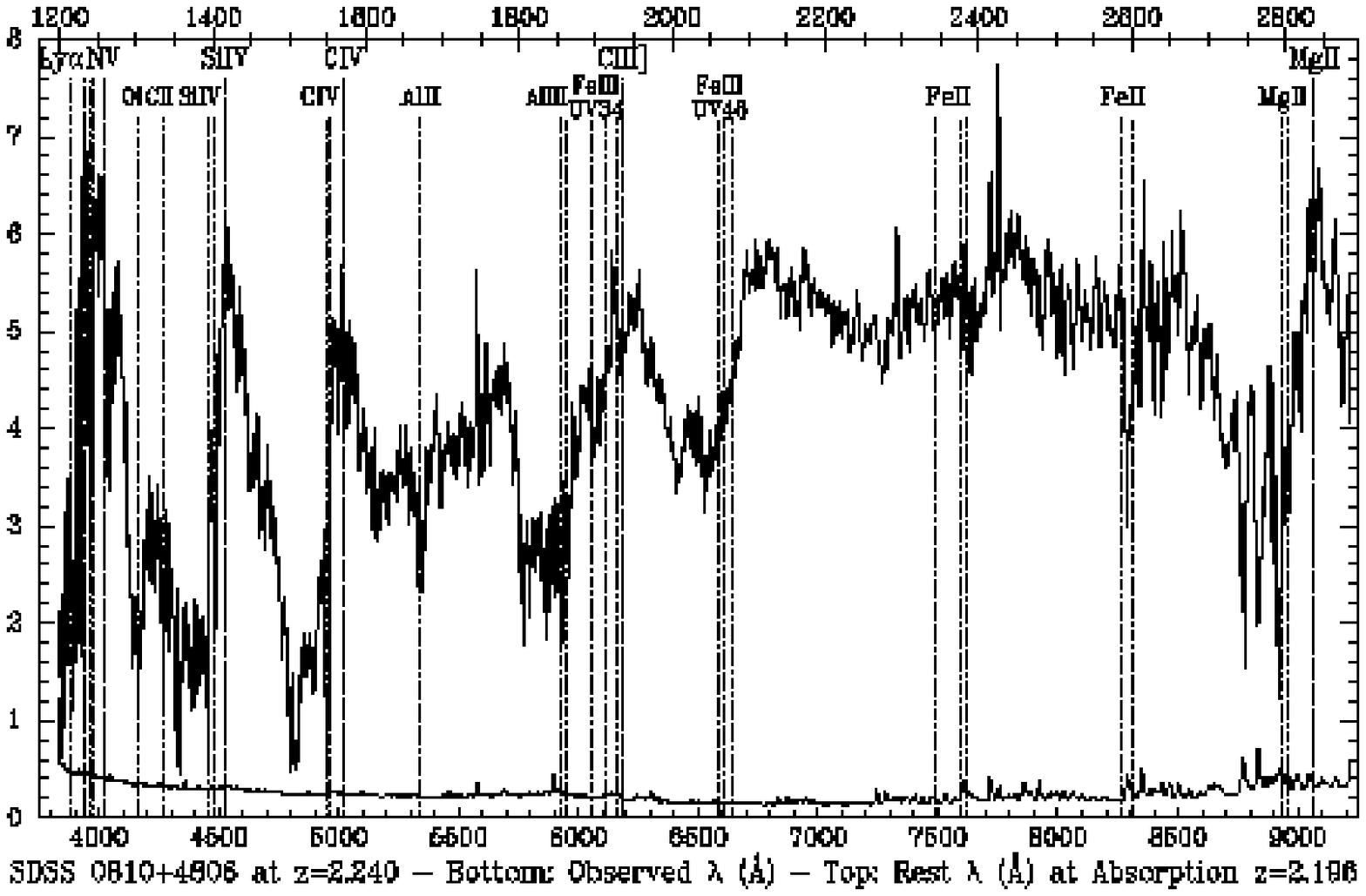}
\caption[]{ \singlespace 
Other SDSS BAL quasars with possible \feiii\,UV48 absorption:
a) SDSS~0149$-$0114 at $z=2.10\pm0.01$.
b) SDSS~0810+4806 at $z=2.240\pm0.005$.
The error array for each object is plotted along the bottom of each graph.
}\label{crii}
\end{figure}

SDSS~0149$-$0114 at $z=2.10\pm0.01$ (Figure~\ref{crii}a)
has low SNR but does appear to have a narrow \feiii\,UV48 trough as well as
a broader \mgii\ trough and weak \feii\ UV1,2,3 absorption.
Higher SNR data are needed to confirm the strong \feiii\,UV48 line and
to measure the strength of \feiii\,UV34.

\paragraph{SDSS~0810+4806}	\label{sdss0810}	

SDSS~0810+4806 at $z=2.240\pm0.005$ (Figure~\ref{crii}b) may have weak \feii\ 
UV1,2,3 absorption, but the 2070\,\AA\ trough is much stronger than any such
absorption, meaning that it is almost certainly \feiii\ instead of \crii.
However, 
higher SNR data are needed to say anything beyond that.  

\subsection{The Mysterious Objects SDSS~0105$-$0033 and SDSS~2204+0031} \label{GRADUAL} 

The SDSS has discovered two objects to date with strange but very similar
continuum shapes that include dropoffs shortward of \mgii.  
Both are FIRST sources with fluxes of a few mJy.

\begin{figure}
\epsscale{1.75}
\plottwo{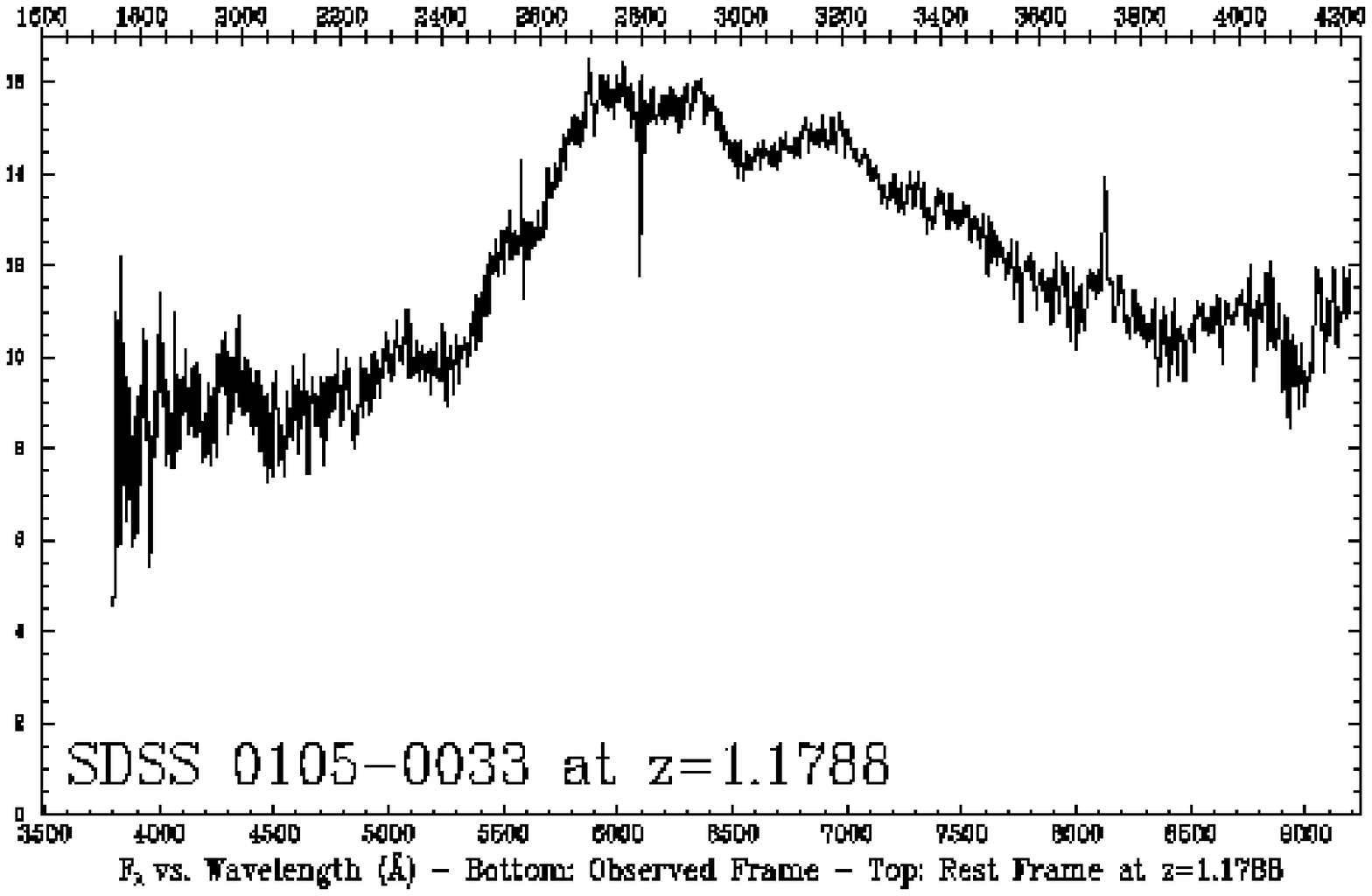}{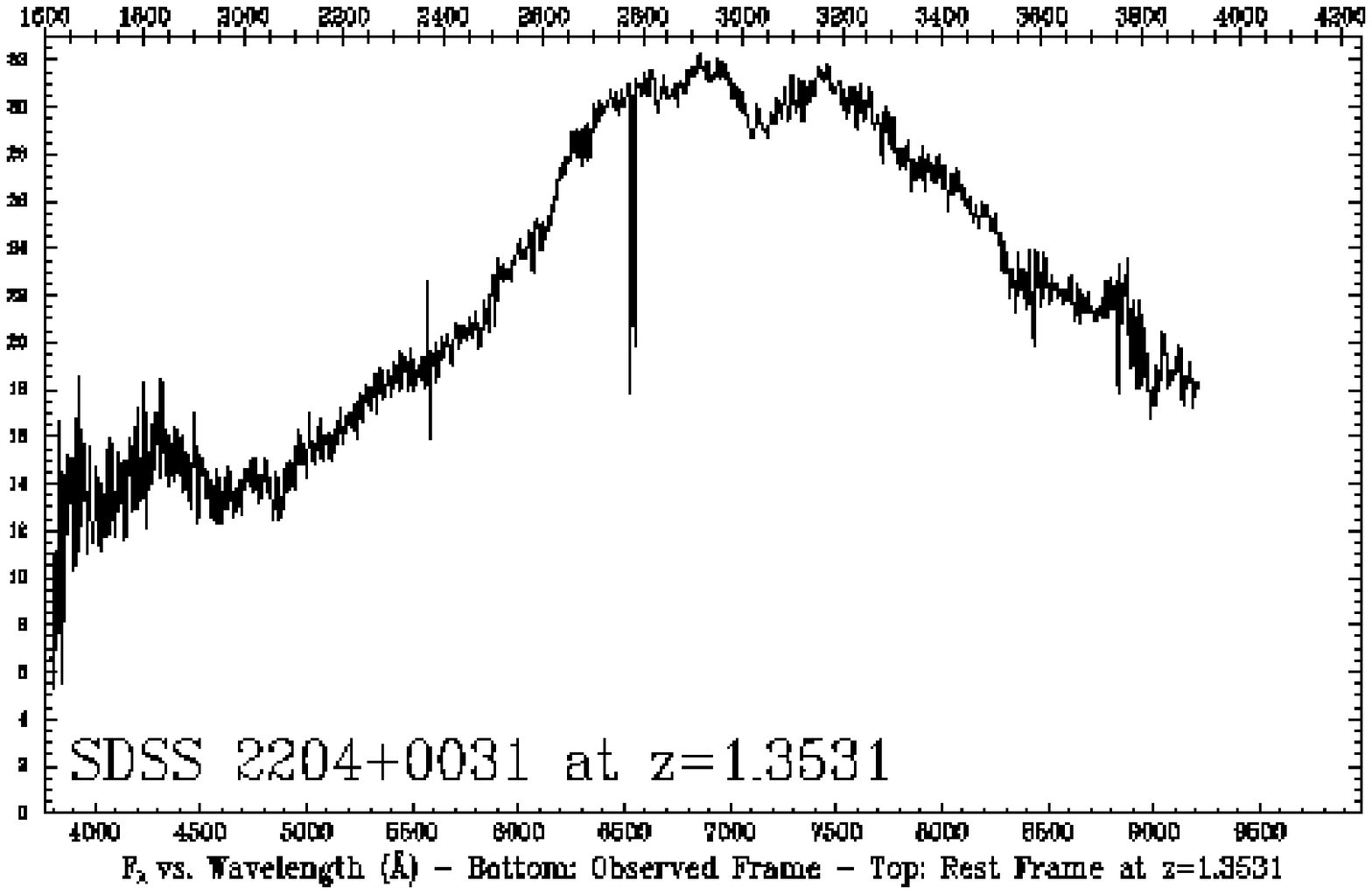}
\caption[]{ \singlespace 
a) SDSS~0105$-$0033 at $z=1.1788$ from \mgii\ absorption and \oii\ emission.
b) SDSS~2204+0031 at $z=1.3531$ from cross-correlation with the spectrum of
SDSS~0105$-$0033.  
Rest wavelengths are shown at the top and observed wavelengths at the bottom
of each plot; both plots have the same rest frame wavelength limits.
Both spectra have been smoothed by seven pixels everywhere except at the
wavelengths of the associated \mgii\ absorption, in order to reduce the noise
while illustrating 
the depth of the \mgii\ absorption at the original resolution of $R\sim1950$.
}\label{gradualbals}
\end{figure}

SDSS~0105$-$0033 has \OII\ emission at $z=1.1787\pm0.0007$ 
(Figure~\ref{gradualbals}a).  No other narrow emission lines are present between
1800\,AA\ and 3900\,\AA\ (rest-frame), but there may be broad emission features 
near 4000\,\AA\ and 4200\,\AA, possibly due to \feii.
There is also unresolved associated \mgii\ absorption at $z=1.1788\pm0.0001$.
Given that the \oii\ redshift agrees with this within its errors,
we adopt the more accurate \mgii\ redshift as systemic.
SDSS~0105$-$0033 is detected by 2MASS, with $J=16.31\pm0.10$, $H=15.81\pm0.15$
and $K_s=14.89\pm0.13$.  Its near-IR colors are typical for quasars.

SDSS~2204+0031 does not show any narrow emission, but it has associated
\mgii\ absorption at $z=1.3354\pm0.0001$ (Figure~\ref{gradualbals}b).
However, comparison of its spectrum with that of SDSS~0105$-$0033 shows
that the various continuum features do not match up if this \mgii\ redshift
is assumed to be the systemic redshift.  Cross-correlation of the two spectra
yields a redshift $z=1.3531\pm0.0009$ for SDSS~2204+0031.  
The SNR of the cross-correlation peak is $R=2.21$, which is low.
However, given the similarity of the continuum features in the two objects,
we adopt the cross-correlation redshift for SDSS~2204+0031.
This places the \mgii\ absorption 2260$\pm$120\,\kms\ shortward of systemic.

Longward of about 3200\,\AA\ rest frame, the continuum in both objects is blue
and mostly featureless. As seen in 
Figure~\ref{gradualbals}, moving shortward in the rest frame, both observed
(unreddened) spectra show a peak near 3200\,\AA, a local mininum near 3000\,\AA,
a relatively flat (in $F_\lambda$) region with associated \mgii\ absorption,
a slight drop near 2600-2625\,\AA, another drop near 2500\,\AA\ 
(stronger in SDSS~0105$-$0033), and finally a continuum which decreases
slowly to shorter wavelengths down to at least 1750\,\AA.
The dropoffs shortward of the \mgii\ absorption appear too steep to be due to
reddening and are not due to obvious BAL troughs.

The only objects with similar spectra that we are aware of are FBQS~1503+2330
($z=0.492$) and FBQS~1055+3124 ($z=0.404$) both of which have
narrow \mgii\ absorption and strong broad \feii\ and Balmer line emission at the
same redshift \markcite{wea00}({White} {et~al.} 2000).  FBQS~1503+2330 
has no obvious \mgii\ emission, while FBQS~1055+3124 
has a spectrum near \mgii\ closely resembling that of SDSS~0105$-$0033.
Just like our two SDSS objects,  
these objects have dropoffs shortward of \mgii\ which are too abrupt to be
caused by reddening and are not obviously due to BAL troughs.  Since these
features are at the blue ends of the spectra, however, they could be due to
detached \mgii\ BAL troughs or to flux calibration errors (cf. the difference
between the spectra of FBQS~1044+3656 in \markcite{wea00}{White} {et~al.} 2000 and \markcite{dek01}{de Kool} {et~al.} 2001).

We consider possible explanations for the spectra of these objects in
\S\ref{GRADEXP}.

\section{Discussion}  \label{DISCUSSION}  

Having 
presented five categories of unusual BAL quasars discovered to date in
the SDSS, we now 
discuss their implications 
for models of BAL outflows and quasars in general.
We do so in a somewhat unorthodox reverse order so that our lengthy discussion
of quasars with longward-of-systemic absorption comes last.

\subsection{Trying to Explain the Mysterious Objects SDSS~0105$-$0033 and SDSS~2204+0031} \label{GRADEXP}

\subsubsection{Reddening} \label{GRADRED} 
SDSS~0105$-$0033 and SDSS~2204+0031 (\S\ref{GRADUAL}) 
are not simply reddened normal quasars, since they lack
obvious emission lines.  However, reddened BAL quasars can have weak
emission lines (\S\ref{EXTRED}), and these high-redshift radio sources must be
AGN since they are more luminous than any galaxy.
Since reddening seems to be present, we attempt to account for it and to see
if reddening of unusual AGN can explain the spectra.  
%
To estimate the reddening in these objects, we followed the procedure outlined
in \S\ref{GETRED}, except that we interpolated over all strong emission lines
in the composite SDSS quasar 
before comparing to it.
%

\begin{figure}
\epsscale{0.63}
\plotone{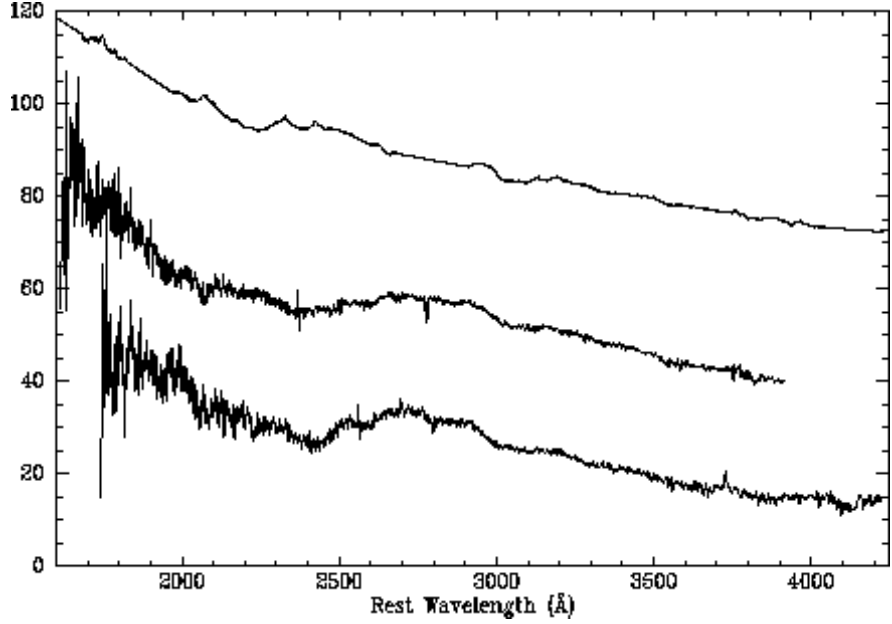}
\caption[]{ \singlespace 
From top to bottom: the SDSS composite quasar, with emission lines interpolated
over; the spectrum of SDSS~2204+0031, dereddened by \ebv=0.32; and the spectrum
of SDSS~0105$-$0033, dereddened by \ebv=0.25 (using the SMC extinction curve).
The spectra have been scaled to a common mean and offset vertically in equal
steps of 30 flux units.  
The object spectra have here been smoothed by seven pixels everywhere,
including at the wavelengths of the associated \mgii\ absorption systems.
}\label{gradered}
\end{figure}

Even before dereddening, both objects are roughly as blue as the
composite when all three spectra are measured at rest wavelengths
3000$-$3900\,\AA. Therefore, if reddening is present, these objects must
be intrinsically bluer than the average quasar.\footnote{This is not
surprising.  Consider a parent sample of unreddened quasars with identical
redshifts and bolometric luminosities, but an intrinsic dispersion
of UV/optical spectral slopes and thus 
of absolute UV magnitudes.  
If these quasars are then reddened with a range of \ebv\ values, quasars with 
bluer intrinsic UV-optical colors and lower \ebv\ will be overrepresented 
at brighter fluxes.}
%
By considering the entire 1800-3900\,\AA\ region, we estimate minimum 
reddenings of \ebv\,$=0.25\pm0.05$ for SDSS~0105$-$0033  
and \ebv\,$=0.32\pm0.03$ for SDSS~2204+0031  
using the SMC extinction curve at the quasar redshift.
The dereddened spectra are compared to the composite spectrum in
Figure~\ref{gradered}.

Some quasars have very weak \mgii\ emission but normal \feii\ emission,
and some have very strong \feii\ emission (e.g., \S\ref{WEIRD1}).
Are the strange continua of SDSS~0105$-$0033 and SDSS~2204+0031
consistent with reddened versions of such quasars?
Some \feii\ emission may be present in these two objects --- the peak near
3200\,\AA\ seen in many quasar spectra is usually ascribed to \feii\,Opt6,7
and the local minimum near 3000\,\AA\ to a gap
between \feii\ multiplets UV78 and Opt6,7+\oiii\ \markcite{sdss73}({Vanden Berk} {et~al.} 2001) --- but their
spectra shortward of \mgii\ do not show emission at the wavelengths 
of \feii\ multiplets seen in normal and strong \feii\ emitters
(compare Figure~\ref{gradered} to Figure~\ref{f_weird1}b).

Thus, while reddening may be present in these objects, it cannot 
fully explain their unusual spectra.

\subsubsection{BAL Troughs with Spatially Distinct Partial Covering?} \label{GRADPC}

The two FBQS objects similar to these two SDSS objects (\S\ref{GRADUAL}) 
have strong \feii\ emission.  Since strong \feii\ emission and the presence of
BAL troughs are correlated \markcite{bm92}({Boroson} \& {Meyers} 1992), this suggests that all four objects are
indeed BAL quasars.  If so, shorter wavelength spectra of the two FBQS objects
should reveal detached but otherwise normal BAL troughs.
For the two SDSS objects a more unusual type of BAL trough is required,
as follows, but even that does not provide a very satisfactory explanation.

We can achieve barely plausible fits for the two SDSS objects as BAL quasars
with \mgii, \feii\,$\lambda$2750, 
and \feii\,$\lambda$2600\ troughs $\gtrsim$20,000\,\kms\ wide and detached 
by $\sim$12300\,\kms.
However, 
\feii$\lambda$2400 seems weaker than expected.
%
Moreover, most BAL quasars have troughs which are saturated in most transitions
even if there is only partially covering of the continuum source \markcite{ara01}({Arav} {et~al.} 2001b).
Thus the absorption does not appear twice as strong when two troughs overlap 
(i.e., when two species have different outflow velocities such that they
produce absorption at the same observed wavelength).  
In contrast, explaining these objects as BAL quasars
requires absorption that increases in strength when troughs overlap.
Unsaturated absorption would do this, but 
the putative \feii\,$\lambda$2600 absorption here is much too strong relative
to \mgii\ for the troughs to be unsaturated.

\begin{figure}
\epsscale{0.63}
\plotone{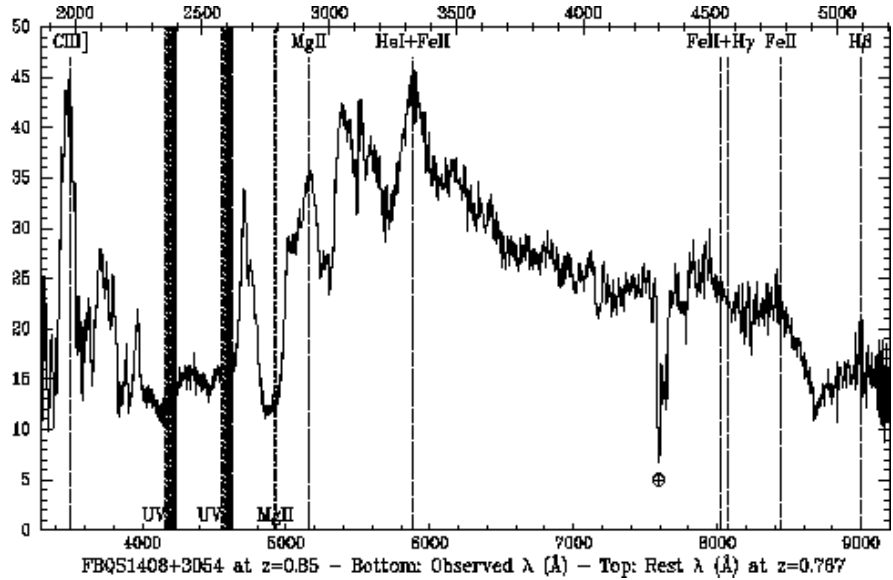}
\caption[]{ \singlespace 
Spatially distinct velocity-dependent partial covering in 
FBQS~1408+3054 \markcite{wea00}({White} {et~al.} 2000).  
Emission lines are listed along the top, and absorption lines
along the bottom.  Where high velocity \feii\,UV1 absorption overlaps with
low velocity \feii\,UV2 absorption, the absorption increases in strength.
For saturated BAL troughs arising from the same terms of the same ion, 
that cannot happen unless the flow partially covers different source regions
as a function of velocity.
}\label{f_1408}
\end{figure}

The only other way to produce absorption which increases in strength when 
troughs overlap
is to have partial covering of {\em different} regions of the continuum source
as a function of velocity.  Partial covering of the {\em same} region of the
source with velocity is what is seen or assumed in most BAL quasars 
\markcite{dek01}(e.g., \S3.3 of {de Kool} {et~al.} 2001).  
The only object known to definitely exhibit spatially distinct
velocity-dependent partial covering is FBQS~1408+3054
(Figure \ref{f_1408}; spectrum from \markcite{wea00}{White} {et~al.} 2000).
The \feii\,UV1 trough in this object has smaller partial covering than the
\mgii\ trough.  However, where the high velocity end of the \feii\,UV1 trough 
overlaps with the low velocity end of the \feii\,UV2,3 trough, at 4300\,\AA\ 
observed, the absorption increases in depth.  
(Contrast this with the top panel of Figure\,\ref{f_scalloped3}, where all
\feii\ troughs have the same apparent partial covering even when they overlap.)
Different partial covering of the same source region cannot explain this,
since \feii\ UV1, UV2, and UV3 absorption arise from the same term of \feii.
Nor can the gas be optically thin since \feii\ absorption would not be
nearly as strong relative to \mgii\ in that case.  The \feii\ gas must partially
cover different regions of the continuum source as a function of velocity.

Spatially distinct velocity-dependent partial covering may not be all that rare,
just difficult to recognize.
Unless the spatial regions covered change rapidly with velocity, such partial
covering would be obvious only when two troughs overlap at substantially
different velocities.  Troughs that wide are uncommon and not very well studied.
Nonetheless, given that even with this type of partial covering the fits for
these objects as BAL quasars are barely plausible, we consider other possible
explanations for these objects.

\subsubsection{BAL Troughs Plus Double-Shouldered \mgii\ Emission?} 

The region from 2500--3200\,\AA\ in these objects could be a region of
continuum emission plus extremely broad `double-peaked' or `double-shouldered'
\mgii\ emission produced by an accretion disk
\markcite{arp102b}(e.g., Arp\,102B, cf. Figure 1 of {Halpern} {et~al.} 1996).
If so, the full width at zero intensity (FWZI) in both objects would be about 
40,000\,\kms, twice the FWZI of the double-shouldered lines in Arp 102B.
\markcite{arp102b}{Halpern} {et~al.} (1996) model Arp 102B as an accretion disk inclined by 32\arcdeg\ to the
line of sight.  The same disk viewed edge-on 
would have a FWZI 1.9 times larger, so a FWZI of 40,000\,\kms\ is possible.
The observed reddening is probably consistent with an edge-on line of sight.

However, as seen in Figure~\ref{gradered}, to explain the dereddened continuum
shape shortward of \mgii,
the double-shouldered emission line hypothesis may require shallow,
smooth absorption troughs from $\sim$2050-2500\,\AA\ rest frame in
SDSS~0105$-$0033 and $\sim$1900-2500\,\AA\ rest frame in SDSS~2204+0031.
There are \feii\ transitions in this wavelength range, but they should be
accompanied by \mgii\ absorption.  We therefore must identify the start of the
trough at 2500\,\AA\ with \mgii\ absorption detached by 35,000\,\kms.
The absorption must also reach an outflow velocity of $\geq$53,000\,\kms\ so
that the \mgii\ and 
\feii\,UV1,2,3 troughs 
all overlap to produce a single smooth trough.
Both the detachment and peak outflow velocities are very large for
low-ionization BAL troughs, but not unprecedented (\S\ref{ABRUPT}).
Alternatively, the difference between the dereddened continuum shapes and the
composite may be due to the absence of \feii\ emission in these two objects.
Similarly, adding a small 2200\,\AA\ bump to the SMC extinction curve
might eliminate the need for these broad, shallow absorption troughs,
and the drop near 2500\,\AA\ could be explained as a moderately broad detached
\mgii\ trough (cf. SDSS~2215$-$0045; \S\ref{sdss2215}) with little or no
accompanying \feii\ absorption.

\subsubsection{Other Possible Explanations} \label{GRADOTHER} 

Our two SDSS objects are FIRST sources, but radio-quiet.  
Thus the peaks in the spectra near \mgii\ in these objects cannot be due to
the turnover of a red synchrotron component \markcite{fww00}(e.g., {Francis}, {Whiting}, \& {Webster} 2000).
These objects are not supernovae, 
as supernovae have redder colors at these rest wavelengths.
They are not galaxies at the redshifts of the \mgii\ absorption which are
gravitationally lensing higher-redshift quasars, since there are no signs of
emission lines or Ly$\alpha$ forest absorption from such quasars.
Neither object shows any sign in its images or spectrum of being
a superimposition of two quasars or a quasar and a star or galaxy.
One very speculative possible explanation is emission from an accretion disk
with a gap in the regions hot enough to produce significant emission at
$\lambda_{rest}<2800$\,\AA.  Such a gap is unlikely to be long-lived, 
so future observations of these objects could test this hypothesis.

One last possibility 
is a reddened version of PG~1407+265
\markcite{mcd95}({McDowell} {et~al.} 1995), which has 
very weak emission lines shortward of \mgii, and 
emission lines 
increasingly blueshifted with ionization potential,
up to a maximum 
of 10,000\,\kms\ in \civ\ relative to \ha.  Reddened
quasars with weak \mgii\ and weak, 
blueshifted \ciii\ might explain these
objects, though their detailed continuum structure would still be puzzling.

\subsubsection{Summary} \label{GRADSUM} 

None of the three viable explanations for these objects --- BAL quasars with
spatially distinct partial covering, double-shouldered emission line objects,
or reddened PG~1407+265 analogs --- are particularly satisfactory.
The spatially distinct partial covering hypothesis predicts these objects
should have other properties similar to BAL quasars: e.g., weak X-ray
emission, strong \feii\ and weak \oiii\ emission, and high polarization,
especially in the BAL troughs.
The double-shouldered hypothesis can be tested by studying the line profiles
of \hb\ or \ha\ in the near-IR or of \civ\ in the atmospheric UV;
our spectra are unfortunately too noisy to determine if there is
double-shouldered 
H$\delta$ emission in either object.
The reddened PG~1407+265 hypothesis can be tested with high-SNR spectra of
\ciii\ and \civ, as well as \hb\ and \ha.

Finally, note that it is possible that both the \oii\ emission and \mgii\ 
absorption in SDSS~0105$-$0033 are from an intervening galaxy.  If this is the 
case then the redshifts of {\em both} objects are unknown, since our redshift
for SDSS~0105$-$0033 comes from the \oii\ emission in its spectrum and our
adopted redshift for SDSS~2204+0031 comes from cross-correlation with
SDSS~0105$-$0033.  The only remaining constraint for both objects is that
$z\lesssim2.15$ since we do not see Ly$\alpha$ or the Ly$\alpha$ forest.

\subsection{Implications of BAL Quasars with Strong \feiii\ Absorption}  \label{IMPFE3}

SDSS~2215$-$0045 (\S\ref{sdss2215}) has a nearly unique spectrum, in which
\feiii\ absorption is much stronger than \feii\ absorption.  There are several
other possibly similar objects with lower SNR (\S\ref{MOREFE3}), but until
better data on them are available we discuss only SDSS~2215$-$0045 in detail.
A full photoionization calculation of the conditions in the BAL gas in this 
object is beyond the scope of the present work, so we merely 
speculate on what conditions 
might be needed to produce such a strange spectrum.

In SDSS~2215$-$0045, \feiii\ absorption is present 
from terms with different excitation potentials (EP), and absorption in
\feiii\,UV48 (EP 5.08\,eV) is as strong or stronger than in
\feiii\,UV34 (EP 3.73\,eV) or in \feiii\,UV50 (EP 7.86\,eV).
Using line strengths from \markcite{np96}{Nahar} \& {Pradhan} (1996), it is easy to show that this
can never happen in conditions of local thermodynamic equilibrium (LTE).
Nor can it happen if the particle densities are too low for LTE to apply, 
since then the population of the lower term of \feiii\,UV48 would be
reduced even further.
%
%
The lower term of the \feiii\,UV48 multiplet must be overpopulated 
in SDSS 2215$-$0045.  This is a rare occurrence;
in most BAL quasars where \feiii\ is detected, absorption from \feiii\,UV34 is
stronger than from \feiii\,UV48. 
For example, in SDSS~1723+5553 (Figure\,\ref{f_hiscal},
top panel), UV34 at $\sim$1910\,\AA\ is strong but UV48 at $\sim$2070\,\AA\ must
be much weaker since the 2070\,\AA\ feature can be explained entirely by \crii.
SDSS~J132139.86$-$004152.0 \markcite{sdss76}({Menou} {et~al.} 2001) also shows \feiii\,UV34 but not UV48.

According to \markcite{zha96}{Zhang} (1996), there are no resonances in the collision strengths
to the lower terms of these multiplets strong enough to explain why the lower
term of the UV48 multiplet ($a^5S$) is overpopulated relative to that of the
UV34 multiplet ($a^7S$) and the UV50 multiplet ($a^5G$).  Thus we conclude that
in this object the population of the lower term of the \feiii\,UV48 multiplet 
is governed by a fluorescence resonance due to absorption of an
emission line (or possibly by recombination from \feiv\ to excited \feiii).
%
%
%
%
In either case the lower term of UV48 could be populated directly or indirectly.
The latter would be via population of an even more excited term followed by a
cascade of emission down to the lower term of UV48, though the lack of UV50
absorption means that any cascade does not involve the lower term of that
multiplet.  
\markcite{jea00}{Johansson} {et~al.} (2000) show that there is a direct resonance with \lya\ from the ground
term of \feiii\ ($a^5D$) to the lower term of \feiii\,UV34, but that the
transitions from the ground term to the lower term of \feiii\,UV48 are weak and
forbidden.  Thus, if a fluorescence resonance is at work, it is probably a
indirect one.


The other unusual aspect of the absorption in this quasar is that \feiii\ is
much stronger than \feii, and probably stronger than \mgii\ as well.
In most BAL regions, where the ionization parameter is large
and the densities may not be high enough for LTE to apply, the ratio
\feiii/\feii\ (as well as \mgiii/\mgii\ and \aliii/\alii) will be determined
by photoionization equilibrium rather than by the Saha equation.
Thus, without detailed modeling, we can only conclude that 
the gas producing the observed BAL trough in SDSS~2215$-$0045 must be
of moderately high ionization, and with a high enough column density to
produce appreciable \feiii\ absorption.  Given this requirement, and the
observed strength of its absorption relative to \feii,
\feiii\ is probably the dominant Fe ion.
Note that such a highly ionized BAL region should also have a high temperature;
if so, the observed reddening must be produced by dust outside the BAL region.
%

In the \markcite{mc98}{Murray} \& {Chiang} (1998) and \markcite{elv00}{Elvis} (2000) disk wind models for BAL quasars, 
detached troughs such as those in SDSS~2215$-$0045 are
seen along lines of sight which skim the upper surface of the outflow,
which is more highly ionized than the rest of the flow.
The unique highly ionized HiBAL SBS 1542+541 \markcite{tea98}({Telfer} {et~al.} 1998) might be seen along
such a line of sight.
Similarly, the uniqueness of SDSS~2215$-$0045 might stem from its being unusual
in three ways: a FeLoBAL quasar seen along a line of sight through the highly
ionized upper surface of a disk wind, and with a column density through the
wind high enough so that considerable \feiii\ absorption is seen. 
Alternatively, since at
least some quasars experience a phase of high-covering-factor LoBAL absorption
in their youth \markcite{cs01}({Canalizo} \& {Stockton} 2001), SDSS~2215$-$0045 might be a
`teenage' quasar captured in the act of ionizing away its youthful BAL cocoon.
Lastly, the unusual absorption might be due to an unusual spectral energy
distribution for this quasar.
Photoionization modeling of the BAL gas in SDSS~2215$-$0045 and other
strong-\feiii-absorption BAL quasars, especially ones with narrower troughs,
is needed to understand what is special about such objects, and why they are
so rare.

\subsection{Implications of Heavily Reddened BAL Quasars}	\label{IMPRED}

We are aware of only three BAL quasars with extinctions greater than those seen
in our extremely reddened BAL quasars 
(\S\ref{EXTRED}).\footnote{The `dusty warm absorber' Seyfert\,1 galaxy 
MCG-6-30-15 has 0.61$<$\ebv$<$1.09, but it is not known if this object has 
UV absorption substantial enough to qualify as a BAL quasar \markcite{rey97}({Reynolds} {et~al.} 1997).}
Hawaii~167 has \ebv$\simeq$0.54--0.7 \markcite{ega96}({Egami} {et~al.} 1996) and
FIRST~0738+2750 has \ebv$\simeq$0.7 \markcite{gre01}({Gregg} {et~al.} 2002).
FIRST~1556+3517 has \ebv$\simeq$0.6 along the direct line of sight
to its central engine, but its observed UV flux is dominated by a scattered
component with \ebv$\simeq$0.1 \markcite{ndb00}({Najita}, {Dey}, \& {Brotherton} 2000).  Its UV spectrum therefore does
not resemble the extremely reddened BAL quasars of \S\ref{EXTRED}; in fact,
it is very similar to the overlapping-trough BAL quasars of \S\ref{ABRUPT}.

The detection of numerous heavily reddened quasars in the SDSS lends weight to
the growing consensus that a population of red quasars does exist
\markcite{gre01}(e.g., {Gregg} {et~al.} 2002).
Quasars do not have to have BAL troughs to be reddened \markcite{sdss55}({Richards} {et~al.} 2001),
but BAL quasars are more often reddened than normal QSOs are \markcite{ric01}({Richards} 2001),
and there is evidence for dust in the absorbing region of at least one 
BAL quasar \markcite{dek01}({de Kool} {et~al.} 2001).
In this section we note a few constraints our reddened BAL quasars
(\S\ref{REDDENED}) place on reddening of quasars by dust.

Dust reddening does not affect equivalent widths, yet many of these objects lack
broad emission lines.  This may indicate that the broad line region is more
heavily reddened than the continuum, which could happen if there is a range of
extinctions among the sightlines to the emission regions \markcite{hw95}({Hines} \& {Wills} 1995).
Alternatively, the majority of the observed flux could be scattered continuum
light, a hypothesis which can easily be tested via polarization measurements.

The five extremely reddened BAL quasars we have presented (\S\ref{EXTRED}) all
have $2<z<2.5$.  Similar objects at $z>2.5$ will usually be too faint for the
SDSS to detect or target, but it should be able to find lower-redshift analogs.
However, to date we have found no reddened $z<2$ BAL quasars which have
emission lines as weak as these objects.  Simulating such objects' colors
would help determine if this absence is real or due to their being
overlooked by the quasar target selection algorithm.

\subsubsection{Extinction Curves in Reddened BAL Quasars} \label{EXTCURVERED}

We have seen evidence in several reddened BAL quasars for spectral breaks which
might require an extinction curve steeper than that of the SMC.  The spectra of
hot stars on the spectroscopic plates on which these objects were discovered are
unremarkable, so the spectral breaks are not reduction artifacts.
We now examine alternative explanations for the spectral breaks
as well as possible origins for such extinction curves.

We have not considered the contaminating effects of host galaxy emission,
but it could be important only for objects which are very low luminosity. 
SDSS~0342+0045 is the lowest luminosity object, but the
presence of \feii\ absorption indicates it is a BAL quasar.
The troughs in it and the other objects are also very deep.  This limits the
contribution of host galaxy light to typically $\lesssim$20\%, even assuming
that the troughs are resolved and host galaxy light is entirely responsible
for the remaining emission in the troughs.

The apparently different reddenings shortward and longward of 2000\,\AA\ in 
the quasar rest frame are not more easily explained by extinction in a galaxy
between us and the quasar instead of by extinction at the quasar redshift.  
In the ultraviolet through the optical,
plausible extinction curves have greater extinctions at shorter wavelengths 
(except near the 2200\,\AA\ bump, for which there is no evidence in the 
spectra), and the {\em slope} of the extinction curve is also greater at 
shorter wavelengths.  Thus the intervening reddening needed to match the 
composite spectrum at $>$2000\,\AA\ would be larger than the intrinsic reddening
needed, and even then, the discrepancy at $<$2000\,\AA\ would be worse.

BAL quasars may be preferentially viewed edge-on \markcite{gsf94}(e.g., {Glenn}, {Schmidt}, \& {Foltz} 1994). 
Limb darkening and relativistic effects 
should make edge-on accretion disk spectra look different
\markcite{ce87,ln89,sm89,hub00}({Czerny} \& {Elvis} 1987; {Laor} \& {Netzer} 1989; {Sun} \& {Malkan} 1989; {Hubeny} {et~al.} 2000).  However, in all these models, at
$\lambda$$\gtrsim$912\,\AA\ the dominant effect is limb darkening with a smooth
wavelength dependence which does not change the spectra appreciably.
Only for one choice of parameters considered by 
\markcite{ce87}{Czerny} \& {Elvis} (1987, model 1.5 in their Figure 4) does a change in viewing angle
produce a break in the spectrum similar to those observed here.
Thus viewing angle effects are unlikely to be the explanation for our spectra.

\markcite{tea94}{Turnshek} {et~al.} (1994) show that PG~0043+039 appears to have a continuum break at
2400\,\AA\ (their Figure 4), but is actually a reddened extreme \feii\ emitter
(their Figure 5).  Among our objects, however, only SDSS~1453+0029 
shows the {\em extreme} \feii\ emission required to mimic
such a continuum break.  

The objects might be intrinsically much bluer than the typical quasar and even
more heavily reddened than we estimate, as small differences in the apparent
slopes in different spectral regions are exaggerated by very large reddenings.
However, the distribution of spectral slopes of unreddened quasars limits the
reddening which can reasonably be inferred.  For example, if SDSS~0318$-$0600
is reddened by \ebv$\sim$0.5, it would have an intrinsic $\alpha_\nu=1.75$,
bluer than essentially all known quasars.

Most of these objects do not match the composites exactly even when dereddened.
This is probably just because quasar continuum and emission line properties vary
from object to object.
However, if strong luminosity or redshift dependencies in quasar spectra exist 
\markcite{gfk01}(e.g., {Green}, {Forster}, \&  {Kuraszkiewicz} 2001b) then the SDSS composite quasar spectrum might be an
incorrect representation of quasars at $z\sim2$ because the objects that 
contribute to the composite spectrum at $<$2000\,\AA\ are of higher average 
redshift and luminosity than those that contribute at 2000-3000\,\AA.  
Any such dependencies would have to be improbably strong to bias the SDSS
composite significantly, 
as the sample of quasars used to construct
the template has a mean $M_{r^*}\simeq-24$ at $z=1$ and $M_{r^*}\simeq-25$ at
$z=2.3$, with only a 
small number of objects contributing at $z>2.3$.

It is also conceivable that the apparent spectral break in the dereddened
spectra of these objects is an intrinsic property, rather than an artifact of
dereddening with an inappropriate extinction curve.
However, we can think of only two ways to produce such a break, both of which
are rather {\em ad hoc}.
An accretion disk with a gap at radii where $\lambda$$<$2000\,\AA\ emission is
produced (cf. \S\ref{GRADOTHER}) might produce a continuum shape with the broad
peak required ($\Delta\lambda/\lambda\sim0.5$).  Such a gap would be unstable,
and so this idea can be ruled out if no variability is seen.  
A spectral break could also conceivably arise if the
continuum source has a spatial temperature gradient and its hotter regions are
seen through more dust than its cooler regions.  An accretion disk viewed nearly
edge on through a dusty coplanar torus is one example.  If the inclination angle
is such that the line of sight to the black hole has $\tau\sim1$, 
and thus lines of sight to the near side $\tau>1$, then the
emission from the far side of the disk will dominate, with decreasing optical
depth for lower temperature regions farther out in the disk.  

None of the explanations discussed above seem particularly likely,
so we conclude that
the extinction curve in these objects are steeper than the SMC extinction curve
at $<$2000\,\AA.  The SMC extinction curve is itself steeper than those of 
the LMC, the MW, and the starburst extinction curve of \markcite{cks94}{Calzetti} {et~al.} (1994).  
Smaller average grain sizes are needed to produce steeper extinction curves,
since small particles do not scatter efficiently at long wavelengths.
The theoretical extinction curve of 0.01\,\micron\ grains of `astronomical 
silicate' constructed by \markcite{sf92}{Sprayberry} \& {Foltz} (1992) from the work of \markcite{dra85}{Draine} (1985) is steeper 
than the SMC curve.
Spherical Si or Si/SiO$_{\rm 2}$ grains of $\sim$10\,\AA\ radius can also
produce steep extinction at 1400\,\AA$<$$\lambda$$<$2000\,\AA\ 
\markcite{ld01}({Li} \& {Draine} 2002).\footnote{\markcite{wel01}{Welty} {et~al.} (2001) present evidence that Si may be undepleted
in the SMC.  Thus it may be that steeper extinction occurs when Si is depleted,
but if Si is never significantly depleted then models for steeper extinction
must consider other explanations, perhaps oxides and/or metallic grains.}
Observations of our objects at higher SNR should be able to better constrain the
shape of the observed extinction curve for comparison to theoretical models.

Observationally, \markcite{ckbr01}{Crenshaw} {et~al.} (2001a) find that the Seyfert~1 NGC~3227 has 
the same extinction at $\lambda\gtrsim3500$\,\AA\ as in the SMC 
but greater extinction at $\lambda\lesssim3000$\,\AA.
However, the slope of its extinction curve at $\lambda\lesssim3000$\,\AA\ 
is very similar to that of the SMC, so it cannot explain our objects.
The heavily reddened gravitationally lensed quasar MG~0414+0534
may also show evidence for a very steep UV extinction curve \markcite{lea95,fea99}({Lawrence} {et~al.} 1995; {Falco} {et~al.} 1999).
However, \markcite{aw99}{Angonin-Willaime} {et~al.} (1999) find substantial differential reddening between the
different components of MG~0414+0534, which makes very uncertain any conclusions
about extinction curves reached from study of its combined spectrum.

Our result is not necessarily in conflict with 
\markcite{mmo00}{Maiolino} {et~al.} (2001b), who found
evidence for flat extinction curves in the circumnuclear regions of Seyferts,
which \markcite{mmo01}{Maiolino}, {Marconi}, \&  {Oliva} (2001a) interpret as being produced by dust dominated by very large
grains.  
The circumnuclear environments of quasars span a wide range of conditions,
and it is 
possible that dust might
exist in different dominant forms at different locations, especially if dust
can actually be created in quasar outflows \markcite{emk02}({Elvis}, {Marengo}, \& {Karovska} 2002).

On a related note, the presence of substantial reddening with detected narrow
\mgii\ but not \feii\ absorption, as seen in SDSS~1453+0029 and SDSS~0127+0114
(Figure~\ref{f_w1dered}), is anomalous compared to the SMC 
\markcite{wel01}(e.g., {Welty} {et~al.} 2001).
In the SMC, for reddenings as large as those inferred here, \feii\ as well as
\mgii\ absorption from gas associated with the dust is consistently detected.
This anomaly is also seen in 
SDSS~2204+0031 and SDSS~0105$-$0033 (Figure~\ref{gradered}).
It may be that the dust causing the reddening and the gas causing the absorption
lines are not located in the same region.  But if they are, we may be seeing gas
which is ionized such that \feii\ and \mgii\ are not the dominant ions of Fe and
Mg (and so only the strong \mgii\ absorption line is seen) but which is not so
hot or dense that all the dust has not been destroyed (cf. SDSS~2215$-$0045,
\S\ref{IMPFE3}).  This idea can be tested with photoionization modeling
and by looking for narrow \aliii, \civ\ or \ovi\ absorption.

To determine how common very steep UV extinction curves are around quasars will
require large samples.  
For example, using photometry of 104 reddened SDSS quasars,
not just BAL quasars, \markcite{ric01}{Richards} (2001) estimate a typical UV extinction curve
at least as steep as that of the SMC \markcite{smc84}({Prevot} {et~al.} 1984).

\subsection{Implications of BAL Quasars with Overlapping Troughs}\label{IMPABRU}

Despite their impressively absorbed spectra, overlapping-trough BAL quasars 
(\S\ref{ABRUPT}) do not necessarily have absorbing gas with column densities
much 
larger than other FeLoBALs, just gas with a larger velocity range.
The low-ionization gas must have optical depth $\tau>1$
over a velocity range of at least 5000\,\kms, 
and possibly $>$19000\,\kms, with typical covering factor $\sim$90\%.
Whether or not a given model can produce absorption meeting these requirements
is a question for detailed modeling beyond the scope of this paper.

\subsubsection{BAL Trough Variability}	\label{VAR}

How unusual is the variable absorption in SDSS~0437$-$0045 (\S\ref{WEIRD2})?
The only systematic work on variability in BAL trough strengths is that of
\markcite{bar94}{Barlow} (1994), who monitored 23 BAL quasars on rest frame timescales of one to
several years.\footnote{The other literature on this subject consists 
of reports of variability in individual objects:
the HiBALs Q~1303+308 \markcite{fol87}({Foltz} {et~al.} 1987), Q~1413+117 \markcite{tea88}({Turnshek} {et~al.} 1988), Q~1246$-$057
\markcite{sp88}({Smith} \& {Penston} 1988), Q~1303+308 \markcite{vi01}({Vilkoviskij} \& {Irwin} 2001) and PG~0946+301 \markcite{ara01}({Arav} {et~al.} 2001b),
the mini-BAL Q~1159+0128 \markcite{abf97}({Aldcroft}, {Bechtold}, \& {Foltz} 1997), 
the LoBALs Mrk~231 \markcite{bea91}({Boroson} {et~al.} 1991) and LBQS 0103$-$2753A \markcite{jea01}({Junkkarinen} {et~al.} 2001), 
and the possible very high ionization BAL PG~1115+080 \markcite{mon96}({Michalitsianos}, {Oliversen}, \&  {Nichols} 1996).
}
He found that four showed large BAL trough changes (20-40\% in normalized 
intensity).  Results for three of these have been published:
UM~232 \markcite{bjb89}({Barlow}, {Junkkarinen}, \&  {Burbidge} 1989), H~0846+1540 \markcite{bjb92}({Barlow}, {Junkkarinen}, \&  {Burbidge} 1992a) and CSO~203 \markcite{bea92}({Barlow} {et~al.} 1992b).
Five other objects showed smaller but still significant BAL trough changes,
and a further six showed marginal changes. Thus the best available estimate
is that on a rest frame timescale of 1-3 years, 
17$_{-8}^{+12}$\% of BAL quasars show trough intensity variations of 20-40\% 
and at least 21$_{-9}^{+12}$\% more show $<$20\% variations.

Given our sample of
$\sim$20 unusual BAL quasars, for about half of which we have multiple spectra,
finding one strongly variable object (up to $\sim$15\% in two months) is broadly
consistent with the results of \markcite{bar94}{Barlow} (1994).  However, SDSS~0437$-$0045 is still
very unusual because of both the rapidity with which it has varied and the
fact that at least three wide regions have varied, not just one trough.

%
Creating even a few lines of sight with the high optical depth, high covering
factor absorption extending over the large velocity range seen in these LoBALs
might require a large global covering factor for the BAL gas.
Such covering factors are expected if these LoBALs are recently
(re)fueled quasars seen expelling a dense shroud of dust and gas (\S\ref{FRAC}).
It is unclear if variability in trough amplitude and velocity in such LoBALs
would be stronger and more common than if the outflow arises from a disk wind.
Stable disk winds are often invoked to explain the observed lack of velocity
variability in BAL outflows, but disk winds may not be particularly stable
\markcite{psk00}(e.g., {Proga} {et~al.} 2000).  
Observational monitoring of all varieties of BAL quasars is needed to
establish the range of velocity and trough strength variability seen in each.
Theoretical modeling of `quasar breakout' from a dense shield of dust and gas
is needed to determine whether that model can reproduce the detailed properties
of LoBAL troughs (variability, partial covering, velocity structure, etc.)
and LoBALs in general (e.g., strong \feii\ and weak \oiii\ emission).
The work of \markcite{wbp99}{Williams} {et~al.} (1999) is an important start in this respect.

\subsection{Implications of BAL Quasars with Many Narrow Troughs}\label{IMPSCAL}

The two low-redshift BAL quasars with many narrow troughs (SDSS~1125+0029 and
SDSS~1128+0113; \S\ref{SCALLOPED}) were identified as unusual because of those 
troughs, but the most interesting thing about them is that they seem to show 
\mgii\ absorption longward of the systemic redshift.
SDSS~1125+0029 has \mgii\ absorption ranging from 800\,\kms\ shortward
of the host galaxy absorption-line redshift to $-$1400\,\kms\ longward, 
while SDSS~1128+0113 has \mgii\ absorption ranging from 1400\,\kms\ shortward
to $-$1700\,\kms\ longward of the narrow-line redshift
(all velocities are accurate to $\pm$100\,\kms).
In both objects the absorption is deepest near the high-velocity ends of the 
troughs, which is the opposite of what is usually seen for low-ionization lines
\markcite{vwk93}({Voit} {et~al.} 1993).  
The \feii\ trough shapes in both these objects look similar to the \mgii\ 
trough shapes (best seen at the red end of the UV1 multiplet near 2630\,\AA).  

For SDSS~1128+0113, it is remotely possible that the longward-of-systemic
absorption could be spurious, but only if the narrow \oii\ emission line
is blueshifted by $\sim$1400\,\kms\ from the true systemic redshift.
In nearby AGN, the narrow forbidden line redshifts are within 
$\pm$100\,\kms\ of the host galaxy stellar absorption and \hi\,21~cm emission
redshifts (see discussions in \markcite{tf92}{Tytler} \& {Fan} 1992 and \markcite{dmac99}{McIntosh} {et~al.} 1999).
However, outflow velocities of up to $\sim$1000\,\kms\ have been seen in the
narrow-line regions in the central $\sim$1\,kpc around even low-luminosity AGN
\markcite{vea94}({Veilleux} {et~al.} 1994), so more luminous objects could in 
principle have somewhat higher outflow velocities in their narrow line regions.
A combination of disk obscuration and patchy line-emitting gas in the outflow 
could then explain why only blueshifted line emission is seen  
\markcite{vsm01,cea01}(e.g., NGC 3079; {Veilleux}, {Shopbell}, \& {Miller} 2001; {Cecil} {et~al.} 2001). 
However, for SDSS~1125+0029 we know that the \oii\ is blueshifted from the 
host galaxy absorption lines by only 310$\pm$130\,\kms.  Thus blueshifted 
narrow lines cannot explain the longward-of-systemic absorption in that object.

\begin{figure}
\epsscale{1}
\plotone{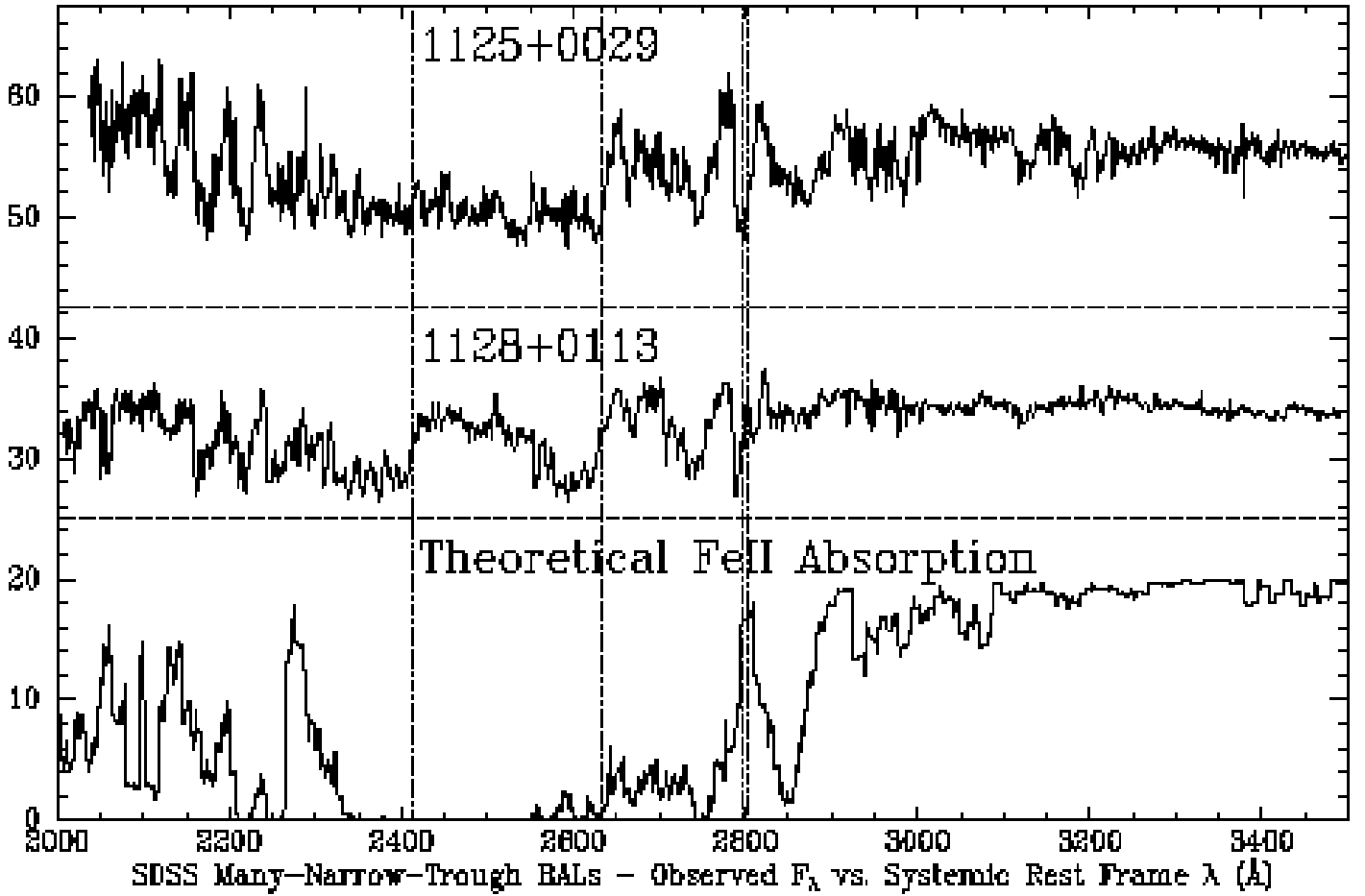}
\caption[]{ \singlespace 
Comparison of SDSS~1125+0029 and SDSS~1128+0113 (each smoothed by a 3-pixel
boxcar) to a theoretical \feii\ absorption spectrum constructed using data from
\markcite{ver99}{Verner} {et~al.} (1999).  Note that a covering factor of unity has been assumed for the
theoretical spectrum but that the objects have partial covering, so that even
saturated troughs do not reach zero flux.
The dotted vertical lines show the positions of the red ends of the strong
\feii\ multiplets at 2414 (UV2) and 2632\,\AA\ (UV1) plus the \mgii\ doublet.
There is no strong narrow \feii\ trough near 2820\,\AA\ in the theoretical
spectrum which could explain the apparent longward-of-systemic \mgii\ absorption.
}\label{f_scallopedcomp}
\end{figure}

Could the longward-of-systemic troughs be due to \feii\ absorption?
This could be the case if \feii\ 
transitions exist between 2820\,\AA\ to 2800\,\AA. 
To test this possibility, we have used the \feii\ line lists of \markcite{ver99}{Verner} {et~al.} (1999)
to construct a toy model of expected \feii\ absorption at the systemic redshift.
We use only the Einstein A coefficients to estimate
the strength of each line, thus assuming that all levels are equally populated.
In reality, 
the higher levels which might give rise to absorption near \mgii\ will be
less populated than we assume, and thus the strength of such absorption lower.
We created a toy absorption spectrum equal to exp($-S$), where $S$ is the
arbitrarily normalized line strength `spectrum', smoothed by the spectral 
resolution.  Figure\,\ref{f_scallopedcomp} compares this theoretical spectrum 
to the two objects in question.  
There is no strong narrow \feii\ trough near 2820\,\AA\ which could explain
the apparent longward-of-systemic \mgii\ absorption.  
Blueshifted absorption from \feii\,UV195,196 at 2855\,\AA\ cannot explain it
either:  there is no corresponding blueshifted \feii\,$\lambda$2750\,\AA\ 
absorption at $\sim$2695$-$2715\,\AA\ in SDSS~1125+0029, and no corresponding
strong
blueshifted \mgii\ absorption at $\sim$2745$-$2765\,\AA\ in SDSS~1128+0113.

The only previously recognized 
case of a possible longward-of-systemic BAL trough,
UN~J1053$-$0058 \markcite{bro98b}({Brotherton} {et~al.} 1998a), has \civ, \aliii\ and \mgii\ troughs extending 
$\sim$1500\,\kms\ longward of the \ciii\ redshift.  
However, this redshift could be biased because
\CIII\ can be blended with emission from \AlIII, \SiIII\ and \feiii\,UV34,61,68.
It is also possible that \ciii\ is shifted shortward of the systemic redshift
in this object: two of eighteen quasars
studied by \markcite{dmac99}{McIntosh} {et~al.} (1999) have \ciii\ shifted $>$1000\,\kms\ shortward of \oiii.
A literature search uncovered one previously unrecognized possible
longward-of-systemic trough: the mini-BAL in 3C~288.1 \markcite{hns00}({Hamann}, {Netzer}, \& {Shields} 2000),
which lies within $\pm$1000\,\kms\ of the broad emission line $z\simeq0.961$.
A narrow-line $z$ is clearly needed for both these objects.

An indisputable host galaxy redshift for SDSS~1128+0113 and detailed \feii\ 
absorption line modeling for it and SDSS~1125+0029 are needed to firmly rule
out other possibilities, but the simplest explanation seems to be that the BAL
troughs in these objects extend longward of the systemic host galaxy redshifts.

\subsubsection{Some Possible Explanations for Longward-of-Systemic BAL Troughs}	\label{EXPSCAL}

Infalling gas crossing our line of sight would produce absorption longward of 
the systemic redshift.  If the gas is in individual clouds, high-resolution 
spectra should show that the apparently smooth troughs around the systemic 
redshifts in these objects break up into narrow absorption lines.  If the gas is
not in individual clouds, then both a low-ionization inflow and a low-ionization
outflow are present.  The inflow must be optically thick and have the same
maximum local covering factor as the outflow, since the absorption at
$z>z_{sys}$ joins up smoothly to that at $z<z_{sys}$ both in \mgii\ and in
\feii\ lines with much lower optical depths.  \markcite{wbp99}{Williams} {et~al.} (1999) show that inflows
and outflows with velocities and possibly column densities comparable to those
observed here can coexist in the central regions of AGN.  The gas in the flows
seen in their isothermal models is too highly ionized to produce the absorption
seen here, but it remains to be seen if more refined models explicitly
including the thermal balance of the gas can reproduce the observations.

The longward troughs might be due to a freely expanding outflow.
The problem with this is that the maximum trough velocities shortward and
longward of the systemic $z$ are roughly equal.  The free expansion velocity 
is therefore much larger than the outflow velocity, leaving us with 
no explanation for what has accelerated the gas to such high velocities.  
A similar problem arises in the model of \markcite{sn95}{Scoville} \& {Norman} (1995).  In that model BALs
arise in debris trails from mass-losing stars in the nuclear star cluster,
but the observed redshifted velocities in these objects are larger
than the central velocity dispersions of any galaxy.

We might instead be viewing an outflow that arises from 
a distribution of clouds with a substantial virial velocity dispersion.  
In this picture, the extent of the absorption longward of the systemic redshift
would be the same for troughs of different ionization,
but the absorption shortward would depend on the ionization present at different
velocities in the outflow, and so could be different for different ions.
(Recall that high-ionization absorption typically extends to
higher terminal velocities than low-ionization absorption.)
To keep the velocity dispersion $\sigma$ from being too large in this case,
the clouds must be located far away from the quasar.  
Assuming a $10^8 M_{\odot}$ black hole and an isotropic velocity
field, a distance of $5\times10^{17}$ cm (7 light-months) is required
to match the observed $\sigma_{LOS} = \sigma/\sqrt3 \sim 1000$\,\kms.
An isotropic velocity dispersion for the cloud population seems most likely: 
the clouds 
cannot all be on plunging orbits as
they would soon be destroyed after approaching too close to the central engine,
and it is unlikely that they have a disklike distribution given the observed
radial velocity dispersion.  
An isotropic velocity dispersion implies that the clouds 
have a global covering factor of essentially unity, which has been 
predicted for at least some LoBALs from other considerations \markcite{cs01}({Canalizo} \& {Stockton} 2001).  
However, longward-of-systemic absorption should not be rare in this model.
Its apparent rarity must be attributed to the difficulty of obtaining
narrow-line redshifts given the weakness of \oiii\ in LoBALs \markcite{bm92}({Boroson} \& {Meyers} 1992).
It is also not obvious if this model can explain why BAL troughs do not show
variability in velocity 
as well as amplitude, or if it can explain the similar
trough shapes in our two objects, which the next model we discuss might.

\subsubsection{An Extended Continuum Source Seen Through a Rotation Dominated Disk Wind?}	\label{ROTSCAL}

A possibly more promising explanation arises from geometry and rotation.
A wind arising from a rotating accretion disk will share the rotation velocity
of the disk, and in fact the tangential velocity may initially dominate the
vertical and radial velocities.
If we have a grazing line of sight to such an accretion disk which passes
through a section of the wind where the velocity is predominantly rotational,
and if the continuum source is no more than an order of magnitude smaller than
the inner radius of the wind, the wind can have both approaching {\em and}
receding velocities along our line of sight (cf. \markcite{pea02}({Proga} {et~al.} 2002), where this
possibility is mentioned for disk winds in cataclysmic variables).

\begin{figure}
\epsscale{1.2}
\plottwo{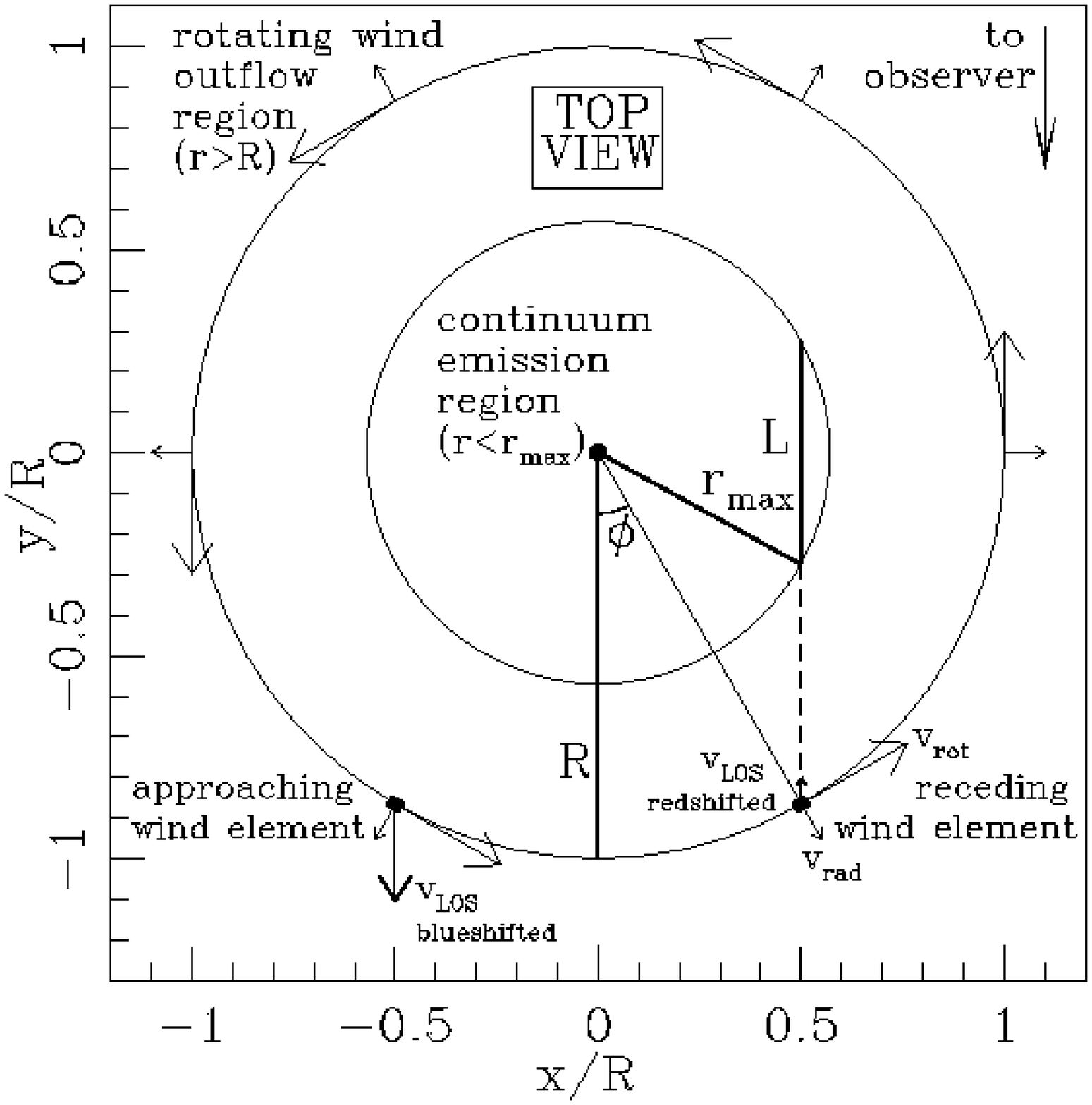}{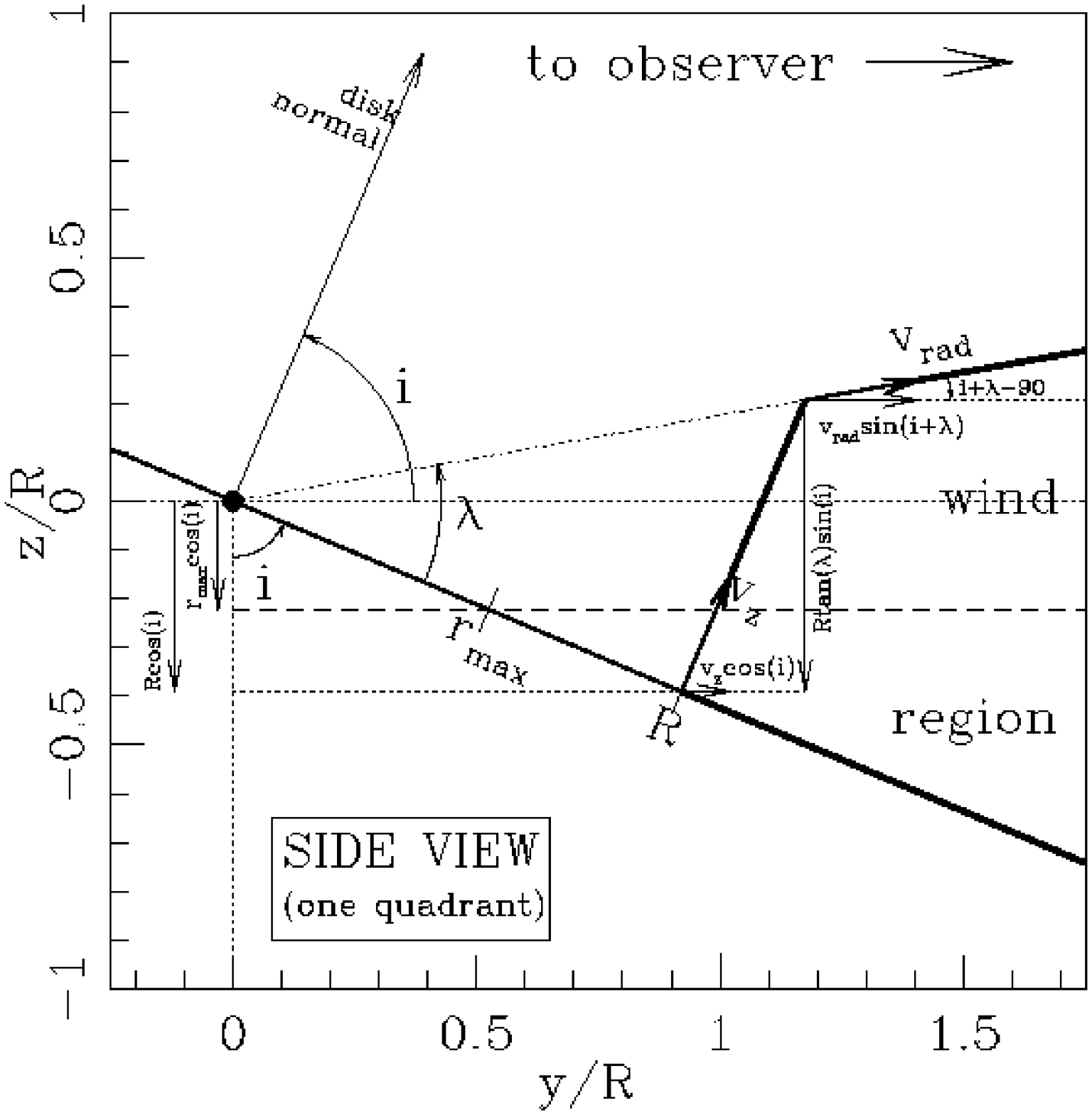}
\caption[]{ \singlespace 
How an extended continuum source seen through a rotating disk wind can give
rise to redshifted BAL troughs.  See \S\ref{IMPSCAL} for a detailed discussion.
a. Top view of an accretion disk.  The black hole is the black dot at (0,0).
The y axis is the line of sight to the observer; the x and z axes are in the
plane of the sky.  The units on all axes are relative to the innermost wind
radius $R$.  The continuum emission at a given wavelength arises at $r<r_{max}$.
For an edge-on disk, a wind element at ($R,\phi$) can have a redshifted line of
sight velocity $v_{LOS}$ if the tangential velocity dominates the radial
velocity.
b. Side view of one quadrant of the same disk, shown as the thick diagonal line
crossing the figure.  The disk shown has inclination angle $i=67\arcdeg$, wind
opening angle $\lambda=33\arcdeg$, and $r_{max}/R$=0.57, closer to the
\markcite{elv00}{Elvis} (2000) model than the \markcite{mur95}{Murray} {et~al.} (1995) model.
The wind will occupy at most the region outlined by the three thick lines on
the right-hand side; in reality the transition from vertical flow ($v_z$) to
radial flow ($v_{rad}$) will not be as abrupt as shown.  The vertical velocity
$v_z$ will also contribute to the line of sight outflow velocity.
The wind will shadow at least part of the continuum emitting region if 
$R \cos(i) - r_{max} \cos(i) \leq R \tan(\lambda) \sin(i)$.
}\label{f_view}
\end{figure}

This 
is demonstrated in Figure~\ref{f_view}a, which shows a top view of
such a disk.  The inner edge of the 
wind is at radius $R$, and $\phi$ is the angular coordinate along the surface
of the disk measured from our line of sight.
The continuum at some absorption wavelength of interest is assumed to arise
predominantly from radii $r<r_{max}$, 
since the disk beyond $r_{max}$ will be too cool to contribute significant
emission at that wavelength.  
A wind element at ($R,\phi$) will have velocity $v_{LOS}$ and will absorb
continuum emission from a region of length $L$ at that velocity.  As seen in 
the Figure, $v_{LOS}$ can be redshifted if the rotational velocity dominates.

Figure~\ref{f_view}b is a side view of one quadrant of the same disk.  
The disk has inclination angle $i$ measured from the line of sight toward the
disk normal; i.e., $i=90$\arcdeg\ for an edge-on disk and $i=67\arcdeg$ in 
Figure~\ref{f_view}b.
The wind opening angle is $\lambda$, measured in the $\theta$ direction from
the plane of the disk.  The disk, which is optically thick,
is shown as the heavy line running diagonally across the entire plot.
The wind arises at radius $R$ (and beyond), and fills at most the region 
outlined by the three thick lines.  The wind may initially be dominated 
by its vertical velocity $v_z$, 
but at the terminal opening angle $\lambda$ the radial velocity $v_{rad}\gg v_z$
and the wind has essentially no vertical velocity component.

The wind can absorb flux from a continuum emitting region of size $r_{max}$
only if the wind rises far enough above the disk to be seen in projection
against this region before becoming purely radial.
Geometrically, $i$ must be large enough to satisfy the inequality
\begin{equation}                 \label{e_casi}
R \cos(i) - r_{max} \cos(i) \leq R \tan(\lambda) \sin(i) 
\end{equation}
which simplifies to 
\begin{equation}                 \label{e_cosi}
1 - \tan(i) \tan(\lambda) \leq r_{max}/R
\end{equation}
This constraint is a weak one which will be satisfied for any line of sight
to the black hole which passes through the BAL wind.  However, if the
continuum source is extended enough, this constraint can also be satisfied for
some inclination angles $i<90\arcdeg-\lambda$ where the line of sight to the
black hole itself does not pass through the wind.  (This complicates the
conversion from wind covering factor to BAL covering factor, but provides
an additional way to achieve partial covering of the continuum source.)

For disks with inclination angles $i$ large enough to satisfy 
Equation~\ref{e_cosi}, the wind will be seen in absorption
over a range of velocities given by
\begin{equation}                 \label{e_vLOS}
v_{LOS} = v_{rot} \sin(i) \sin(\phi) + v_z \cos(i)
+ [0,v_{rad}]\times[\cos(i),\sin(i+\lambda)]\times \cos(\phi) 
\end{equation}
for $-\Phi < \phi < \Phi$.
Here $v_{rot}$ is the rotational velocity at $R$, 
$\Phi$ is the maximum tangential angle $\phi$ for which the wind shadows
the continuum emission region ($\sin\Phi=r_{max}/R$),
and $v_z$ refers only to the initial vertical velocity of the wind and not the
component of the radial velocity $v_{rad}$ perpendicular to the disk.  The terms
in brackets indicate possible ranges.  Acceleration will produce a range of
radial velocities $v_{rad}(\theta)$ as wind elements move from $\theta=0$ to
$\theta=\lambda$, with $v_{rad}(0)=0$ for an initially vertical wind.  Geometry
produces a range of line of sight projection factors for those radial
velocities, from cos($i$) at $\theta=0$ to sin($i$$+$$\lambda$) at
$\theta=\lambda$, passing through a maximum of unity if $i+\lambda>90\arcdeg$.

The $v_{rot}$ term in Equation\,\ref{e_vLOS} is the only one which can be
negative.  This equation demonstrates that longward-of-systemic absorption can
occur for lines of sight where the wind velocity is dominantly rotational.  The
maximum longward-of-systemic velocity will occur if our line of sight passes
through BAL gas which has not yet experienced significant radial acceleration.

\subsubsection{Consistency with Literature Models} \label{CONSCAL}

The quasar models of Murray \etal\ (\markcite{mc98}{Murray} \& {Chiang} 1998, and references therein) and
\markcite{elv00}{Elvis} (2000) both incorporate disk winds, and so longward-of-systemic
absorption is possible in both scenarios.  For both these models,
we can estimate the maximum possible redshifted velocity 
and the maximum angle over which longward-of-systemic absorption might be seen
by assuming a purely vertical flow with $v_{rad}=0$. 
We also assume Keplerian rotation ($v_{rot}=\sqrt{2GM_{BH}/R}$)
and that the 2800\,\AA\ continuum arises from the accretion disk at radii out
to $r_{max}=5.7\times10^{15}$\,cm \markcite{pet97}(pp. 37 and 45 of {Peterson} 1997).  This is a
conservative assumption because we use the radius of peak emission at
2800\,\AA\ as $r_{max}$, rather than the largest radius at which significant
2800\,\AA\ emission is produced, and so we slightly underestimate $r_{max}/R$.

In the Elvis model, the wind begins vertically and slowly becomes radial,
forming a funnel with a effective wind opening angle of $\lambda$=27\arcdeg.
\markcite{elv00}{Elvis} (2000) gives $R=10^{16}$\,cm for NGC~5548, which has
$M_{BH}=5.9\pm2.5~10^7 M_{\odot}$ \markcite{pw00}({Peterson} \& {Wandel} 2000).  
We take $R=1.5\times10^{16}$\,cm since we are primarily concerned with
low-ionization absorption, which in this model arises from radii up to twice as
large as that of the high-ionization absorption (Figure 5 of \markcite{elv00}{Elvis} 2000).
Thus $v_{rot}$=8800\,\kms\ and $\sin\Phi=(r_{max}/R)=0.285$. 
Equation\,\ref{e_cosi} then requires $i>48$\arcdeg\ for shadowing to occur;
however, the wind is only visible for $i>63$\arcdeg\ since the radial part 
of the flow is optically thick.  Using $v_z$=6000\,\kms\ 
(\S3.2 of \markcite{elv00}{Elvis} 2000), Equation\,\ref{e_vLOS} gives
$v_{LOS}=\pm2500$sin($i$)+6000cos($i$)\,\kms.
This yields possible line of sight velocity ranges from
$500<v_{LOS}<5000$\,\kms\ for $i=63\arcdeg$ to $-2400<v_{LOS}<2600$\,\kms\ for
$i=89\arcdeg$ (assuming a disk opening angle of 1\arcdeg).
%
%
The maximum negative velocity and the range of velocities for $i=89\arcdeg$
are only a factor of 1.3--2 higher than the observed velocities in the two SDSS
quasars with longward-of-systemic absorption.\footnote{Regardless of the
presence of longward-of-systemic absorption, however, this points out a
potential flaw in the original Elvis model for explaining narrow absorption line
systems using a vertical wind which shares the disk rotational velocity.  If the
radius of the background continuum source is within a factor of a few of the
wind launch radius, absorption from this vertical wind may be seen at a range
of velocities up to the rotational velocity, which will generally be much larger
than the range of velocities spanned by a single narrow absorption line system.}

In the Murray \etal\ model, the wind streamlines rapidly become purely
radial, resulting in a wind opening angle of only $\lambda$=6\arcdeg.
For $M_{BH}=10^8 M_{\odot}$ the wind begins at $R=3.7\times10^{16}$\,cm.
Thus, $\sin\Phi=r_{max}/R=0.154$, 
and shadowing occurs for $i>83$\arcdeg.  We adopt $v_z$=5000\,\kms\ since
\markcite{mur95}{Murray} {et~al.} (1995) state that the wind streamlines are more vertical than radial
at such outflow velocities.  Equation\,\ref{e_vLOS} then yields
$v_{LOS}=\pm$1300sin($i$)+5000cos($i$) \kms.  
With a disk opening angle of 1\arcdeg\ \markcite{mur95}({Murray} {et~al.} 1995), the ranges of possible
velocities range from $-600<v_{LOS}<1900$\,\kms\ for $i=83\arcdeg$ to
$-1200<v_{LOS}<1300$\,\kms\ for $i=89\arcdeg$.  
The $i=89\arcdeg$ numbers are in good agreement with the observations of the
two SDSS quasars.

Thus, both these models produce estimated ranges of absorption velocities which
are qualitatively consistent with those observed in the two SDSS quasars with
longward-of-systemic absorption.  However, are the $M_{BH}$ values used in
those estimates ($5.9-10\times10^7~M_{\odot}$) appropriate for those quasars?
Converting the two quasars' absolute $i^*$ magnitudes to $M_B$ using 
$M_B\simeq M_{i^*}+0.35$, adopting a bolometric correction of 
$-2.7\pm0.4$ magnitudes and assuming a typical quasar value of $L/L_{edd}=0.1$
\markcite{mcl99}({McLeod}, {Rieke}, \&  {Storrie-Lombardi} 1999), we estimate $M_{BH}$ values of 4.2 and 3.3
$\times 10^7~M_{\odot}$ for the two SDSS quasars.
This is within a factor of two to three of the values used in the estimates,
which is adequate given the uncertainties.

Finally, we note that the \markcite{dkb95}{de Kool} \& {Begelman} (1995) magnetically confined disk wind model
cannot explain longward-of-systemic absorption.  Their wind arises so far from
the continuum source that the source no longer appears appreciably extended.
In that model $R\simeq10^{18}$\,cm, so $\sin\Phi=r_{max}/R<0.001$ and the
potentially negative first term in Equation~\ref{e_vLOS} is negligible.

\subsubsection{Testing Our Explanation}	\label{TESTSCAL}

We have shown that incorporating an extended continuum source in either the
Murray \etal\ or Elvis disk wind models appears capable of explaining redshifted
BAL absorption.  Extended continuum sources are a generic result of accretion
disk models, though such models are not without their problems
\markcite{pet97}(e.g., pp. 45-55 of {Peterson} 1997).  Redshifted absorption requires a very
nearly edge-on geometry so that the rotation-dominated base of the wind shadows
much of the continuum source.
Nearly edge-on geometries are of course intrinsically rare, and many such 
objects will be hidden by dust if accretion disks are roughly coplanar with
the torii of gas and dust thought to exist at larger radii \markcite{sh99}({Schmidt} \& {Hines} 1999).
Thus the observed rarity of redshifted absorption troughs is to be expected.
Nonetheless, other longward-of-systemic BAL troughs may have gone unrecognized
due to confusion with \mgii\ emission and to the scarcity of narrow-line host
galaxy redshifts for BAL quasars, let alone stellar absorption-line redshifts.
In particular, some strong $z_{abs}>z_{em}$ associated absorption systems
in radio-loud quasars \markcite{aea87}({Anderson} {et~al.} 1987) could be redshifted mini-BALs.
%

There are several ways of testing our model.
Shorter continuum wavelengths are produced at smaller $r_{max}$ in standard
disk models.  Since a smaller $r_{max}$ will decrease the maximum observable
redshifted velocity, troughs at shorter wavelengths should have smaller longward
extents if standard disk models are correct.  In fact, near-simultaneous 
UV-optical variability suggests that at least part of the UV-optical continuum
emission
is reprocessed high-energy emission \markcite{pet97}({Peterson} 1997, p. 48), in which case all
troughs might have the same longward extents.  We cannot determine which is the
case from existing data.  In SDSS~1125+0029 and SDSS~1128+0113 the SNR and/or
resolution are insufficient to study any trough besides \mgii, given the huge
number of \feii\ and other transitions in the wavelength range observed.
%
High-resolution data for higher-ionization transitions in the two low-redshift
objects would also help test the virial velocity dispersion model.  The
longward-of-systemic absorption should be the same for all lines in that model, 
while confirmation in {\em both} objects of O-star-like absorption profiles
(deepest at high outflow velocities)
would be difficult to understand because the absorption profiles in different
objects should be uncorrelated since the gas is on random orbits.

More detailed comparisons with the spectra might be possible if some effects
we have neglected are considered (e.g., disks so thick that the disk opening
angle is comparable to or greater than the wind opening angle).
Elliptical and/or warped disks might be needed to explain why
the observed \mgii\ troughs extend farther longward of systemic than shortward;
in the model presented herein any such asymmetry should be to the blue.
Also, the continuum emission
region may be more of an annulus than a disk, since the inner disk will be very
hot and may not contribute significant emission at the wavelength in question.
For a narrow annulus, the path length $L$ shadowed by the BAL flow may be
smaller at $\phi=0$ than near the maxima $\phi=\pm\Phi$.
Similarly, the column density through the BAL flow will have
a weak minimum at $\phi=0$ if its gas density is constant with $\phi$.
This may explain the local minimum in the absorption at 2800\,\AA\ in 
SDSS~1128+0113.  
Finally, limb darkening \markcite{hub00}({Hubeny} {et~al.} 2000) and other radiative
transfer effects \markcite{mc97}(e.g., {Murray} \& {Chiang} 1997) will affect the disk emission source
function $S_{\nu}(\theta,\phi)$ and thus the amount of emission available for
absorption by a wind element at ($R,\phi$).  For example, relativistic beaming
will brighten the approaching side of the disk and dim the receding side
\markcite{hub00}(e.g., {Hubeny} {et~al.} 2000). If this effect is the dominant one and the BAL region is
optically thick, 
the absorption profiles will be deeper at shorter wavelengths. Such profiles are
seen in both low-redshift objects and possibly in SDSS~1723+5553, though in that
case the shape may be due to unabsorbed broad \mgii\ emission \markcite{aea99b}({Arav} {et~al.} 1999a). 
Nonetheless, if the BAL regions in these objects are probing different regions
of the quasar accretion disk as a function of velocity, then they have the
potential to provide powerful constraints on accretion disk models.

\subsection{How Common Are Unusual BAL Quasars?}  \label{PERCENT} 

Having selected different types of unusual BAL quasars from
early SDSS data, we can ask what fraction of the quasar population they form.
In the redshift range $1.485<z<3.9$ where selection of both HiBALs (via \civ)
and LoBALs (via \mgii\ or \aliii) is possible, there are 1807 
quasars in the SDSS EDR quasar sample \markcite{sdssedrq}({Schneider} {et~al.} 2002).  
In this redshift range our sample of BAL quasars \markcite{sdssBAL}({Reichard et al.} 2002)
includes 284-306 BAL quasars (15.7\%-16.9\%) by the AI definition,
or 181-219 (10.0\%-12.1\%) by the BI definition.  
The lower numbers are firm figures, while the higher numbers include borderline
cases (low SNR, preliminary indices less than required by only 1$\sigma$, etc.)
For comparison, at somewhat lower average redshifts the LBQS survey finds a
BAL quasar fraction of 8-11\% \markcite{wey97}({Weymann} 1997) and the FBQS survey 14-18\%
\markcite{bea00}({Becker} {et~al.} 2000).
The number of LoBALs at $1.485<z<3.9$ in the SDSS EDR is 28 
by the AI definition or 24.5 (the 0.5 is a borderline case) 
by the BI definition.  
Thus LoBALs make up 1.5$\pm$0.4\% of the SDSS EDR sample at $1.485<z<3.9$,
in agreement with canonical numbers (\S\ref{INTRO}).  

There are four FeLoBALs in the EDR in this redshift range,
and three or four more at lower $z$.  
There are two unusual FeLoBALs in the EDR in this redshift range:
SDSS~1723+5553 (\S\ref{SCALLOPED}) 
and SDSS~1730+5850 (\S\ref{ABRUPT}), 
and two or three more at lower $z$:
SDSS~1125+0029 (\S\ref{SCALLOPED}), 
SDSS~0300+0048 (\S\ref{ABRUPT}), 
and possibly SDSS~0105$-$0033 (\S\ref{GRADUAL}). 
The other unusual BAL quasars in the EDR are all at $z<1.485$: 
SDSS~1214$-$0001 (\S\ref{FE3}), 
SDSS~1453+0029 and
SDSS~0127+0114 (\S\ref{WEIRD1}). 
%
Thus, preliminary numbers from the (incomplete) SDSS EDR quasar sample indicate
that unusual LoBAL quasars are about equal in number to FeLoBALs and 
that 50\% of all FeLoBALs fall into one of our categories of unusual objects.
Since most of these unusual BAL quasars are considerably redder than the targets
of most quasar surveys, and since FeLoBALs comprise $\sim$15\% of LoBALs, most
previous samples of LoBAL quasars are likely to be incomplete by $\gtrsim$15\%.
This incompleteness
would have little effect on the overall numbers of quasars in any survey,
since even a factor of two increase in the number of LoBALs would translate to
only a few percent more quasars overall, 
but it has probably biased our view of the range of 
column densities and velocity widths spanned by BAL outflows.

Note that our numbers are in agreement with \markcite{mb01}{Meusinger} \& {Brunzendorf} (2001), 
who used an optical variability and zero proper motion 
survey to limit the fraction of `unusual' quasars to $<5$\% for $B\leq19.7$.

%
%

\section{Summary and Conclusions}  \label{CONCLUSIONS}

We have presented over twenty unusual broad absorption line quasars selected
from $\lesssim$15\% of the eventual Sloan Digital Sky Survey database.  These
objects confirm that several {\em populations} of BAL quasars with unusual
properties exist, and that the range of parameter space spanned by BAL
outflows is larger than previously realized.  Even so, these objects are all
low-ionization BAL quasars (LoBALs).  A corresponding population of HiBALs with
extensive absorption from only high-ionization transitions could exist, but the
rest-frame optical-UV region probed by SDSS spectra is not very sensitive to
such absorption.  The total
population of unusual LoBAL quasars is at least as extensive as that of LoBAL
quasars with absorption from excited-state \feii\ or \feiii\ --- about 15\% of
LoBALs --- and about half of such FeLoBALs qualify as unusual (\S\ref{PERCENT}).
Unusual BAL quasars are thus not an enormous population, but their properties
may provide more stringent tests of BAL outflow models than do the more numerous
`normal' BAL quasars.

The unusual BAL quasars we have found to date can be divided into four or five
categories.\\
$\bullet$ Two objects with many narrow troughs show \mgii\ absorption extending
longward of their systemic host galaxy redshifts by $\sim$1200\,\kms.  
We favor an explanation where this is due
to absorption of an extended continuum source by the rotation-dominated base of
a disk wind, but other explanations are possible (\S\ref{IMPSCAL}).\\
$\bullet$ Five objects have absorption which removes an unprecedented $\sim$90\%
of all flux shortward of \mgii\ (\S\ref{ABRUPT}).  The absorption in one of them
has varied across the ultraviolet with an amplitude and rate of change
among the greatest ever seen in BAL quasars (\S\ref{IMPABRU}).
This same object may also show broad H$\beta$ absorption (\S\ref{WEIRD2}).
These objects may not be fundamentally new, but they do show that low-ionization
absorption can have outflow velocities as large as any ever seen in
high-ionization lines.\\
$\bullet$ Numerous heavily reddened BAL quasars have been found, including two
reddened mini-BALs with very strong \feii\ emission (\S\ref{WEIRD1}).  
The five reddest objects are reddened by \ebv\,$\simeq0.5$ (\S\ref{EXTRED}),
and in two of them we find strong evidence that the extinction curve is 
even steeper than that of the SMC (\S\ref{EXTCURVERED}).\\
$\bullet$ We have found at least one object with absorption from \feiii\ but not
\feii, which may be due to an unusually high column density of moderately
high-ionization BAL gas (\S\ref{FE3}).  Also, the relative strengths of
excited term \feiii\ absorption in these objects is different than in
most BALs with \feiii\ absorption, and cannot be achieved in LTE,
so some sort of resonance must be at work (\S\ref{IMPFE3}).\\
$\bullet$ Lastly, we have found two luminous, probably reddened high-redshift
objects which may be BAL quasars whose troughs partially cover
different regions of the continuum source as a function of velocity
(\S\ref{GRADEXP}).

Many of these objects show
absorption from 
neutral atoms such as \mgi\ and \hei, which can be used to
constrain the density and ionization in the BAL gas.  Sometimes the distance
of the BAL gas from the central ionizing source can also be constrained,
assuming photoionization equilibrium with that source \markcite{aea01}({Arav} {et~al.} 2001a).
%
We have presented three new BAL quasars with \hei\ absorption
(Table~\ref{t_info}), and possibly two more (SDSS~1730+5850 and
SDSS~0318$-$0600).  Also, all five of our overlapping-trough BAL quasars 
(including SDSS~1730+5850) have \mgi\ absorption, 
and SDSS~0318$-$0600 may have it as well.  
This substantially increases the number of objects known to exhibit absorption
in these valuable transitions.\footnote{Only four 
previously known AGN show both \MgI\ and \hei\ absorption:
the Seyfert\,1 / LoBAL Mrk~231 \markcite{rsf85}({Rudy}, {Stocke}, \& {Foltz} 1985) and the FeLoBALs Q~2359$-$1241
\markcite{aea01}({Arav} {et~al.} 2001a), FBQS~1044+3656 \markcite{dek01}({de Kool} {et~al.} 2001) and PSS~1537+1227 \markcite{djo01b}({Djorgovski} {et~al.} 2001).
The only other previously known AGN with \hei\ absorption are 
the Seyfert\,1 NGC~4151 \markcite{and74}({Anderson} 1974) 
and perhaps the reddened radio-loud quasar 3CR~68.1 \markcite{bro98}({Brotherton} {et~al.} 1998b).
The other previously known AGN with \mgi\ absorption are the FeLoBALs
FIRST~1556+3517 \markcite{bec97}({Becker} {et~al.} 1997), FBQS~1408+3054 \markcite{wea00,bea00}({White} {et~al.} 2000; {Becker} {et~al.} 2000) and possibly
Tol~1037$-$2703 \markcite{sp01}({Srianand} \& {Petitjean} 2001),
plus the mini-BALs Arp 102B \markcite{arp102b}({Halpern} {et~al.} 1996), 3C~191 \markcite{hea01}({Hamann} {et~al.} 2001), and possibly
FBQS~1427+2709 and FBQS~1055+3124 \markcite{wea00}({White} {et~al.} 2000).  
%
}

We close with a mention of some potentially fruitful avenues of research
on these unusual LoBAL quasars.
Near-IR spectroscopy and spectropolarimetry of all these objects would help
determine if they fit into a geometrical `dusty torus' picture of BAL quasars 
\markcite{sh99}({Schmidt} \& {Hines} 1999) or whether they represent an early `cocoon' phase in the evolution
of individual quasars \markcite{bea00}({Becker} {et~al.} 2000).
High resolution spectroscopy of the longward-of-systemic absorption troughs in
SDSS~1125+0029 and SDSS~1128+0113 could differentiate between several models
for BAL outflows.
Despite the broad and blended absorption in overlapping-trough BAL quasars,
targeted high resolution spectroscopy of \caii\ in SDSS~0300+0048 and \hei\ in 
SDSS~1730+5850 might determine if the column densities as well as the
outflow velocities are very large in these objects.
Photoionization modeling including detailed treatment of \feii\ and \feiii\ may
be needed to understand \feiii-dominant BAL quasars, particularly the resonance
at work in at least SDSS~2215$-$0045.
Lastly, detailed modeling of disk winds and turbulent outflows
is needed to determine if such models can explain the wide range of properties
seen in these objects, and where such models need to be revised if they cannot.

\acknowledgements

The Sloan Digital Sky Survey (SDSS) 
is a joint project of The University of Chicago, Fermilab,
the Institute for Advanced Study, the Japan Participation Group,
The Johns Hopkins University, the Max-Planck-Institute for Astronomy, 
the Max-Planck-Institute for Astrophysics, 
New Mexico State University, Princeton University, 
the United States Naval Observatory, and the University of Washington.  
Apache Point Observatory, site of the SDSS telescopes, is operated by the 
Astrophysical Research Consortium. 
Funding for the project has been provided by the Alfred P. Sloan Foundation,
the SDSS member institutions, the National Aeronautics and Space Administration,
the National Science Foundation, the U.S. Department of Energy, the Japanese
Monbukagakusho, and the Max Planck Society.
The SDSS web site is \url{http://www.sdss.org/}.

We thank an anonymous referee, 
F. Barrientos, M. Brotherton, C. Carilli, D. Clements, B. Draine, D. Morton, M. Rupen,
J. Veliz, K. Verner, M. Vestergaard, B. Wills and H. Yee for discussions and assistance.
%
This research has made use of NASA's Astrophysics Data System, 
the Atomic Line List v2.04 at \url{http://www.pa.uky.edu/$\sim$peter/atomic/},
and the Astronomical Data Center at NASA Goddard Space Flight Center.
The Apache Point Observatory 3.5-meter telescope is owned and operated by the
Astrophysical Research Consortium.
The W. M. Keck Observatory is operated as a scientific partnership among the
California Institute of Technology, the University of California and the 
National Aeronautics and Space Administration, made possible by the generous
financial support of the W. M. Keck Foundataion.
UKIRT is operated by the Joint Astronomy Centre on behalf of the UK Particle
Physics and Astronomy Research Council.  
The CFHT is a joint facility of the National Research Council of Canada, the
Centre National de la Recherche Scientifique of France and the University of
Hawaii.
The Two Micron All Sky Survey (2MASS) is a joint project of the University of
Massachusetts and the Infrared Processing and Analysis Center/California 
Institute of Technology, funded by the National Aeronautics and Space 
Administration and the National Science Foundation.
The FIRST Survey is suppported by grants from the National Science Foundation
(grant AST-98-02791), NATO, the National Geographic Society, Sun Microsystems,
and Columbia University.
PBH acknowledges financial support from Chilean 
grant FONDECYT/1010981 and a Fundaci\'{o}n Andes grant, 
MAS from NSF grant AST-0071091, and DPS and GTR from NSF grant 99-00703.

\begin{appendix}
\section{The Balnicity Index Reconsidered}	\label{BI}

\subsection{The Standard Balnicity Index (BI)}	\label{STANDARD}

\markcite{wea91}{Weymann} {et~al.} (1991) defined a `balnicity index' or BI 
with the aim of providing a continuous measure of the strength of the broad
absorption while excluding intervening and associated absorption systems.  
The most physically relevant quantity for gauging BAL strength is the column
density in the BAL system(s), but this can be measured only in some objects,
and even then requires high-resolution spectroscopy and detailed modeling.
The BI is a useful substitute, being an observable
quantity calculable from low-resolution data.  

The BI is calculated from
\civ\ (treating it as a single transition at 1549\,\AA) as follows.  Define the
quasar $z$ using \mgii, \ciii, or \civ\ (in order of preference), fit a
continuum between \sio\ and \civ, and interpolate over non-\civ\ absorption in
that wavelength region.  Between 3000\,\kms\ and 25000\,\kms\ shortward of 
1549\,\AA\ at the systemic redshift, measure the modified equivalent width
(in \kms) of the portions of contiguous absorption troughs exceeding 
2000\,\kms\ in velocity.
The equivalent width is `modified' because `absorption' is defined as only those
parts of the troughs which dip at least 10\% below the adopted continuum level.
Thus the maximum BI is 20000\,\kms.

The criterion of a 10\% dip for absorption was established to ensure that false
BAL troughs are not identified by placing the continuum too high;
the 2000\,\kms\ contiguity condition to avoid 
strong intervening narrow-line absorption complexes;
the 3000\,\kms\ red integration limit to avoid 
strong associated narrow-line \CIV\ complexes;
and the 25000\,\kms\ blue integration limit to avoid 
confusion with \SiO\ emission and \SIiv\ BAL troughs.

\subsection{The Ideal Balnicity Index}	\label{IDEAL}

It is worth considering exactly what the `ideal' balnicity index is that we are
trying to measure.  The ideal BI would be found by measuring the
standard equivalent width (in rest-frame \kms\ at the quasar systemic redshift)
of all confirmed \civ\ BAL troughs shortward of the systemic redshift.

The ideal BI would have the following advantages over the standard BI:\\
$\bullet$~It would not exclude BAL quasars with broad, shallow troughs such as
UM~660 \markcite{tur88}({Turnshek} 1988).\\
$\bullet$~It would not exclude BAL troughs within 3000\,\kms\ of the
systemic velocity.\\
$\bullet$~It would not exclude `mini-BAL' troughs \markcite{ham99}({Hamann} 2000) which have
contiguous width $<$2000\,\kms\ but otherwise share all the characteristics of
BAL troughs.\\
$\bullet$~It can be calculated 
for high-velocity BAL quasars with troughs that extend to $>$25000\,\kms\ but 
are not confused with troughs from other transitions 
(e.g., PG~2302+029; \markcite{jea96}{Jannuzi} {et~al.} 1996),

The drawback of the ideal BI, of course, is that it requires exact
knowledge of the quasar's systemic velocity and continuum shape, and of which
absorption troughs arise in BAL outflows as opposed to being intervening systems
(of \civ\ or other species) or intrinsic systems unrelated to the BAL outflow.

\subsection{A Suggested Compromise: The Intrinsic Absorption Line Strength Index (AI)}	\label{COMPROMISE}

\markcite{wea91}{Weymann} {et~al.} (1991) note that ``...better resolution data, coupled with a more 
refined definition of balnicity, are required to treat ... borderline cases."
Since SDSS spectra have a wavelength resolution good enough to identify
many intervening absorption systems, we have defined a provisional revised
balnicity index that makes more optimal use of SDSS data.  
Since our goal is to measure the strength of all intrinsic absorption, not just
broad intrinsic absorption, we refer to this revised balnicity index as
the {\em intrinsic absorption index} or AI.
The final values of some of the parameters used in the AI may change based on
the results of ongoing detailed studies of SDSS BAL quasars, but the general
framework for choosing those parameters is presented here.
The specific issues addressed in the definition of the AI are as follows.

\paragraph{The systemic velocity:}
In some objects the \mgii\ emission line $z$ is uncertain, let alone that of
\ciii\ or \civ.  We adopt the host galaxy stellar absorption-line redshift
whenever possible; otherwise, we follow \markcite{sdss73}{Vanden Berk} {et~al.} (2001) and adopt as systemic the
redshift of \oiii\,$\lambda$5008 (vacuum $\lambda$), or \OII\ if \oiii\ is
unavailable.  We correct the $z$ measured from other lines to this
systemic $z$ using the velocity shift for that line as measured in the SDSS
composite quasar \markcite{sdss73}({Vanden Berk} {et~al.} 2001).  These velocity shifts may change slightly as
more objects are added to this composite, but a greater concern to be addressed
when more quasars are in hand is whether different quasar subtypes have
different velocity shifts.  

\paragraph{The continuum shape:}
We begin with a fifth-order polynomial fit to the windows discussed in
\markcite{wea91}{Weymann} {et~al.} (1991): 1575-1625\,\AA, 1800-1820\,\AA, 1975-2000\,\AA, 2140-2155\,\AA,
2190-2200\,\AA, 2240-2255\,\AA, 2665-2695\,\AA\ plus a window longward of \mgii.
The windows do not always sample the continuum, so they are adjusted
where necessary to reach an acceptable final fit.  Note that in most cases
it is not possible to accurately include broad emission lines in the fit,
and so the depth of the absorption will be underestimated where the absorption
overlaps in wavelength with the broad emission.  It may be possible to automate
the continuum determination for even heavily absorbed BAL quasars by comparison
with weaker BAL quasars as well as non-BAL quasars, perhaps along the lines of
\markcite{cs99}{Connolly} \& {Szalay} (1999).  It may also be worthwhile to use a 3$\sigma$ significance
criterion for determining the starting and ending points of absorption troughs, 
which would also allow the minimum detectable AI for a given spectrum to
be defined without requiring an arbitrary 10\% dip below the continuum level.
For now, we use the same modified equivalent width as the BI since empirically
we find it reduces the identification of random fluctuations as BAL troughs.
A 10\% systematic uncertainty in continuum placement is reasonable in
most cases, so we adopt that as a characteristic systematic
error until a larger number of SDSS quasar spectra are available to study
the issue statistically.  

\paragraph{Intrinsic vs. Intervening Absorption:}
\markcite{ham99}{Hamann} (2000) points out that mini-BAL troughs and confirmed intrinsic narrow
absorption line (NAL) troughs are found over a range of ejection velocities 
comparable to those of BAL troughs.
We wish to measure the strength of such intrinsic, outflowing
systems\footnote{Of course it could be that intrinsic NAL and even mini-BAL
troughs form in similar ways to BAL troughs but not in the very same flow.}
while ignoring intervening systems and associated systems
from gas in or near the quasar host galaxy and its environs.
%
Confirmation that a given system is intrinsic requires detection of 
nonblack saturation, time variability on timescales of years,
well-resolved smooth profiles which are broad compared to thermal line widths, 
densities $>$100\,cm$^{-3}$ inferred from excited-state absorption, 
or possibly very high metallicities \markcite{hea97b}(e.g., {Hamann} {et~al.} 1997).
Most of these measurements are beyond the capabilities of SDSS spectra, 
but with an instrumental resolution of $\simeq150$\,\kms\ we can resolve 
individual narrow \civ\ and \mgii\ systems 
(doublet separations 498\,\kms\ and 769\,\kms, respectively).
Thus in many cases we will be able to directly determine whether an absorption
feature $<$2000\,\kms\ wide is a BAL trough or a complex of narrow absorption
systems.  
%
For the AI we therefore adopt a 450\,\kms\ contiguity criterion instead of
the 2000\,\kms\ criterion used for the BI.  We also slightly revise the
definition of contiguous to include {\em all} absorption from troughs
$\geq450$\,\kms\ wide, rather than only the absorption beyond the first
2000\,\kms.  We feel the odds of finding non-BAL systems in such close proximity
to BAL systems are small enough to make an occasional overestimate of the AI
acceptable.
However, we do consider the 450\,\kms\ value of the contiguity criterion
to be preliminary and subject to revision 
pending more detailed future studies of SDSS BAL quasars.


\paragraph{Which Transitions To Use:}
The AI can be calculated for any transition, but for consistency with the BI
we use \civ\ whenever possible.  When another transition is used, it should be
noted in parentheses: e.g., AI(\mgii).
For objects with no \civ\ data, 
\SIiv\ is the best alternative for HiBALs, and \mgii\ or \aliii\ for LoBALs.
\mgii\ or \aliii\ absorption
can also be used for objects where the continuum around
\civ\ is very difficult to estimate.  
Low-ionization troughs such as \mgii\ and \aliii\ are typically narrower than
high-ionization troughs, 
but eventually it may be possible to correct for this effect statistically.


\paragraph{Definition of the AI:}
After defining the systemic redshift and the continuum level
and interpolating over all absorption features besides the relevant trough,
the AI and statistical uncertainty are calculated as 
\begin{eqnarray}
{\rm AI} = \int_{0}^{(25,000)} [1-f(v)/0.9] ~C' ~dv \\
\sigma_{\rm AI}^2 = \int_{0}^{(25,000)} (\sigma_{f(v)}/0.9)^2 ~C' ~dv
\end{eqnarray}
where $f(v)$ and $\sigma_{f(v)}$ are the normalized flux and uncertainty
(unsmoothed whenever possible) as a function of velocity in \kms\ from the
systemic $z$.  Thus, like the BI, the AI has units of \kms.
The integral begins at the systemic redshift ($v=0$) at the wavelength of the
shortest-wavelength line of any multiplet involved
(e.g., 1548.20\,\AA\ for \civ\ or 2796.35\,\AA\ for \mgii).
The integral extends beyond the highest velocity intrinsic \civ\ system, but in
practice the maximum velocity will often be limited by confusion with \SIiv\ 
troughs at 25000\,\kms, so we have noted this in parentheses as a typical limit.
The value of $C'$ is unity in contiguous intervals of width 450\,\kms\ or
greater where the quantity in brackets is everywhere positive; otherwise $C'=0$.
The AI is therefore linear from 450\,\kms\ to $>$25,000\,\kms, the latter
being a lower limit when 
the \civ\ absorption appears to extend
beyond \SIiv\ but is confused with the \SIiv\ BAL troughs.  The AI can also be
extended to $<$450\,\kms\ using higher resolution spectra.
Finally, the presence of BAL troughs longward of the systemic redshift
(\S\ref{ROTSCAL}) can be indicated with a superscripted +, e.g., 3000$^+$.
If our explanation for this phenomenon is correct (\S\ref{IMPSCAL}),
the AI in such cases should still be measured from the systemic redshift.

In practice the statistical uncertainty is dwarfed by the systematic uncertainty
in the continuum placement.  Nonetheless we quote errors for the AI since they
are useful for marginal objects, and since detailed study of large BAL and
non-BAL quasar samples may eventually reduce the systematic uncertainty to
a level comparable to the statistical uncertainty.

IRAF and SM code for determining a continuum, normalizing a spectrum by it,
and calculating its AI or BI value is available, along with all spectra 
of unusual BAL quasars presented in this paper (\S\ref{SELECT}),
from the contributed data 
section of the SDSS Archive at {\tt http://archive.stsci.edu/sdss/}.
\end{appendix}

\footnotesize 


\clearpage

\begin{deluxetable}{rlrrlll}
\tablecaption{Transitions with $\lambda\geq1215$~\AA\ Seen to Date in BAL Quasars\label{t_lines}}
\tabletypesize{\scriptsize}
\tablehead{
\colhead{} & \colhead{}
& \multicolumn{2}{c}{Ionization Potentials for...} 
& \colhead{BAL} & \colhead{Frequency} \\[.2ex]
\colhead{Transition} & \colhead{Vacuum Wavelength (Multiplet)}
& \colhead{Creation\tablenotemark{a}} & \colhead{Destruction}
& \colhead{Subtype} & \colhead{in Subtype} 
}
\startdata
\lya	& 1215.67	  	& 0.0	   & 13.6 & Hi & common \\ 
\Nv	& 1238.82,1242.80 	& 77.5     & 97.9 & Hi & common \\ 
\SIii	& 1260.42~(UV4)	  	& 8.1	   & 16.3 & Lo & unusual \\ 
\SIii*	& 1264.74,1265.00~(UV4) & (0.05)   & 16.3 & Lo & rare \\ 
\oi	& 1302.17	  & 0.0	   & 13.6 & Lo & unusual \\ 
\SIii   & 1304.37~(UV3)	  & 8.1	   & 16.3 & Lo & unusual \\ 
\SIii*  & 1309.28~(UV3)	  & (0.05) & 16.3 & Lo & rare \\ 
\cii	& 1334.53~(UV1)	  & 11.3   & 24.4 & Lo & unusual \\ 
\cii* & 1335.66,1335.71~(UV1) & (0.01) & 24.4 & Lo & rare \\ 
\SIiv	& 1393.76,1402.77 & 33.5   & 45.1 & Hi & common \\ 
\SIii   & 1526.73~(UV2)	  & 8.1    & 16.3 & Lo & unusual \\
\SIii*  & 1533.43~(UV2)	  & (0.05) & 16.3 & Lo & rare \\ 
\civ	& 1548.20,1550.77    & 47.9 & 64.5 & Hi	& (defines) \\ 
\alii	& 1670.79	      & 6.0	& 18.8	& Lo	& unusual \\ 
\NIii(*) & 1709.6,1741.5,1751.9,1773.9~(UV5,4,3) & 7.6 & 18.2 & FeLo? & rare \\
P\,{\sc i} & 1774.95,1782.83,1787.65~(UV1) & 0 & 10.5 & FeLo? & rare \\ 
\SIii	& 1808.01~(UV1)		& 8.1	& 16.3	& Lo	& unusual \\ 
\aliii	& 1854.72,1862.79	& 18.8	& 28.4	& Lo	& common \\ 
\feiii  & 1895.46,1914.06,1926.30~(UV34) & (3.7) & 30.7 & FeLo & common \\ 
\mgi    & 2026.48               & 0.0   & 7.6   & FeLo? & rare   \\ 
\znii   & 2026.14,2062.66              &  9.4 & 18.0 & FeLo? & rare \\ 
\crii   & 2056.25,2062.23,2066.16     &  6.8 & 16.5 & FeLo? & rare \\ 
\feiii  & 2062.21,2068.90,2079.65~(UV48) & (5.1) & 30.7 & FeLo & unusual \\ 
\fei    & 2167.45~(UV21)	& 0.0	& 7.9	& FeLo?	& rare	 \\ 
\NIii    & 2225~(UV12,13) & 7.6	& 18.2	& FeLo? & rare \\ 
\coii\tablenotemark{b} & 2300~(UV9) & 7.9 & 17.1 & FeLo? & rare \\ 
\NIii    & 2300~(UV11)	& 7.6	& 18.2	& FeLo?	& rare	\\ 
\mnii	 & 2576.88,2594.50,2606.46 & 7.4 & 15.6 & FeLo? & rare \\ 
\crii    & 2680~(UV7,8)	  & (1.48-1.55)	& 16.5	& FeLo?	& rare	\\ 
\mgii	 & 2796.35,2803.53       & 7.6	& 15.0	& Lo	& (defines) \\ 
\mgi	 & 2852.96		& 0.0	& 7.6	& FeLo?	& rare	 \\ 
\crii    & 2860~(UV5)	& (1.48-1.55) & 16.5	& FeLo?	& rare	\\ 
\hei & 2945.97,3188.67,3889.74 & (19.8) & 24.6	& Lo	& rare	 \\ 
\Oiii 	& 3133.70		& (36.9) & 54.9	& Lo    & rare   \\ 
\caii	& 3934.78,3969.59	& 6.1	& 11.9	& Lo	& rare	 \\ 
\smallskip 
\nai	& 5891.58,5897.56	& 0.0	& 5.1	& Lo	& rare	 \\ 
\tableline 
\multicolumn{6}{c}{\feii\ Multiplets}\\
\tableline \\
\feii  & 1570~(UV44,45,46) & (0.15-0.25) & 16.2 & FeLo  & rare \\ 
\feii(*) & 1608.45~(UV8)    & 7.9	& 16.2	& Lo	& rare	\\ 
\feii  & 1710~(UV38)	    & (0.23-0.38) & 16.2  & FeLo & rare \\ 
\feii] & 1781.70~(UV67)        & (1.07) & 16.2  & FeLo  & rare \\ 
\feii  & 1785.27,1786.75,1787.996~(UV191) & (2.88) & 16.2 & FeLo & rare \\ 
\feii & 2151.8,2153.0,2177.7~(UV106) & (2.27-2.33) & 16.2 & FeLo & rare \\ 
\feii & 2164.34,2173.72~(UV79) & (1.66-1.69) & 16.2 & Lo & unusual \\ 
\feii(*) & 2249.88~(UV5)	& 7.9	    & 16.2 & Lo	& unusual \\ 
\feii(*) & 2260.78~(UV4)	& 7.9	   & 16.2 & Lo	& unusual \\ 
\feii	 & 2298.93~(UV133)	& (2.63)  & 16.2 & FeLo & rare \\ 
\feii(*) & 2344.21~(UV3)	 & 7.9	 & 16.2	& Lo	& unusual \\ 
\feii(*) & 2374.46,2382.77~(UV2) & 7.9	& 16.2	& Lo	& unusual \\ 
\feii & 2382.90,2388.39~(UV117) & (2.51-2.57) & 16.2 & FeLo & rare \\ 
\feii* & 2400,2600~(UV2,1) & (0.05-0.12) & 16.2 & FeLo & (defines) \\ 
\feii & 2420~(UV35,36) & (0.23-0.39) & 16.2 & FeLo & rare \\ 
\feii	& 2460~(UV209)	& (3.14-3.22) & 16.2 & FeLo & rare \\ 
\feii	& 2580~(UV64)        & (0.98-1.09) & 16.2 & FeLo  & rare \\ 
\feii(*) & 2586.65,2600.17~(UV1) & 7.9 & 16.2 & Lo & unusual \\ 
\feii	& 2750~(UV62,UV63)   & (0.98-1.09) & 16.2 & FeLo & common \\ 
\feii	& 2880~(UV61)        & (0.98-1.07) & 16.2 & FeLo  & rare \\ 
\feii	& 2950~(UV60)        & (0.98-1.09) & 16.2 & FeLo  & rare \\ 
\feii	& 2985~(UV78)        & (1.66-1.72) & 16.2 & FeLo  & rare \\ 
\feii	& 3180~(Opt7)        & (1.66-1.72) & 16.2 & FeLo  & rare \\ 
\feii	& 3200~(Opt6)        & (1.66-1.72) & 16.2 & FeLo  & rare \\ 
\enddata
\tablenotetext{a}{For transitions from excited levels or excited terms,
instead of the Ionization Potential for Creation we list, in parentheses,
the range of Excitation Potentials for all transitions in the multiplet,
relative to the ground level of the ion.}
\tablenotetext{b}{\markcite{wcp95}{Wampler} {et~al.} (1995) mention the presence of other (blended)
multiplets of \coii\ in the spectrum of Q~0059$-$2735, but only UV9 by name.}
\tablecomments{\tiny We list only firmly identified transitions seen in one or
more BAL quasars 
(in papers up to, but not including, \markcite{dek02}{de Kool} {et~al.} 2002).
Thus, for example, we do not list \crii\,UV6 or \feii\,UV144-149,158-165
absorption even though at least some of those multiplets are certainly present
in SDSS~1125+0029 because we cannot verify any individual lines from those
multiplets in our low-resolution spectra.
Under Transitions, 
`*' refers to excited level absorption from ground term multiplets, 
and `(*)' denotes ground term multiplets 
where absorption from both ground and excited levels has been seen 
(however, the BAL Subtype refers only to the ground level lines).
Ionization Potentials are given in eV.
Wavelengths given only to the nearest \AA\ refer to multiplet absorption.
Under `BAL Subtype', 
FeLo? means that the transitions have only been seen in FeLoBALs to date,
but that there is no {\em a priori} reason they could not be seen in LoBALs.
Transitions seen in HiBALs are also seen in the other two subtypes,
and transitions seen in LoBALs are also seen in FeLoBALs.
The `Frequency in Subtype' entries (common, unusual, or rare) refer to the
frequency within the listed BAL subtype.
`Defines' mean that the presence of that transition defines that BAL subtype:
\civ\ for HiBALs, \mgii\ for LoBALs, and \feii* (excited \feii) for FeLoBALs.
Note that absorption from any excited level or term of \feii\ qualifies the
object as a FeLoBAL; we explicitly list only the UV1 and UV2 multiplets as
\feii* simply because they contain the strongest excited-level transitions.
}
\end{deluxetable}

%

\begin{deluxetable}{lclllccrrcccccc}
\setlength{\tabcolsep}{0.02in}
\tablecaption{SDSS Unusual BAL Quasars\label{t_info}}
\rotate
\tabletypesize{\scriptsize}
\tablewidth{620.00000pt}
\tablehead{
\colhead{J2000} & \colhead{Redshift} & \colhead{BAL} &
\colhead{$BI$,} & \colhead{$AI\pm\sigma_{AI}$,} &
\colhead{} & \colhead{Target} & \colhead{20cm,} & \colhead{} &
\colhead{} &
\colhead{} &
\colhead{} &
\colhead{} &
\colhead{} &
\colhead{\tiny Galactic}\\[.2ex]
\colhead{Coordinates} & \colhead{$z$$\pm$$\sigma_z$} & \colhead{Type} &
\colhead{\kms} & \colhead{\kms} &
\colhead{\tiny EDR} & \colhead{Code} & \colhead{mJy} & \colhead{$M_{i^*}$} &
\colhead{$u^*\pm\sigma_{u^*}$} &
\colhead{$g^*\pm\sigma_{g^*}$} &
\colhead{$r^*\pm\sigma_{r^*}$} &
\colhead{$i^*\pm\sigma_{i^*}$} & 
\colhead{$z^*\pm\sigma_{z^*}$} &
\colhead{\tiny $E$$($$B$$-$$V$$)$} 
}
\startdata
010540.75$-$003314.0 & 1.1788$\pm$0.0001 & ? & 0 & 0 & Y & 11011 & 4.59 & $-$26.59 & 20.42$\pm$0.06 & 19.29$\pm$0.02 & 18.01$\pm$0.02 & 17.69$\pm$0.01 & 17.41$\pm$0.03 & 0.034 \\ 
012702.52$+$011412.5 & 1.1571$\pm$0.0002 & red,\hei & 0 & 50$\pm$1 & Y & 01000 & 1.06 & $-$26.84 & 20.40$\pm$0.07 & 18.92$\pm$0.02 & 17.87$\pm$0.02 & 17.38$\pm$0.02 & 17.08$\pm$0.03 & 0.027 \\ 
014905.28$-$011404.9 & 2.10$\pm$0.01 & \feiii? & 56 & 2930$\pm$5 & Y & 10000 & $<$1.05 & $-$25.76 & 23.14$\pm$0.96 & 21.29$\pm$0.05 & 20.13$\pm$0.03 & 19.59$\pm$0.02 & 18.78$\pm$0.04 & 0.038 \\ 
\smallskip
030000.57$+$004828.0 & 0.89191$\pm$0.00005 & ot & $>$11520 & $>$15347$\pm$3 & Y & 10000 & $<$0.93 & $-$27.23 & 20.04$\pm$0.04 & 19.48$\pm$0.02 & 16.79$\pm$0.01 & 16.57$\pm$0.01 & 16.19$\pm$0.02 & 0.089 \\ 
031856.62$-$060037.7 & 1.9668$\pm$0.0015 & red,Fe & 4017 & 6294$\pm$6 & N & 10000 & \nodata & $-$28.15 & 21.66$\pm$0.18 & 19.31$\pm$0.02 & 17.60$\pm$0.02 & 17.23$\pm$0.01 & 16.81$\pm$0.02 & 0.050 \\ 
033810.85$+$005617.6 & 1.627$\pm$0.002 & red,Fe & 360 & 476$\pm$6~(\mgii) & Y & 10000 & \nodata & $-$26.74 & 20.72$\pm$0.08 & 19.65$\pm$0.02 & 18.68$\pm$0.01 & 18.36$\pm$0.01 & 18.43$\pm$0.03 & 0.104 \\ 
034258.00$+$004539.1 & 2.418$\pm$0.001 & red,Fe & 0 & 437$\pm$4 & N & \nodata & \nodata & $-$24.63 & 23.84$\pm$0.84 & 24.02$\pm$0.36 & 22.18$\pm$0.15 & 21.33$\pm$0.10 & 19.39$\pm$0.07 & 0.123 \\ 
\smallskip
043742.81$-$004517.6 & 2.8183$\pm$0.0009 & ot & 16330 & 20767$\pm$3 & N & \nodata & $<$0.071\tablenotemark{a} & $-$26.40 & 24.23$\pm$0.90 & 22.77$\pm$0.15 & 21.45$\pm$0.10 & 19.69$\pm$0.05 & 20.73$\pm$0.15 & 0.034 \\ 
081024.75$+$480615.5 & 2.240$\pm$0.005 & \feiii? & 5670 & 6699$\pm$5 & N & 10000 & $<$1.45 & $-$26.93 & 21.71$\pm$0.15 & 20.44$\pm$0.03 & 19.51$\pm$0.02 & 18.72$\pm$0.01 & 18.37$\pm$0.03 & 0.047 \\ 
081948.91$+$420930.0 & 1.9258$\pm$0.0006 & ot & $>$13050 & $>$17151$\pm$1~(\mgii) & N & \nodata & $<$0.95 & $-$23.53 & 23.60$\pm$0.29 & 23.75$\pm$0.22 & 22.86$\pm$0.16 & 21.82$\pm$0.12 & 19.61$\pm$0.08 & 0.057 \\ 
083413.91$+$511214.6 & 2.3907$\pm$0.0002 & red & 760 & 4084$\pm$12 & N & 11010 & 2.23 & $-$27.01 & 23.61$\pm$1.03 & 22.07$\pm$0.12 & 19.80$\pm$0.03 & 18.75$\pm$0.02 & 18.08$\pm$0.04 & 0.035 \\ 
\smallskip
094736.70$+$620504.6 & 2.1254$\pm$0.0006 & red & 0 & 1630$\pm$80~(\mgii) & N & 10010 & \nodata & $-$26.70 & 24.77$\pm$0.88 & 21.74$\pm$0.05 & 19.77$\pm$0.02 & 18.80$\pm$0.01 & 18.27$\pm$0.03 & 0.028 \\ 
112526.13$+$002901.3 & 0.8654$\pm$0.0001 & mnt,\hei & 0 & 491$^+$$\pm$1 & Y & 10011 & $<$0.99 & $-$25.54 & 19.43$\pm$0.03 & 19.29$\pm$0.05 & 18.35$\pm$0.03 & 18.07$\pm$0.03 & 17.75$\pm$0.03 & 0.031 \\ 
112828.31$+$011337.9 & 0.8931$\pm$0.0001 & mnt & 0 & 696$^+$$\pm$1 & S & 10010 & $<$0.98 & $-$25.30 & 19.89$\pm$0.03 & 19.60$\pm$0.02 & 18.61$\pm$0.03 & 18.38$\pm$0.02 & 17.95$\pm$0.04 & 0.034 \\ 
115436.60$+$030006.4 & 1.458$\pm$0.008 & ot & $>$13840 & $>$13840$\pm$10~(\mgii) & N & 10010 & $<$0.99 & $-$26.97 & 21.90$\pm$0.14 & 20.26$\pm$0.10 & 20.28$\pm$0.07 & 17.74$\pm$0.06 & 17.48$\pm$0.04 & 0.027 \\ 
\smallskip
121441.43$-$000137.9 & 1.0448$\pm$0.0004 & \feiii & 2880 & 3149$\pm$4~(\mgii) & Y & 01011 & 1.79 & $-$25.19 & 20.53$\pm$0.05 & 19.41$\pm$0.01 & 18.83$\pm$0.01 & 18.80$\pm$0.01 & 18.57$\pm$0.03 & 0.020 \\ 
132444.11$-$021746.6 & 2.264$\pm$0.001 & red & 0 & 1360$\pm$280~(\mgii) & N & 10000 & $<$1.01 & $-$26.02 & 24.15$\pm$0.46 & 22.78$\pm$0.13 & 20.59$\pm$0.04 & 19.63$\pm$0.03 & 19.18$\pm$0.07 & 0.039 \\ 
145333.01$+$002943.7 & 1.297$\pm$0.001 & red,\hei & 0 & 253$-$477~(\mgii) & Y & 10000 & $<$0.054\tablenotemark{a} & $-$24.95 & 23.53$\pm$0.42 & 21.46$\pm$0.03 & 19.82$\pm$0.02 & 19.54$\pm$0.01 & 18.99$\pm$0.03 & 0.042 \\
145603.08$+$011445.5 & 2.363$\pm$0.008 & red & 1350 & 4525$\pm$10 & S & 11010 & 8.56 & $-$26.21 & 24.77$\pm$1.00 & 22.48$\pm$0.09 & 20.55$\pm$0.04 & 19.54$\pm$0.03 & 18.62$\pm$0.04 & 0.045 \\ 
\smallskip
172341.10$+$555340.5 & 2.1127$\pm$0.0006 & mnt & 4350 & 7120$\pm$20 & Y & 10010 & $<$0.93 & $-$26.96 & 20.53$\pm$0.05 & 20.13$\pm$0.02 & 18.36$\pm$0.02 & 18.54$\pm$0.02 & 17.45$\pm$0.02 & 0.035 \\ 
173049.11$+$585059.5 & 2.035$\pm$0.005 & ot,\hei? & $>$10900 & $>$15500$\pm$100~(\mgii) & Y & 00010 & \nodata & $-$24.54 & 22.72$\pm$0.39 & 23.72$\pm$0.32 & 21.39$\pm$0.07 & 20.82$\pm$0.06 & 18.41$\pm$0.04 & 0.030 \\ 
220445.26$+$003142.0 & 1.3531$\pm$0.0009 & ? & 0 & 0 & N & 11010 & 2.94 & $-$27.83 & 20.04$\pm$0.08 & 18.78$\pm$0.01 & 17.39$\pm$0.01 & 16.78$\pm$0.01 & 16.64$\pm$0.01 & 0.055 \\ 
221511.94$-$004549.9 & 1.4755$\pm$0.0002 & \feiii & 1160 & 1293$\pm$3~(\mgii) & N & 10001 & $<$0.94 & $-$28.42 & 18.38$\pm$0.02 & 17.36$\pm$0.04 & 16.53$\pm$0.03 & 16.47$\pm$0.03 & 16.58$\pm$0.10 & 0.102 \\ 
\enddata
\tablenotetext{a}{For these two objects, instead of FIRST 20\,cm flux densities
we give VLA 6\,cm flux density limits provided by C. Carilli and M. Rupen.}
\tablecomments{Redshift determinations are discussed in the text
for each object.  
Under BAL Type, 
\feiii? or \feiii\ means it is a candidate or confirmed \feiii$>$\feii\ FeLoBAL,
respectively (\S\ref{FE3});
\hei\ means it shows neutral helium absorption;
mnt means a many-narrow-trough FeLoBAL (\S\ref{SCALLOPED});
ot means an overlapping-trough FeLoBAL, all of which also show \MgI\ absorption
(\S\ref{ABRUPT});
? means it is a mystery object which may be a BAL quasar (\S\ref{GRADUAL});
and red or red,Fe means a heavily reddened LoBAL or FeLoBAL, respectively
(\S\ref{REDDENED}).
See Appendix \ref{BI} for details on the balnicity index (BI) and intrinsic
absorption index (AI).
Note that their typical systematic error is at least 10\%.
Under the column heading EDR (Early Data Release),
Y means both the spectrum and photometric data are in the
primary EDR area, S means the spectrum was released in the EDR but the object
is in the secondary rather than the primary EDR area \markcite{sdss85}({Stoughton} {et~al.} 2002),
and N means neither the spectrum nor the photometry are available in the EDR.
All Y objects except SDSS~1453+0029 are included in the EDR quasar sample
\markcite{sdssedrq}({Schneider} {et~al.} 2002).
The Target Code is a 5-digit number that shows whether the object was
targeted by Quasar, FIRST, ROSAT, Serendipity, and star selection, in that
order: a digit of 0 means not targeted by that method, while 1 means targeted.
The 20\,cm column gives the 20\,cm flux density in millijanskies (mJy) from the
FIRST survey \markcite{bwh95}({Becker} {et~al.} 1995). In that column, `\nodata' means that the object lies
outside the current FIRST survey area; none of those objects are detected by the
NVSS \markcite{con98}({Condon} {et~al.} 1998).
The $M_{i^*}$ are computed using \ho=50, $\Omega_M$=1, $\Omega_{\Lambda}$=0
for comparison with the quasar literature.
Galactic \ebv\ values are from \markcite{sfd98}{Schlegel}, {Finkbeiner}, \&  {Davis} (1998) and are accurate to $\pm$15\%.
}
\end{deluxetable}


\clearpage

\end{document}